\documentclass[twocolumn]{aastex62}
\usepackage[T1]{fontenc}
\usepackage{ae,aecompl}
\usepackage{epsfig,graphics}	
\usepackage{amsmath}
\usepackage{amssymb}	

\usepackage{color}
\usepackage{multirow}
\usepackage{booktabs}
\usepackage{tabularx}
\usepackage{varioref}
\usepackage{ulem}

\usepackage{natbib}
\pdfoutput=1

 \newcommand{\kms}{\,km\,s$^{-1}$} 

 \definecolor{darkspringgreen}{rgb}{0.09, 0.45, 0.27}
 \definecolor{amber(sae/ece)}{rgb}{1.0, 0.49, 0.0}

 \def\Fig{\mbox{Figure~}}
 \def\Figs{\mbox{Figures~}}
 \def\Tab{\mbox{Table~}}
 
 \def\Sec{\mbox{Section~}}

 \def\UCMG{\mbox{{\textsc{ucmg}}}}
 \def\UCMGs{\mbox{{\textsc{ucmg}}s}}

 \def\KiDSfull{\mbox{KiDS\_{\textsc{full}}}}
 \def\KiDSspec{\mbox{{KiDS\_{\textsc{spec}}}}}
 \def\UCMGfull{\mbox{{\textsc {ucmg\_full}}}}
 \def\UCMGPhotSpec{\mbox{{\textsc{ucmg\_phot\_spec}}}}
 \def\UCMGSpec{\mbox{{\textsc{ucmg\_spec}}}}

 \def\Mauto{\mbox{{\tt MAG\_AUTO}}}
 \def\mst{\mbox{$M_{\star}$}}
 \def\Re{\mbox{$R_{\rm e}$}}
 \def\magapsix{\mbox{{\tt MAGAP\_6}}}

 \def\MErrautor{\mbox{{\tt MAGERR\_AUTO\_r}}}
 \def\Zsun{\mbox{$Z_{\rm \odot}$}}
 \def\Msun{\mbox{$M_\odot$}}
 \def\SN{\mbox{$S/N$}}
 \def\zp{\mbox{$z_{\rm phot}$}}
 \def\zs{\mbox{$z_{\rm spec}$}}

 \def\lsim{\mathrel{\rlap{\lower3.5pt\hbox{\hskip0.5pt$\sim$}}
   \raise0.5pt\hbox{$<$}}}      
 \def\gsim{~\rlap{$>$}{\lower 1.0ex\hbox{$\sim$}}}
 \def\Re{\mbox{$R_{\rm e}$}}
 \def\mst{\mbox{$M_{\star}$}}

 \def\CF{\mbox{$\mathcal{C}_F$}}
 \def\IF{\mbox{$\mathcal{I}_F$}}

 \def\Zsun{\mbox{$Z_{\rm \odot}$}}
 \def\lephare{\mbox{\textsc{le phare}}}
 \def\twodphot{\mbox{\textsc{2dphot}}}

 \submitjournal{ApJ}

 \shorttitle{ {\textsc{ucmg}s} in KiDs}

 \shortauthors{Scognamiglio et al.}

\submitjournal{ApJ}

\begin{document}

\title{Building the largest spectroscopic sample of ultra-compact massive galaxies with the Kilo Degree Survey}

\correspondingauthor{Nicola~R.~Napolitano}
\email{napolitano@mail.sysu.edu.cn, dianasco@astro.uni-bonn.de}

\author{Diana~Scognamiglio }
\affiliation{INAF -- Osservatorio Astronomico di Capodimonte, Salita Moiariello 16, 80131 - Napoli, Italy} \affiliation{Argelander-Institut f\"ur Astronomie, Auf dem H\"ugel 71, D-53121 - Bonn, Germany}

\author{Crescenzo~Tortora}
\affil{INAF - Osservatorio Astronomico di Arcetri, L.go E. Fermi 5, 50125 - Firenze, Italy}

\author{Marilena~Spavone}
\affil{INAF -- Osservatorio Astronomico di Capodimonte, Salita Moiariello 16, 80131 - Napoli, Italy}

\author{Chiara~Spiniello}
\affil{INAF -- Osservatorio Astronomico di Capodimonte, Salita Moiariello 16, 80131 - Napoli, Italy}
\affil{European Southern Observatory, Karl-Schwarschild-Str. 2, 85748 - Garching, Germany}

\author{Nicola~R.~Napolitano}
\affil{School of Physics and Astronomy, Sun Yat-sen University Zhuhai Campus, Daxue Road 2, 519082 - \\ Tangjia, Zhuhai, Guangdong, P.R. China}
\affil{INAF -- Osservatorio Astronomico di Capodimonte, Salita Moiariello 16, 80131 - Napoli, Italy}

\author{Giuseppe~D`Ago}
\affil{Instituto de Astrof\'isica
Pontificia Universidad Cat\'olica de Chile, Avenida Vicu\~na Mackenna, 4860 - Santiago, Chile}

\author{Francesco~La~Barbera}
\affil{INAF -- Osservatorio Astronomico di Capodimonte, Salita Moiariello 16, 80131 - Napoli, Italy}

\author{Fedor~Getman}
\affil{INAF -- Osservatorio Astronomico di Capodimonte, Salita Moiariello 16, 80131 - Napoli, Italy}

\author{Nivya~Roy}
\affil{School of Physics and Astronomy, Sun Yat-sen University Zhuhai Campus, Daxue Road 2, 519082 - \\ Tangjia, Zhuhai, Guangdong, P.R. China}

\author{Maria~Angela~Raj}
\affil{INAF -- Osservatorio Astronomico di Capodimonte, Salita Moiariello 16, 80131 - Napoli, Italy}

\author{Mario~Radovich}
\affil{INAF -- Osservatorio Astronomico di Padova, Vicolo Osservatorio 5, 35122 - Padova, Italy}

\author{Massimo~Brescia}
\affil{INAF -- Osservatorio Astronomico di Capodimonte, Salita Moiariello 16, 80131 - Napoli, Italy}

\author{Stefano~Cavuoti}
\affil{INAF -- Osservatorio Astronomico di Capodimonte, Salita Moiariello 16, 80131 - Napoli, Italy}
\affil{Dipartimento di Scienze Fisiche, Universit\`{a} di Napoli Federico II, Compl. Univ. Monte S. Angelo, 80126 - Napoli, Italy}

\author{L\'{e}on~V.E.~Koopmans}
\affil{Kapteyn Astronomical Institute, University of Groningen, P.O. Box 800, 9700  AV - Groningen, the Netherlands}

\author{Konrad~H.~Kuijken}
\affil{Leiden Observatory, Leiden University, P.O. Box 9513, 2300 RA - Leiden, the Netherlands}

\author{Giuseppe~Longo}
\affil{Dipartimento di Scienze Fisiche, Universit\`{a} di Napoli Federico II, Compl. Univ. Monte S. Angelo, 80126 - Napoli, Italy}

\author{Carlo~E.~Petrillo}
\affil{Kapteyn Astronomical Institute, University of Groningen, P.O. Box 800, 9700  AV - Groningen, the Netherlands}

\begin{abstract}
Ultra-compact massive galaxies \UCMGs, 
i.e. galaxies with stellar masses $\mst > 8 \times 10^{10} M_{\odot}$
and effective radii $\Re < 1.5\,\rm kpc$, are very rare systems, in particular at low 
and intermediate 
redshifts. Their origin as well as their number density across cosmic time are still under scrutiny, especially because of the paucity of spectroscopically confirmed samples.
We have started a systematic census of \UCMG\ candidates within the ESO Kilo Degree Survey, together with a large spectroscopic follow-up campaign to build the largest possible sample of confirmed \UCMGs. This is the third paper of the series and the second based on the spectroscopic follow-up program.
Here, we present photometrical and structural parameters of 33 new candidates at redshifts $0.15 \lesssim z \lesssim 0.5$ and confirm 19 of them as \UCMGs, based on their nominal spectroscopically inferred \mst\ and \Re. This corresponds to a success rate of $\sim 58\%$, nicely consistent with our previous findings. 
The addition of these 19 newly confirmed 
objects, allows us to fully assess the systematics on the system selection, and finally reduce the number density uncertainties. 
Moreover, putting together the results from our current and past observational campaigns and some literature data, we build the largest sample of \UCMGs\ ever collected, comprising $92$ spectroscopically confirmed objects at $0.1 \lesssim z \lesssim 0.5$. 

This number raises to 116, allowing for a 3$\sigma$ tolerance on the \mst\ and \Re\ thresholds for the \UCMG\ definition. 
For all these galaxies we have estimated the velocity dispersion values at the effective radii which have been used to derive a preliminary mass--velocity dispersion correlation.
\end{abstract}

\keywords{galaxies: evolution - galaxies: general - galaxies: elliptical and lenticular,
cD - galaxies: structure}

\section{Introduction} \label{sec:intro}
The discovery that massive, quiescent galaxies at redshift $z > 2$ are extremely compact with respect to their local counterparts \citep{Daddi+05, Trujillo+06, vanDokkum+10, Damjanov+09, Damjanov+11} has opened a new line of investigation within the context of galaxy formation and evolution. In particular, the strong galaxy size growth \citep{Daddi+05, Trujillo+06} needed to account for the difference in compactness from high- to low-$z$, finds the best explanation in the so-called two-phase formation model \citep{Oser 2010}. 
First of all, massive and very compact gas-rich disky objects are created due to dissipative inflows of gas. These so-called ``blue nuggets'' form stars ``in situ'' at high rate, and this causes 
a gradual stellar and halo mass growth \citep{Dekel_Burkert14}. Subsequently, the star formation in the central region quenches and the blue nuggets quickly (and passively) evolve into compact ``red nuggets''.

In many cases, the masses of these high-$z$ red nuggets are similar to those of local giant elliptical galaxies, which indicates that almost all the mass is assembled during this first formation phase. However, their sizes are only about a fifth of the size of local ellipticals of similar mass (\citealt{Werner+18}). Thus, during the second phase of this scenario, at lower redshifts, red nuggets undergo dry mergers with lower mass galaxies growing in size (but only slightly increasing their masses) and becoming, over billions of years, present-day ETGs. 

Nevertheless, given the stochastic nature of mergers, a small fraction of the red nuggets slips through the cosmic time untouched and without accreting any stars from satellites and mergers: the so-called ``relics'' \citep{Ferre-Mateu+17}. 
These galaxies have assembled early on in time and have somehow missed completely the size growth. They are therefore supposedly made of only ``in situ'' stellar population and as such they provide a unique opportunity to track the formation of this specific galaxy stellar component, which is mixed with the accreted one in normal massive ETGs.

Indeed, very massive, extremely compact systems 
have been already found at intermediate to low redshifts, also including the local Universe (\citealt{Trujillo+09_superdense,Trujillo+14}; \citealt{Taylor+10_compacts}; \citealt{Valentinuzzi+10_WINGS}; \citealt{Shih_Stockton11};
\citealt{Lasker+13_IMF_compact}; \citealt{Poggianti+13_evol, Poggianti+13_low_z}; \citealt{Hsu+14_compacts}; \citealt{Stockton+14_compacts}; \citealt{Damjanov+15_compacts,Damjanov+15_env_compacts}; \citealt{Ferre-Mateu+15}; \citealt{Saulder+15_compacts}; \citealt{Stringer+15_compacts}; \citealt{Yildirim+15};
\citealt{Wellons+15_lower_z};  \citealt{Gargiulo+17_dense};  \citealt{Tortora+16_compacts_KiDS, Tortora+18_UCMGs};
\citealt{Charbonnier+17_compact_galaxies}; \citealt{Beasley+18};
\citealt{Buitrago+18_compacts}).  
Ultra-Compact Massive Galaxies (\UCMGs\ hereafter), defined here as objects  with stellar mass $M_{*} > 8 \times 10^{10} M_{\odot}$ and effective radius $R_{\text{e}} < 1.5$ kpc (although sometimes other stellar mass and effective radius ranges are adopted, see \Sec\ \ref{sec:sample}) are the best relics candidates.

The precise abundance  of relics, and even more generally of \UCMGs, without any age restriction, at low redshifts is an open issue. In fact, at $z\le 0.5$, a strong disagreement exists between simulations and observations and among observations themselves on the number density of \UCMGs\ and its redshift evolution. 
From a theoretical point of view, simulations predict that the fraction of objects that survive without undergoing any significant transformation since $z \sim 2$ is about $1 - 10\%$ \citep{Hopkins+09_DELGN_IV, Quilis_Trujillo13}, and at the lowest redshifts (i.e., $ z \lesssim 0.2$), they predict densities of relics of $10^{-7} - 10^{-5}$ Mpc$^{-3}$. This is in agreement with the lower limit given by NGC 1277, the first discovered local ($z \sim 0.02$) compact galaxy with old stellar population, which is the first prototype of local ``relic" of high-$z$ nuggets \citep{Trujillo+14}, and the most updated estimate of $6 \times 10^{-7} \, \rm Mpc^{-3}$ set by \cite{Ferre-Mateu+17}, who report the discovery of two new confirmed, local ``relics".  
In the near-by Universe, large sky surveys as the Sloan Digital Sky Survey (SDSS\footnote{\url{https://www.sdss.org/}}) show a sharp decline in compact galaxy number density of more than three orders of magnitude below the high-redshift values \citep{Trujillo+09_superdense, Taylor+10_compacts}. In contrast,  \cite{ Poggianti+13_evol, Poggianti+13_low_z} suggest that the abundance of low-redshift compact systems might be even comparable with the number density at high redshift. Moreover, data from the WINGS survey of near-by clusters \citep{Fasano+2006, Valentinuzzi+10_WINGS} estimate, at $z\sim 0$, a number density of two orders of magnitude above the estimates based on the SDSS dataset.

Since the situation in the local Universe is very complex and different studies report contrasting results, it is crucial to increase the \UCMG\ number statistics in the range $0.1\lesssim z \lesssim 0.5$, where these systems should be more common. In recent years different works have contributed to the census of \UCMGs\ in wide-field surveys at these redshifts (\citealt{Tortora+16_compacts_KiDS,Tortora+18_UCMGs}; \citealt{Charbonnier+17_compact_galaxies}; \citealt{Buitrago+18_compacts}). In particular, within the Kilo Degree Survey (KiDS, see \Sec\ \ref{sec:sample}) collaboration, we have undertaken a systematic search for \UCMGs\ in the intermediate redshift range with the aim of building a large spectroscopically-confirmed sample. 
In the first paper of the series \citep[hereafter T16]{Tortora+16_compacts_KiDS}, we collected a sample of $\lsim 100$ candidates in the first $\sim$ 156 deg$^2$ of KiDS (corresponding to an effective area of $\sim 107$ deg$^2$, after masking). 
In the second paper (\citealt[hereafter T18]{Tortora+18_UCMGs}) we updated the analysis and extended the study to the third KiDS Data Release (KiDS--DR3). We have collected a sample of $\sim 1000$ candidates, building the largest sample of \UCMG\ candidates at $z < 0.5$, assembled to date over the largest sky area (333 deg$^2$). 

It is worth noticing that most of all the previously published findings on these peculiar objects are based on photometric samples. However, after identification of the candidates, spectroscopic validation is necessary to obtain precise spectroscopic redshifts and confirm the compactness of the systems. Thus, in T18 we presented the first of such spectroscopic validation, with data obtained at Telescopio Nazionale Galileo (TNG) and at the New Technology Telescope (NTT).

In this third paper of the series we therefore continue the work started in T18 to spectroscopically validate \UCMGs\ and derive their ``true'' \footnote{With the word ``true'', we mean here the number density obtained with a spectroscopically confirmed sample.} number densities at intermediate redshifts. 
In particular, we present here spectroscopic observations for 33 new KiDS \UCMG\ candidates and add to these  all the spectroscopic confirmed \UCMGs\ publicly available in the literature to update the \UCMG\ number density distribution, already presented in T18, at redshift $0.15<z<0.5$. 
Finally, we also obtain and present here the velocity dispersion measurements ($\sigma$) for the new 33 \UCMGs\ and for the 28 \UCMGs\ from T18.  
Finally, we present a preliminary correlation between stellar mass and velocity dispersion of these rare objects, with the aim of starting to fully characterize the properties of these systems. 

This paper represents a further step forward to our final goal, which is to unequivocally prove that a fraction of the red and dead nuggets, which formed at $z > 2$, evolved undisturbed and passively into local ``relics''. 
In particular, to be classified as such, the objects have to 1) be spectroscopically validated \UCMGs, 2) have a very old stellar populations (e.g., assuming a formation redshift $\zp \gsim\ 2$, the stellar population age needs to be $t \gsim\ 10$ Gyrs).
Since we do not derive stellar ages, this paper makes significant progress only on the first part of the full story, as not all the confirmed \UCMGs\ satisfy a stringent criterion on its stellar age. 
We are confident that most of our confirmed \UCMGs\ will likely be old, as we showed in T18 that most of the candidates presented very red optical and near-infrared colours. Moreover, in the spectra we present here (see \Sec \ref{sec:observations_and_analysis}), we find spectral features typical of passive stellar population. 
However, only with higher resolution and high signal-to-noise spectra, which would allow us to perform an in-depth stellar population analysis, it will be possible to really disentangle relics from younger \UCMGs. 
The detailed stellar population analysis is also particularly important as a fraction of our \UCMGs\ also shows some hint of recent star formation or of younger stellar population. This has been already seen in other samples too (\citealt{Trujillo+09_superdense}; \citealt{Ferre-Mateu+12}; \citealt{Poggianti+13_evol}; \citealt{Damjanov+15_compacts,Damjanov+15_env_compacts}; \citealt{Buitrago+18_compacts}),  
but it is not necessarily in contrast with the predictions from galaxy assembly simulations (see e.g. \citealt{Wellons+15_z2}).
In fact, they find that ultra-compact systems host accretion events, but still keep their bulk of stellar population old and the compact structure almost unaltered. Hence, higher quality spectroscopical data will be mandatory to perform a multi-population analysis and possibly confirm also this scenario. 

\medskip
The layout of the paper is as follows. In \Sec\ref{sec:sample}, we briefly describe the KiDS sample of high signal-to-noise ratio (\SN) galaxies, the subsample of our photometrically selected \UCMGs, the objects we followed-up spectroscopically, and the impact of the selection criteria we use. 
In \Sec\ref{sec:observations_and_analysis} we give an overview on observations and data reduction, and we discuss the spectroscopic redshift and velocity dispersion calculation procedures. 
In \Sec\ref{sec:results}, we discuss the main results, i.e. the number density as a function of redshift and the the impact of systematics on these number densities. We also derive a tentative relation between the stellar mass and the velocity dispersion at the effective radius of our sample of \UCMGs, compared with a sample of normal-sized elliptical galaxies at similar masses and redshifts.
Finally, in \Sec\ref{sec:conclusions}, we summarize our findings and discuss future perspectives. 
In the Appendix we report the final validated \UCMGs\ catalog where some redshifts come from our spectroscopic program and others from the literature. For all galaxies we give structural parameters in the $g,~r,~i,$ bands and the $u,~g,~r,~i,$ aperture photometry from KiDS. 

Throughout the paper, we assume $H_{0}$ = 70 km s$^{-1}$ Mpc$^{-1}$, $\Omega_{\text{m}} = 0.3$, and $\Omega_{\Lambda} = 0.7$ (\citealt{Komatsu+11_WMAP7}).

\section{Sample definition}\label{sec:sample}

KiDS is one of the ESO public wide-area surveys (1350 deg$^2$ in total) being carried out with the VLT Survey Telescope (VST; \citealt{Capaccioli_Schipani11}). It provides imaging data with unique image quality (pixel scale of $0.21$/pixel and a median $r$-band seeing of $0.65''$) and baseline ($ugri$ in optical + $ZYJHK$ if combined to VIKING \citep{Edge+14_VIKING-DR1, Wright+18}). These features make the data very suitable for measuring structural parameters of galaxies, including very compact systems, up to $z \sim$ 0.5 (\citealt{Roy+18}; T16; T18). Both image quality and baseline are very important for the selection of \UCMGs\ as they allow us to mitigate systematics that might have plagued previous analyses from the ground.

As baseline sample of our search, we use the data included in the third Data Release of KiDS (KiDS--DR3) presented in \cite{deJong+17_KiDS_DR3}, consisting of 440 survey tiles ($\approx$ 333 deg$^2$, after masking). The galaxy data sample is described in the next \Sec\ref{subsec:selectionprocedure}. 

\subsection{Galaxy data sample}
\label{subsec:selectionprocedure}
From the KiDS multi-band source catalog (\citealt{deJong+15_KiDS_paperI, deJong+17_KiDS_DR3}), we built a catalog of $\sim 5$ million galaxies (\citealt{LaBarbera_08_2DPHOT}) within KiDS--DR3, using SExtractor (\citealt{Bertin_Arnouts96_SEx}). 
Since we mainly follow the same selection procedure of T16 and T18, we refer the interested reader to these papers for more general details. Here we only list relevant physical quantities for the galaxies in the catalog, explaining how we obtain them and highlighting the novelty of the set-up we use in the stellar mass calculation: 

\begin{itemize}
\item {\it Integrated optical photometry}. 
We use aperture magnitudes $\texttt{MAGAP\_6}$, measured within circular apertures of $6''$ diameter, Kron-like $\texttt{MAG\_AUTO}$ as the total magnitude and Gaussian Aperture and PSF (GAaP) magnitudes, $\texttt{MAG\_GAaP}$ (\citealt{deJong+17_KiDS_DR3}) in each of the four optical bands ($ugri$).

\item {\it Structural parameters}. Surface photometry is performed using the \twodphot\ environment (\citealt{LaBarbera_08_2DPHOT}), which fits galaxy images with a 2D S\'ersic model. The model also includes a constant background and assumes elliptical isophotes. In order to take the galaxies best-fitted and remove those systems with a clear sign of spiral arms, we put a threshold on the goodness of the fit, only selecting $\chi^{2} < 1.5$. We also calculate a modified version, $\chi^{\prime 2}$, which includes only the central image pixels, that are generally more often affected by these substructures. \twodphot\ model fitting provides the following parameters: average surface brightness $\mu_{\text{e}}$, major-axis effective radius $\Theta_{\text{e,maj}}$, S\'ersic index n, total magnitude $m_{S}$, axial ratio q, and position angle. In this analysis, we use the circularized effective radius $\Theta_{\text{e}}$, defined as $\Theta_{\text{e}}= \Theta_\text{{e,maj}}\sqrt{q}$. Effective radius is then converted to the physical scale value $R_{\text{e}}$ using the measured (photometric and/or spectroscopic) redshift. Only galaxies with $r$-band $(\SN)_r \equiv 1/\MErrautor > 50$, where \MErrautor\ is the error on the $r$-band \Mauto, are kept for the next analysis (\citealt{LaBarbera_08_2DPHOT, SPIDER-I}; \citealt{Roy+18}, T16; T18). 

\item {\it Photometric redshifts}. Redshifts are determined with the Multi Layer Perceptron with Quasi Newton Algorithm (MLPQNA) method (\citealt{Brescia+13, Brescia+14}; \citealt{Cavuoti+15_PhotoRApToR}), and presented in \cite{Cavuoti+15_KIDS_I, Cavuoti+17_KiDS}, which we refer to for all details. 

\item {\it Spectroscopic redshifts}. We cross-match our KiDS catalog with overlapping spectroscopic surveys to obtain spectroscopic redshifts for the objects in common, i.e. the \KiDSspec\ sample. We use redshifts from the Sloan Digital Sky Survey Data Release 9 (SDSS$-$DR9; \citealt{Ahn+12_SDSS_DR9, Ahn+14_SDSS_DR10}), Galaxy And Mass Assembly Data Release 2 (GAMA$-$DR2; \citealt{Driver+11_GAMA}) and  2dFLenS (\citealt{Blake+16_2dflens}).

\item  {\it Stellar masses}. We run \lephare\ (\citealt{Arnouts+99}; \citealt{Ilbert+06}) to estimate stellar masses. This software performs a simple $\chi^{2}$ fitting  between the stellar population synthesis (SPS) theoretical models and the data. In order to minimize the degeneracy between colours and stellar population parameters, we fix the redshift, either using the \zp\, or \zs, depending on the availability and the sample under exam. It is evident that, when a \zs\ is obtained for a \UCMG\ candidate, the stellar mass needs to be re-estimated  as the ``true'' redshift might produce a different mass that needs to be checked against the criteria to confirm the \UCMG\ nature (see next section). Since the \UCMG\ candidates sample analyzed in this paper has been collected using a slightly different spectral library with respect to the sample presented in T18, we use a partially different set-up to estimate stellar masses. As in T18, we fit multi-wavelength photometry of the galaxies in the sample with Single burst models from \citealt{BC03} (BC03 hereafter). However, here we further constrain the parameter space, forcing metallicities and ages to vary in the range $0.2 \leq Z/\Zsun \leq 2.5$ and $3 \leq t \leq t_{\text{max}}$ Gyr, respectively. The maximum age, $t_{\text{max}}$, is set by the age of the Universe at the redshift of the galaxy, with a maximum value of 13 Gyr at $z=0$. 
The age cutoff of 3 Gyr is meant to minimize the probability of underestimating the stellar mass by obtaining a too young age, following \cite{Maraston+13_BOSS}. Then, as in T18, we adopt a \cite{Chabrier01} IMF and the observed $ugri$ magnitudes \magapsix\ (and related $1\sigma$ uncertainties $\delta u$, $\delta g$, $\delta r$, and $\delta i$), which are corrected for Galactic extinction using the map in \cite{Schlafly_Finkbeiner11}. In order to correct the $M_{*}$ outcomes of \lephare\ for missing flux, we use the total magnitudes derived from the S\'ersic fitting and the formula:
\begin{equation}
\log_{10} \mst = \log_{10} \mst^{\lephare} + 0.4 \times(\magapsix-m_{\rm S})
\end{equation}
where $\log_{10} \mst^{\lephare}$ is the output of \lephare. We consider calibration errors on the photometric zero-point $\delta_{\rm zp} \equiv
(\delta u_{\rm zp}, \, \delta g_{\rm zp}, \, \delta r_{\rm zp}, \, \delta
i_{\rm zp}) = (0.075, \, 0.074, \, 0.029, \, 0.055)$, quadratically added to the SExtractor magnitude errors (see T18). 
\item \textbf{{\it Galaxy classification.}} Using \lephare, we also fit the observed magnitudes with the set of $66$
empirical spectral templates used in \cite{Ilbert+06}, in order to
determine a qualitative galaxy classification. The set is based on
different templates resembling spectra of ``Elliptical", ``Spiral" and ``Starburst" galaxies.
\end{itemize}

We use the above dataset, that we name KiDS\_{\textsc{full}, to collect a complete set of photometrically selected \UCMGs, using criteria as described in the next section. 

Moreover, in order to check what galaxies had already literature spectroscopy, we cross-match the \KiDSfull\ with publicly available spectroscopic samples and define the so-called \KiDSspec\ sample, which comprises all galaxies from our complete photometric sample with known spectroscopic redshifts. 

\subsection{\UCMGs\ selection and our sample}\label{subsec:UCMG_criteria}
To select the \UCMG\ candidates, we use the same criteria reported in T16 and T18:
\begin{enumerate}
\item {\it Massiveness}: A Chabrier-IMF based stellar mass of $M_{*} > 8 \times 10^{10} M_{\odot}$ (\citealt{Trujillo+09_superdense}; T16, T18);
\item {\it Compactness}: A circularized effective radius $\Re < 1.5$ kpc (T18);
\item {\it Best-fit structural parameters}: A reduced $\chi^{2}$ < 1.5 in $g$-, $r$- and $i$- filters (\citealt{SPIDER-I}), and further criteria to control the quality of the fit, as $\Theta_{\rm e} > 0.05\arcsec$, q $> 0.1$ and n $>0.5$;
\item {\it Star/Galaxy separation}: A discrimination between stars and galaxies using the $g-J$ vs. $J-Ks$ plane to minimize the overlap of sources with the typical stellar locus (see e.g., Fig. 1 in T16).
\end{enumerate}

Further details about the above criteria, to select} \UCMGs\ from both \KiDSfull\ and \KiDSspec\ can be found in T16 and T18. In the following we refer to the {\it photometrically selected} and the {\it spectroscopically selected} samples as the ones where \mst\ and \Re\ are calculated using \zp\ or \zs, respectively.\footnote{When the spectroscopic redshift becomes available for a given \UCMG\ candidate, one has to recompute both the \Re\ in kpc (which obviously scales with the true redshift) and the stellar mass (see \Sec\ref{subsec:selectionprocedure}) to check that the criteria of compactness and massiveness hold.}

After applying all the requirements we end up with the following samples at $z < 0.5$:
\begin{itemize}
 \item \UCMGfull: a photometrically selected sample of 1221 \UCMG\ candidates\footnote{In T18 we collected 995 photometrically selected candidates (1000 before the colour-colour cut), which is different from the number of 1221 found here. The difference between these numbers is related to the different sets of masses adopted in T18 and in the present paper. We will discuss the impact of the mass assumption later in the paper, showing the effect on the number density evolution.} (1256 before the colour-colour cut) extracted from \KiDSfull;
\item \UCMGSpec: a spectroscopically selected sample of 55 \UCMGs, selected from the \KiDSspec\ sample, for which  stellar masses and radii have been computed using the spectroscopic redshifts;
\item \UCMGPhotSpec: a sample of 50 photometrically selected \UCMG\ candidates which have spectroscopic redshift available from literature. Practically, these galaxies have been extracted from \KiDSspec\ 
but they resulted to be \UCMG\ on the basis of their \zp.
\end{itemize}

In the \UCMGfull\ sample, that provides the most statistically significant characterization of our \UCMG\ candidates, the objects are brighter than $r \sim 21$. Most of them are located at $\zp > 0.3$, with a median redshift of $\zp = 0.41$. Median values of 20.4 and 11 dex are found for the extinction corrected $r$-band $\texttt{MAG\_AUTO}$ and $\log_{10} (M_*/M_{\odot})$. More than $97$ per cent of the \UCMGfull\ candidates have KiDS photometry consistent with ``Elliptical" templates in \cite{Ilbert+06}, and they have very red colours in the optical-NIR colour-colour plane. The $\Re < 1.5\  \rm kpc$ constraint corresponds to $\Theta_{\rm e} \lsim 0.4$ arcsec, the medians for these parameters are $\Re = 1.22\ \rm kpc$ and $\Theta_{\rm e} = 0.23$ arcsec, respectively. The range of the values for axis ratio and S\'ersic index is wide, but their distributions are peaked around values of $q \sim 0.4$ and $n \sim 4$, with median values of $0.47$ and $4.6$, respectively.

\subsection{The impact of selection criteria}\label{subsec:impact_sel_criteria}
Following the previous papers of this series (T16 and T18), we adopt rather stringent criteria on the sizes and masses to select only the most extreme (and rare) \UCMGs. However, there is a large variety of definitions used in other literature studies. Until there will be no consensus, the comparison among different analyses will be prone to a ``definition bias''. Here in this section we  evaluate the impact of different definitions on our \UCMGfull\ sample (see also a detailed discussion in T18). 
For instance, keeping the threshold on the stellar mass unchanged and releasing the constraint on the size, such as $\Re < 2 \, \rm kpc$ and $<3 \, \rm kpc$, the number of candidates (before colour-colour cut) would increase to 3430 and 12472, respectively.
Instead, decreasing the threshold in mass from $\log_{10} (M_*/M_{\odot}) = 10.9$ to $10.7$, the number of selected galaxies within \UCMGfull\ would not change by more than 3\%, i.e. the size criterion is the one impacting more the \UCMG\ definition.
Besides the threshold in size and mass, another important assumption that might significantly impact our selection is the shape of the stellar Initial Mass Function (IMF).
Here, we assume a universal Chabrier IMF for all the galaxies despite recent claims for a bottom-heavier IMF in more massive ETGs (e.g, \citealt{Conroy_vanDokkum12a}; \citealt{Cappellari+12}; \citealt{Spiniello+12}; \citealt{TRN13_SPIDER_IMF}; \citealt{LaBarbera+13_SPIDERVIII_IMF}, \citealt{Spiniello+14,Spiniello+15_IMF_vs_density}). This choice has been made to compare our results with other results published in the literature, all assuming a Chabrier IMF. If a Salpeter IMF were to be used instead, more coherently with predictions for compact and massive systems (\citealt{Martin-Navarro+15_IMF_relic}; \citealt{Ferre-Mateu+17}), keeping the massiveness and compactness criteria unchanged, we would retrieve 1291 \UCMGs\ instead of 1256. Thus, also the IMF slope has a negligible impact on our selection.

\begin{table*}
\centering \caption{Integrated photometry for the 33 \UCMG\ candidates observed within our spectroscopic program, 13 in $\textsc{ucmg\_int\_2017}$, 11 in $\textsc{ucmg\_tng\_2017}$ and 9 in $\textsc{ucmg\_tng\_2018}$ (within each subsample the galaxies are ordered by Right Ascension). 
From left we give: a) progressive ID number; b) KIDS identification name; c) $r$-band KiDS $\texttt{MAG\_AUTO}$; d-g) $u$-, $g$-, $r$- and $i$-band KiDS magnitudes measured in
an aperture of 6 arcsec of diameter with 1$\sigma$ errors; h) photometric redshift from machine learning.
All the magnitudes have been corrected for galactic extinction using the maps of \citet{Schlafly_Finkbeiner11}. More details are provided in \Sec\ref{sec:sample}.}

\label{tab:photometry}
\begin{tabular}{cccccccc} 
\hline
$\rm ID$ & $\rm name$ &  $\texttt{MAG\_AUTO\_r}$ & $u_{6''}$ & $ g_{6''}$ & $ r_{6''}$ & $i_{6''}$  & \zp\\ 
\hline
\hline
&  \multicolumn{2}{c}{Observation~date:~March~2017} & & 
\multicolumn{2}{c}{Instrument:~INT/IDS} & \\

\hline
$	1	$	&	\rm	KIDS J085700.29--010844.55	&	$	19.21	$	&	$	22.70	\pm	0.21	$	&	$	20.74	\pm	0.01	$	&	$	19.22	\pm	0.003	$	&	$	18.71	\pm	0.01	$	&	$	0.28	$	\\
$	2	$	&	\rm	KIDS J111108.43+003207.00	&	$	19.05	$	&	$	22.49	\pm	0.14	$	&	$	20.46	\pm	0.01	$	&	$	19.04	\pm	0.003	$	&	$	18.61	\pm	0.006	$	&	$	0.26	$	\\
$	3	$	&	\rm	KIDS J111447.86+003903.71	&	$	19.00	$	&	$	22.35	\pm	0.12	$	&	$	20.47	\pm	0.01	$	&	$	19.03	\pm	0.003	$	&	$	18.57	\pm	0.009	$	&	$	0.26	$	\\
$	4	$	&	\rm	KIDS J111504.01+005101.16	&	$	19.21	$	&	$	20.43	\pm	0.02	$	&	$	19.92	\pm	0.006	$	&	$	19.24	\pm	0.003	$	&	$	19.01	\pm	0.014	$	&	$	0.45	$	\\
$	5	$	&	\rm	KIDS J111750.31+003647.35	&	$	19.13	$	&	$	22.80	\pm	0.19	$	&	$	20.74	\pm	0.01	$	&	$	19.12	\pm	0.003	$	&	$	18.69	\pm	0.01	$	&	$	0.37	$	\\
$	6	$	&	\rm	KIDS J122009.53--024141.88	&	$	18.69	$	&	$	21.93	\pm	0.1	$	&	$	20.02	\pm	0.007	$	&	$	18.71	\pm	0.002	$	&	$	18.19	\pm	0.006	$	&	$	0.22	$	\\
$	7	$	&	\rm	KIDS J122639.96--011138.08	&	$	18.59	$	&	$	22.15	\pm	0.11	$	&	$	20.06	\pm	0.008	$	&	$	18.63	\pm	0.003	$	&	$	18.21	\pm	0.008	$	&	$	0.23	$	\\
$	8	$	&	\rm	KIDS J122815.38--015356.06	&	$	18.84	$	&	$	22.17	\pm	0.1	$	&	$	20.26	\pm	0.008	$	&	$	18.84	\pm	0.003	$	&	$	18.37	\pm	0.008	$	&	$	0.24	$	\\
$	9	$	&	\rm	KIDS J140127.77+020509.13	&	$	19.04	$	&	$	21.47	\pm	0.06	$	&	$	20.23	\pm	0.007	$	&	$	19.01	\pm	0.003	$	&	$	18.65	\pm	0.007	$	&	$	0.34	$	\\
$	10	$	&	\rm	KIDS J141120.06+023342.62	&	$	18.85	$	&	$	22.72	\pm	0.17	$	&	$	20.47	\pm	0.01	$	&	$	18.83	\pm	0.003	$	&	$	18.39	\pm	0.007	$	&	$	0.32	$	\\
$	11	$	&	\rm	KIDS J145700.42+024502.06	&	$	18.62	$	&	$	22.17	\pm	0.13	$	&	$	19.95	\pm	0.008	$	&	$	18.67	\pm	0.002	$	&	$	18.23	\pm	0.007	$	&	$	0.24	$	\\
$	12	$	&	\rm	KIDS J150309.55+001318.10	&	$	18.99	$	&	$	22.59	\pm	0.19	$	&	$	20.47	\pm	0.01	$	&	$	19.02	\pm	0.003	$	&	$	18.67	\pm	0.007	$	&	$	0.28	$	\\
$	13	$	&	\rm	KIDS J152844.81--000912.86	&	$	18.56	$	&	$	22.91	\pm	0.25	$	&	$	19.98	\pm	0.01	$	&	$	18.59	\pm	0.002	$	&	$	18.20	\pm	0.005	$	&	$	0.23	$	\\
\hline
&  \multicolumn{2}{c}{Observation~date:~March~2017} & & 
\multicolumn{2}{c}{Instrument:~TNG/DOLORES } & \\
\hline

$	14	$	&	\rm	KIDS J084239.97+005923.71	&	$	19.63	$	&	$	22.95	\pm	1.76	$	&	$	21.14	\pm	0.12	$	&	$	19.58	\pm	0.04	$	&	$	19.02	\pm	0.08	$	&	$	0.35	$	\\
$	15	$	&	\rm	KIDS J090412.45--001819.75	&	$	19.11	$	&	$	22.51	\pm	0.95	$	&	$	20.58	\pm	0.07	$	&	$	19.13	\pm	0.02	$	&	$	18.66	\pm	0.02	$	&	$	0.27	$	\\
$	16	$	&	\rm	KIDS J091704.84--012319.65	&	$	19.21	$	&	$	22.87	\pm	1.03	$	&	$	20.84	\pm	0.08	$	&	$	19.20	\pm	0.02	$	&	$	18.65	\pm	0.02	$	&	$	0.33	$	\\
$	17	$	&	\rm	KIDS J104051.66+005626.73	&	$	19.52	$	&	$	23.27	\pm	0.29	$	&	$	20.97	\pm	0.02	$	&	$	19.54	\pm	0.005	$	&	$	18.52	\pm	0.01	$	&	$	0.33	$	\\
$	18	$	&	\rm	KIDS J114800.92+023753.02	&	$	19.41	$	&	$	23.13	\pm	0.33	$	&	$	20.54	\pm	0.01	$	&	$	19.41	\pm	0.005	$	&	$	18.61	\pm	0.009	$	&	$	0.32	$	\\
$	19	$	&	\rm	KIDS J120203.17+025105.56	&	$	19.43	$	&	$	22.57	\pm	0.18	$	&	$	20.95	\pm	0.02	$	&	$	19.41	\pm	0.005	$	&	$	18.95	\pm	0.01	$	&	$	0.30	$	\\
$	20	$	&	\rm	KIDS J121856.54+023241.69	&	$	19.23	$	&	$	22.75	\pm	0.17	$	&	$	20.79	\pm	0.01	$	&	$	19.23	\pm	0.004	$	&	$	18.70	\pm	0.008	$	&	$	0.30	$	\\
$	21	$	&	\rm	KIDS J140257.62+011730.39	&	$	19.96	$	&	$	23.31	\pm	0.48	$	&	$	21.33	\pm	0.02	$	&	$	19.94	\pm	0.008	$	&	$	19.44	\pm	0.02	$	&	$	0.33	$	\\
$	22	$	&	\rm	KIDS J145656.68+002007.41	&	$	19.46	$	&	$	22.99	\pm	0.23	$	&	$	20.84	\pm	0.02	$	&	$	19.43	\pm	0.005	$	&	$	18.94	\pm	0.006	$	&	$	0.28	$	\\
$	23	$	&	\rm	KIDS J145948.65--024036.57	&	$	18.57	$	&	$	21.96	\pm	0.88	$	&	$	19.92	\pm	0.05	$	&	$	18.58	\pm	0.02	$	&	$	18.10	\pm	0.04	$	&	$	0.25	$	\\
$	24	$	&	\rm	KIDS J152700.54--002359.09	&	$	19.64	$	&	$	24.54	\pm	1.45	$	&	$	21.19	\pm	0.03	$	&	$	19.62	\pm	0.006	$	&	$	19.12	\pm	0.01	$	&	$	0.33	$	\\
\hline
&  \multicolumn{2}{c}{Observation~date:~March~2018} & & 
\multicolumn{2}{c}{Instrument:~TNG/DOLORES} & \\
\hline
$	25	$	&	\rm	KIDS J083807.31+005256.58	&	$	19.29	$	&	$	22.48	\pm	0.14	$	&	$	20.66	\pm	0.01	$	&	$	19.29	\pm	0.004	$	&	$	18.75	\pm	0.009	$	&	$	0.28	$	\\
$	26	$	&	\rm	KIDS J084412.25--005850.00	&	$	19.67	$	&	$	22.76	\pm	0.22	$	&	$	21.16	\pm	0.02	$	&	$	19.64	\pm	0.006	$	&	$	19.10	\pm	0.015	$	&	$	0.32	$	\\
$	27	$	&	\rm	KIDS J084413.29+014847.59	&	$	19.78	$	&	$	23.01	\pm	0.32	$	&	$	21.22	\pm	0.02	$	&	$	19.75	\pm	0.008	$	&	$	19.21	\pm	0.014	$	&	$	0.33	$	\\
$	28	$	&	\rm	KIDS J090933.87+014532.21	&	$	19.55	$	&	$	23.13	\pm	0.35	$	&	$	21.14	\pm	0.02	$	&	$	19.51	\pm	0.005	$	&	$	18.98	\pm	0.01	$	&	$	0.33	$	\\
$	29	$	&	\rm	KIDS J092030.99+012635.38	&	$	19.52	$	&	$	22.70	\pm	0.19	$	&	$	20.96	\pm	0.02	$	&	$	19.51	\pm	0.005	$	&	$	19.04	\pm	0.015	$	&	$	0.29	$	\\
$	30	$	&	\rm	KIDS J092407.03--000350.69	&	$	19.87	$	&	$	24.06	\pm	0.55	$	&	$	21.48	\pm	0.02	$	&	$	19.84	\pm	0.005	$	&	$	19.20	\pm	0.012	$	&	$	0.39	$	\\
$	31	$	&	\rm	KIDS J103951.25+002402.34	&	$	19.63	$	&	$	22.41	\pm	0.15	$	&	$	20.66	\pm	0.01	$	&	$	19.62	\pm	0.006	$	&	$	18.70	\pm	0.013	$	&	$	0.41	$	\\
$	32	$	&	\rm	KIDS J145721.54--014009.02	&	$	19.43	$	&	$	23.12	\pm	0.35	$	&	$	21.03	\pm	0.02	$	&	$	19.47	\pm	0.004	$	&	$	18.97	\pm	0.014	$	&	$	0.29	$	\\
$	33	$	&	\rm	KIDS J152706.54--001223.64	&	$	19.67	$	&	$	23.92	\pm	0.73	$	&	$	21.39	\pm	0.03	$	&	$	19.68	\pm	0.006	$	&	$	19.08	\pm	0.01	$	&	$	0.43	$	\\

\hline
\end{tabular}
\end{table*} 


\begin{figure*}
 \centering
\includegraphics[width=\textwidth]{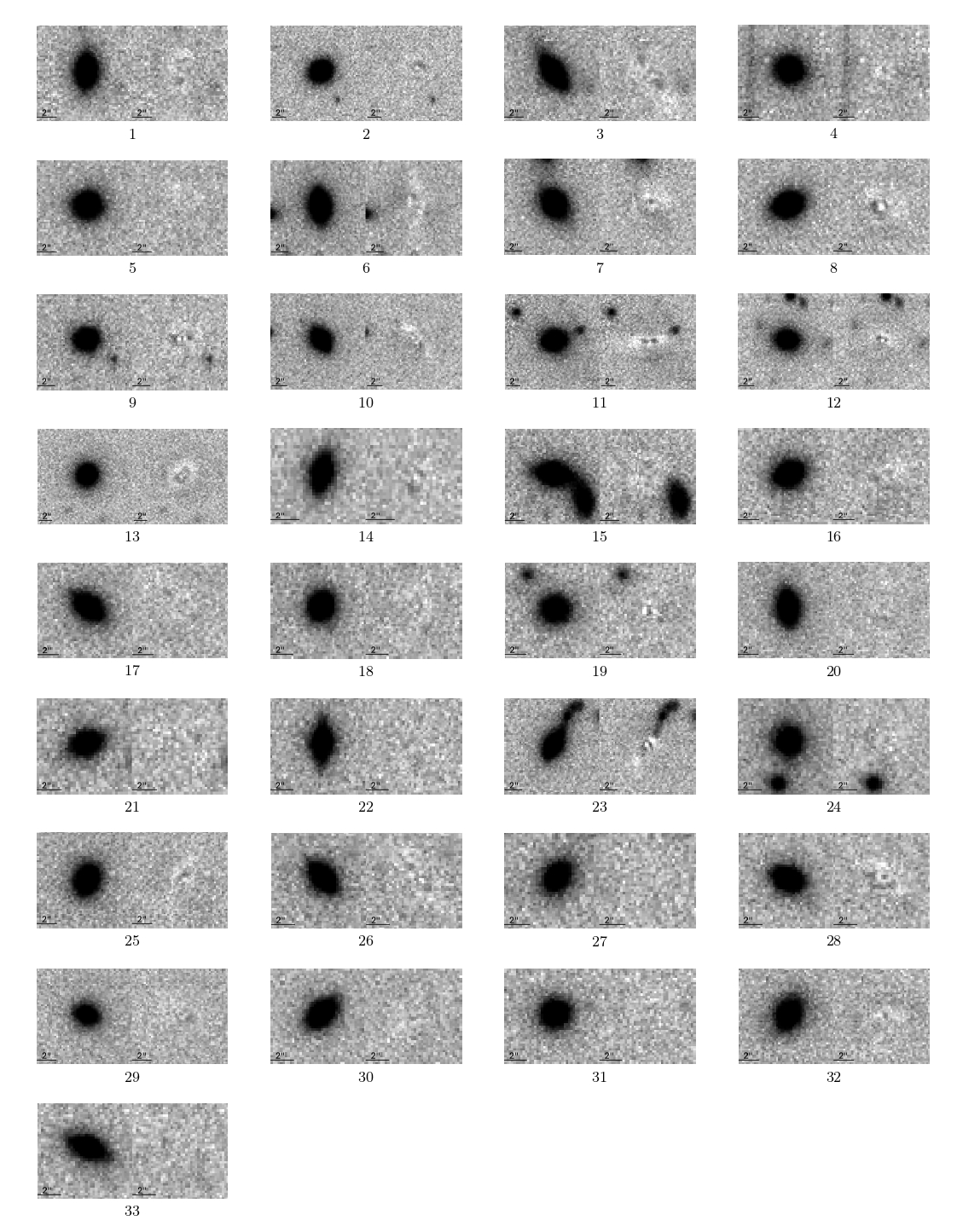}
 \caption{2D fit output from the \twodphot\ procedure on the 33 \UCMG\ candidates for which we obtained new spectroscopic data. For each \UCMG\, the left panel shows the original $r$-band image and the right panel shows the residual after the subtraction of the 2D single S\'ersic PSF convolved model. We also indicate the scale of 2 arcsec in the panels.}
\label{ucmg2}
\end{figure*}


\begin{table*}
\centering 
\caption{Structural parameters derived running \twodphot\ on $g$-, $r$- and $i$-bands. 
For each band we give: a)
circularized effective radius $\Theta_{\rm e}$, measured in arcsec, b)
circularized effective radius $R_{\rm e}$, measured in kpc (calculated
using \zp\ values listed in Table \ref{tab:photometry}), c)
S\'ersic index $n$, d) axis ratio q, e) $\chi^{2}$ of the surface
photometry fit, f) $\chi^{\prime 2}$ of the surface photometry fit
including only central pixels and g) the  signal-to-noise ratio
\SN\ of the photometric images, defined as the inverse of the error on $\texttt{MAG\_AUTO}$.}
\label{tab:struc_parameters}
\hspace{-4.8cm}
\resizebox{1.25\textwidth}{!}{
\begin{tabular}{cccccccccccccccccccccc}
\hline

$ $ & \multicolumn{7}{c}{$g$-band} & \multicolumn{7}{c}{$r$-band} & \multicolumn{7}{c}{$i$-band} \\
\cmidrule(rl){2-8} \cmidrule(rl){9-15} \cmidrule(rl){16-22}

$\rm ID$ & $\Theta_{e}$ & $R_{e}$ & n & q & $\chi^{2}$ & $\chi^{\prime 2}$ & $\SN$ & $\Theta_{e}$ & $R_{e}$ & n & q & $\chi^{2}$ & $\chi^{\prime 2}$ & $\SN$ & $\Theta_{e}$ & $R_{e}$ & n & q & $\chi^{2}$ & $\chi^{\prime 2}$ & $\SN$ \\
\hline
\hline

$1	$	&	$	0.32	$	&	$	1.36	$	&	$	2.94	$	&	$	0.31	$	&	$	1.01	$	&	$	0.92	$	&	$	81	$	&	$	0.37	$	&	$	1.55	$	&	$	2.33	$	&	$	0.33	$	&	$	1.02	$	&	$	0.98	$	&	$	81	$	&	$	0.34	$	&	$	1.43	$	&	$	4.04	$	&	$	0.33	$	&	$	1.01	$	&	$	1.01	$	&	$	98	$	\\
$2	$	&	$	0.40	$	&	$	1.60	$	&	$	3.31	$	&	$	0.74	$	&	$	1.02	$	&	$	0.96	$	&	$	100	$	&	$	0.28	$	&	$	1.11	$	&	$	5.54	$	&	$	0.76	$	&	$	1.02	$	&	$	1.07	$	&	$	100	$	&	$	0.31	$	&	$	1.23	$	&	$	5.83	$	&	$	0.77	$	&	$	1.02	$	&	$	1.02	$	&	$	161	$	\\
$3	$	&	$	0.36	$	&	$	1.45	$	&	$	4.56	$	&	$	0.25	$	&	$	0.99	$	&	$	1.02	$	&	$	94	$	&	$	0.26	$	&	$	1.06	$	&	$	6.08	$	&	$	0.26	$	&	$	1.03	$	&	$	1.20	$	&	$	94	$	&	$	0.34	$	&	$	1.36	$	&	$	4.93	$	&	$	0.24	$	&	$	1.00	$	&	$	1.00	$	&	$	108	$	\\
$4	$	&	$	0.06	$	&	$	0.32	$	&	$	2.96	$	&	$	0.71	$	&	$	1.00	$	&	$	1.02	$	&	$	148	$	&	$	0.06	$	&	$	0.35	$	&	$	6.32	$	&	$	0.87	$	&	$	1.03	$	&	$	1.12	$	&	$	148	$	&	$	0.10	$	&	$	0.55	$	&	$	5.57	$	&	$	0.73	$	&	$	0.97	$	&	$	0.97	$	&	$	62	$	\\
$5	$	&	$	0.16	$	&	$	0.84	$	&	$	7.10	$	&	$	0.81	$	&	$	1.01	$	&	$	0.99	$	&	$	90	$	&	$	0.14	$	&	$	0.71	$	&	$	6.83	$	&	$	0.87	$	&	$	1.07	$	&	$	1.08	$	&	$	90	$	&	$	0.14	$	&	$	0.70	$	&	$	6.00	$	&	$	0.73	$	&	$	1.00	$	&	$	1.00	$	&	$	108	$	\\
$6	$	&	$	0.43	$	&	$	1.52	$	&	$	1.52	$	&	$	0.29	$	&	$	1.02	$	&	$	0.94	$	&	$	134	$	&	$	0.35	$	&	$	1.23	$	&	$	2.15	$	&	$	0.26	$	&	$	1.02	$	&	$	1.16	$	&	$	134	$	&	$	0.41	$	&	$	1.44	$	&	$	2.11	$	&	$	0.31	$	&	$	0.99	$	&	$	0.99	$	&	$	148	$	\\
$7	$	&	$	0.22	$	&	$	0.82	$	&	$	8.46	$	&	$	0.57	$	&	$	1.02	$	&	$	1.07	$	&	$	118	$	&	$	0.31	$	&	$	1.12	$	&	$	7.53	$	&	$	0.68	$	&	$	1.03	$	&	$	1.28	$	&	$	118	$	&	$	0.36	$	&	$	1.32	$	&	$	2.87	$	&	$	0.61	$	&	$	1.00	$	&	$	1.00	$	&	$	123	$	\\
$8	$	&	$	0.39	$	&	$	1.48	$	&	$	2.96	$	&	$	0.53	$	&	$	1.03	$	&	$	0.98	$	&	$	125	$	&	$	0.36	$	&	$	1.36	$	&	$	2.68	$	&	$	0.54	$	&	$	1.03	$	&	$	1.19	$	&	$	125	$	&	$	0.35	$	&	$	1.34	$	&	$	2.87	$	&	$	0.56	$	&	$	1.05	$	&	$	1.05	$	&	$	128	$	\\
$9	$	&	$	0.20	$	&	$	0.97	$	&	$	4.95	$	&	$	0.79	$	&	$	1.04	$	&	$	1.02	$	&	$	161	$	&	$	0.24	$	&	$	1.14	$	&	$	5.19	$	&	$	0.83	$	&	$	1.04	$	&	$	1.20	$	&	$	161	$	&	$	0.22	$	&	$	1.04	$	&	$	5.30	$	&	$	0.72	$	&	$	0.99	$	&	$	0.99	$	&	$	166	$	\\
$10	$	&	$	0.40	$	&	$	1.10	$	&	$	2.49	$	&	$	0.30	$	&	$	1.00	$	&	$	1.01	$	&	$	97	$	&	$	0.21	$	&	$	0.97	$	&	$	2.97	$	&	$	0.30	$	&	$	1.15	$	&	$	1.20	$	&	$	97	$	&	$	0.21	$	&	$	0.98	$	&	$	2.83	$	&	$	0.31	$	&	$	0.99	$	&	$	1.02	$	&	$	156	$	\\
$11	$	&	$	0.39	$	&	$	1.47	$	&	$	7.86	$	&	$	0.51	$	&	$	1.00	$	&	$	0.91	$	&	$	104	$	&	$	0.27	$	&	$	1.02	$	&	$	6.71	$	&	$	0.42	$	&	$	1.04	$	&	$	1.23	$	&	$	377	$	&	$	0.34	$	&	$	1.31	$	&	$	8.40	$	&	$	0.49	$	&	$	0.99	$	&	$	0.99	$	&	$	129	$	\\
$12	$	&	$	0.32	$	&	$	1.37	$	&	$	6.08	$	&	$	0.48	$	&	$	1.00	$	&	$	1.03	$	&	$	79	$	&	$	0.31	$	&	$	1.30	$	&	$	7.16	$	&	$	0.56	$	&	$	1.07	$	&	$	1.14	$	&	$	283	$	&	$	0.30	$	&	$	1.27	$	&	$	6.93	$	&	$	0.52	$	&	$	1.02	$	&	$	0.93	$	&	$	132	$	\\
$13	$	&	$	0.28	$	&	$	1.61	$	&	$	3.94	$	&	$	0.36	$	&	$	1.00	$	&	$	1.07	$	&	$	135	$	&	$	0.39	$	&	$	1.45	$	&	$	4.24	$	&	$	0.77	$	&	$	1.04	$	&	$	1.19	$	&	$	421	$	&	$	0.41	$	&	$	1.50	$	&	$	5.33	$	&	$	0.77	$	&	$	1.01	$	&	$	0.88	$	&	$	175	$	\\

\hline
$14	$	&	$	0.28	$	&	$	1.37	$	&	$	2.22	$	&	$	0.12	$	&	$	1.03	$	&	$	0.94	$	&	$	53	$	&	$	0.23	$	&	$	1.12	$	&	$	3.27	$	&	$	0.29	$	&	$	1.00	$	&	$	1.07	$	&	$	158	$	&	$	0.28	$	&	$	1.40	$	&	$	3.38	$	&	$	0.41	$	&	$	0.98	$	&	$	0.91	$	&	$	105	$	\\
$15	$	&	$	0.43	$	&	$	1.77	$	&	$	4.82	$	&	$	0.32	$	&	$	1.00	$	&	$	1.20	$	&	$	70	$	&	$	0.27	$	&	$	1.13	$	&	$	2.69	$	&	$	0.36	$	&	$	1.04	$	&	$	1.15	$	&	$	297	$	&	$	0.21	$	&	$	0.87	$	&	$	4.37	$	&	$	0.33	$	&	$	1.00	$	&	$	0.99	$	&	$	244	$	\\
$16	$	&	$	0.28	$	&	$	1.35	$	&	$	3.05	$	&	$	0.32	$	&	$	1.02	$	&	$	1.08	$	&	$	70	$	&	$	0.24	$	&	$	1.14	$	&	$	3.03	$	&	$	0.41	$	&	$	1.04	$	&	$	1.18	$	&	$	252	$	&	$	0.27	$	&	$	1.28	$	&	$	4.12	$	&	$	0.41	$	&	$	1.02	$	&	$	1.03	$	&	$	219	$	\\
$17	$	&	$	0.36	$	&	$	1.71	$	&	$	4.57	$	&	$	0.36	$	&	$	1.00	$	&	$	0.93	$	&	$	58	$	&	$	0.31	$	&	$	1.46	$	&	$	6.10	$	&	$	0.38	$	&	$	1.02	$	&	$	1.01	$	&	$	58	$	&	$	0.31	$	&	$	1.47	$	&	$	4.35	$	&	$	0.36	$	&	$	0.99	$	&	$	0.99	$	&	$	91	$	\\
$18	$	&	$	0.27	$	&	$	1.25	$	&	$	2.09	$	&	$	0.58	$	&	$	1.00	$	&	$	0.95	$	&	$	93	$	&	$	0.29	$	&	$	1.36	$	&	$	2.83	$	&	$	0.58	$	&	$	1.03	$	&	$	1.04	$	&	$	93	$	&	$	0.26	$	&	$	1.22	$	&	$	2.75	$	&	$	0.56	$	&	$	1.05	$	&	$	1.05	$	&	$	114	$	\\
$19	$	&	$	0.31	$	&	$	1.38	$	&	$	6.47	$	&	$	0.99	$	&	$	1.04	$	&	$	1.01	$	&	$	59	$	&	$	0.29	$	&	$	1.29	$	&	$	9.54	$	&	$	0.89	$	&	$	1.03	$	&	$	1.09	$	&	$	59	$	&	$	0.36	$	&	$	1.58	$	&	$	5.24	$	&	$	0.87	$	&	$	1.01	$	&	$	1.01	$	&	$	111	$	\\
$20	$	&	$	0.31	$	&	$	1.37	$	&	$	2.05	$	&	$	0.19	$	&	$	1.03	$	&	$	0.93	$	&	$	82	$	&	$	0.33	$	&	$	1.46	$	&	$	2.75	$	&	$	0.30	$	&	$	1.02	$	&	$	1.00	$	&	$	82	$	&	$	0.26	$	&	$	1.15	$	&	$	3.13	$	&	$	0.26	$	&	$	1.03	$	&	$	1.03	$	&	$	132	$	\\
$21	$	&	$	0.17	$	&	$	0.81	$	&	$	6.43	$	&	$	0.44	$	&	$	1.01	$	&	$	0.96	$	&	$	52	$	&	$	0.11	$	&	$	0.50	$	&	$	8.05	$	&	$	0.46	$	&	$	1.03	$	&	$	1.12	$	&	$	52	$	&	$	0.19	$	&	$	0.90	$	&	$	4.08	$	&	$	0.58	$	&	$	1.03	$	&	$	1.03	$	&	$	70	$	\\
$22	$	&	$	0.25	$	&	$	1.04	$	&	$	2.48	$	&	$	0.10	$	&	$	1.04	$	&	$	1.12	$	&	$	74	$	&	$	0.12	$	&	$	0.50	$	&	$	5.60	$	&	$	0.20	$	&	$	1.03	$	&	$	1.11	$	&	$	74	$	&	$	0.11	$	&	$	0.45	$	&	$	5.53	$	&	$	0.31	$	&	$	1.03	$	&	$	1.03	$	&	$	184	$	\\
$23	$	&	$	0.27	$	&	$	1.07	$	&	$	6.15	$	&	$	0.30	$	&	$	1.04	$	&	$	1.39	$	&	$	110	$	&	$	0.31	$	&	$	1.22	$	&	$	4.34	$	&	$	0.30	$	&	$	1.04	$	&	$	2.78	$	&	$	110	$	&	$	0.66	$	&	$	2.57	$	&	$	8.19	$	&	$	0.04	$	&	$	1.00	$	&	$	1.02	$	&	$	146	$	\\
$24	$	&	$	0.39	$	&	$	1.85	$	&	$	10.02	$	&	$	0.94	$	&	$	1.01	$	&	$	1.07	$	&	$	42	$	&	$	0.14	$	&	$	0.67	$	&	$	8.83	$	&	$	0.75	$	&	$	1.01	$	&	$	1.16	$	&	$	42	$	&	$	0.22	$	&	$	1.07	$	&	$	9.16	$	&	$	0.68	$	&	$	1.02	$	&	$	1.02	$	&	$	73	$	\\

\hline

$	25	$	&	$	0.31	$	&	$	1.30	$	&	$	4.08	$	&	$	0.41	$	&	$	0.99	$	&	$	0.92	$	&	$	84	$	&	$	0.35	$	&	$	1.49	$	&	$	4.02	$	&	$	0.45	$	&	$	1.03	$	&	$	1.06	$	&	$	84	$	&	$	0.30	$	&	$	1.27	$	&	$	3.08	$	&	$	0.40	$	&	$	1.03	$	&	$	0.87	$	&	$	106	$	\\
$	26	$	&	$	0.27	$	&	$	1.28	$	&	$	2.00	$	&	$	0.32	$	&	$	1.01	$	&	$	1.01	$	&	$	58	$	&	$	0.29	$	&	$	1.36	$	&	$	2.69	$	&	$	0.36	$	&	$	1.04	$	&	$	1.15	$	&	$	58	$	&	$	0.27	$	&	$	1.26	$	&	$	4.37	$	&	$	0.33	$	&	$	1.02	$	&	$	0.99	$	&	$	75	$	\\
$	27	$	&	$	0.32	$	&	$	1.51	$	&	$	6.83	$	&	$	0.44	$	&	$	1.00	$	&	$	0.98	$	&	$	51	$	&	$	0.23	$	&	$	1.11	$	&	$	4.36	$	&	$	0.52	$	&	$	0.98	$	&	$	0.90	$	&	$	51	$	&	$	0.26	$	&	$	1.26	$	&	$	6.56	$	&	$	0.49	$	&	$	1.01	$	&	$	0.94	$	&	$	78	$	\\
$	28	$	&	$	0.26	$	&	$	1.24	$	&	$	1.74	$	&	$	0.36	$	&	$	1.03	$	&	$	1.04	$	&	$	55	$	&	$	0.24	$	&	$	1.14	$	&	$	2.66	$	&	$	0.48	$	&	$	1.08	$	&	$	1.28	$	&	$	55	$	&	$	0.22	$	&	$	1.03	$	&	$	3.08	$	&	$	0.43	$	&	$	1.01	$	&	$	0.99	$	&	$	109	$	\\
$	29	$	&	$	0.35	$	&	$	1.50	$	&	$	5.72	$	&	$	0.65	$	&	$	1.02	$	&	$	1.04	$	&	$	51	$	&	$	0.33	$	&	$	1.42	$	&	$	6.92	$	&	$	0.68	$	&	$	1.01	$	&	$	0.96	$	&	$	51	$	&	$	0.27	$	&	$	1.17	$	&	$	8.25	$	&	$	0.73	$	&	$	1.01	$	&	$	0.94	$	&	$	70	$	\\
$	30	$	&	$	0.18	$	&	$	0.95	$	&	$	6.19	$	&	$	0.25	$	&	$	1.00	$	&	$	0.99	$	&	$	50	$	&	$	0.26	$	&	$	1.39	$	&	$	2.82	$	&	$	0.32	$	&	$	1.00	$	&	$	1.05	$	&	$	50	$	&	$	0.26	$	&	$	1.35	$	&	$	2.66	$	&	$	0.34	$	&	$	1.02	$	&	$	0.95	$	&	$	95	$	\\
$	31	$	&	$	0.25	$	&	$	1.37	$	&	$	6.14	$	&	$	0.76	$	&	$	1.03	$	&	$	0.99	$	&	$	85	$	&	$	0.23	$	&	$	1.26	$	&	$	5.59	$	&	$	0.80	$	&	$	1.02	$	&	$	1.00	$	&	$	85	$	&	$	0.27	$	&	$	1.47	$	&	$	2.13	$	&	$	0.80	$	&	$	0.99	$	&	$	0.92	$	&	$	83	$	\\
$	32	$	&	$	0.69	$	&	$	3.04	$	&	$	4.60	$	&	$	0.60	$	&	$	1.00	$	&	$	1.00	$	&	$	55	$	&	$	0.34	$	&	$	1.50	$	&	$	8.29	$	&	$	0.53	$	&	$	1.01	$	&	$	1.14	$	&	$	55	$	&	$	0.34	$	&	$	1.48	$	&	$	4.36	$	&	$	0.52	$	&	$	1.01	$	&	$	0.95	$	&	$	63	$	\\
$	33	$	&	$	0.23	$	&	$	1.30	$	&	$	5.77	$	&	$	0.18	$	&	$	1.04	$	&	$	1.04	$	&	$	36	$	&	$	0.27$ &			$1.49	$	&	$	5.46	$	&	$	0.25	$	&	$	1.02	$	&	$	1.05	$	&	$	36	$	&	$	0.23	$	&	$	1.29	$	&	$	6.43	$	&	$	0.23	$	&	$	0.99	$	&	$	0.92	$	&	$	75	$	\\

\hline
\end{tabular}}
\end{table*}

\section{Spectroscopic observations}
\label{sec:observations_and_analysis}
Having obtained a large sample of \UCMG\ candidates, the natural next step is their spectroscopical confirmation. In other terms, a spectroscopic confirmation of their photometric redshifts is crucial to confirm them as \UCMGs\, since both compactness and massiveness are originally based on the \zp\ associated to the photometric sample.
In this work we present the spectroscopic follow-up of 33 objects. Twenty-nine candidates are extracted from \UCMGfull, while the remaining 4 come from the data sample assembled in T16\footnote{The sample in T16 was assembled in the early 2015, applying the same criteria listed in Section \ref{subsec:UCMG_criteria}. It consisted of a mixture of the 149 survey tiles from KiDS--DR1/2 (\citealt{deJong+15_KiDS_paperI}) and few other tiles that have been part of subsequent releases. Although this datasample and the \KiDSfull\ one are partially overlapping in terms of sky coverage, they differ in the photometry, structural parameter values and photometric redshifts.}. The basic photometric properties of these 33 objects are reported in Table \ref{tab:photometry}. The structural parameters and the \emph{r}-band 2D fit outputs derived from \twodphot\ are reported in \Tab\ref{tab:struc_parameters} and the fits themselves are showed in \Fig\ref{ucmg2}\footnote{The r-band KIDS
images sometimes seem to suggest some stripping or interactions with other systems. However the majority of the spectra are typical of a passive, old stellar population. Moreover, we also note that according to the simulations presented in \cite{Wellons+15_lower_z},  compact galaxies can undertake a variety of evolutionary paths, including some interaction with a close-by companion, without changing their compactness.}. 

Data have been collected in the years 2017 and 2018 during three separate runs, two carried out with the 3.6m Telescopio Nazionale Galileo (TNG), and one using the 2.54m Isaac Newton Telescope (INT), both located at Roque de los Muchachos Observatory (Canary Islands). We thus divide our sample into three sub-groups, according to the observing run they belong to: \textsc{ucmg\_int\_2017}, \textsc{ucmg\_tng\_2017} and \textsc{ucmg\_tng\_2018}. They 
consist of 13, 11 and 9 \UCMG\ candidates, respectively, with $\texttt{MAG\_AUTO\_r}\lesssim 20.5$ and $\zp \lesssim 0.45$.

In the following sections, we discuss the instrumental and observational set-up as well as the data reduction steps for the two different instrumentation. Then we describe the \SN\ determination and the redshift and velocity dispersion calculation, obtained with the new Optimised Modelling of Early-type Galaxy Aperture Kinematics pipeline (OMEGA-K, D'Ago et al., in prep). 

\begin{figure*}
\centering
\includegraphics[width=17.2cm]{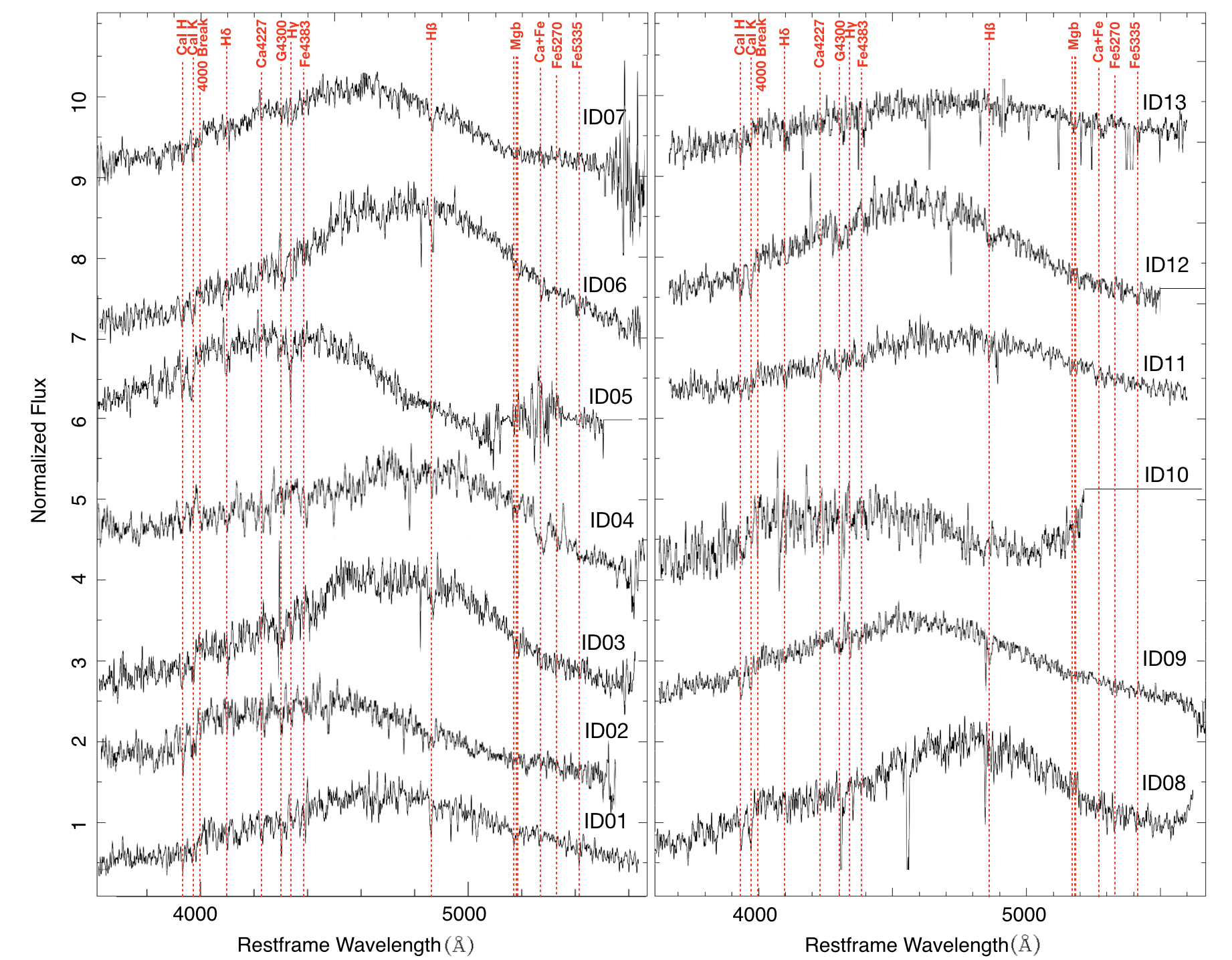}
\caption{Spectra of the 13 candidates observed in our spectroscopic campaign with INT ($\textsc{ucmg\_int\_2017}$), for which we obtain a spectroscopic redshift estimation. The spectra are plotted in ascending order of ID, which is reported above each corresponding spectrum, and refers to the IDs in \Tab\ref{tab:redshift_sigma1}. We only show the wavelength region that was used to derive the redshift and to compute the velocity dispersion. This region includes some of the most common stellar absorption lines such as Ca-H, Ca-K, Balmer lines (H$_{\delta}$, H$_{\gamma}$ and H$_{\beta}$), Mgb, and Fe lines.  The spectra are plotted in rest-framed wavelength, in unit of normalized flux (each spectrum has been divided by its median) and they are vertically shifted for better visualization. In some cases, when the red part of the spectrum was particularly noisy, we cut it out to improve the figure layout.}
\label{fig:INT}
\end{figure*} 

\begin{figure*}
\centering
\includegraphics[width=17.2cm]{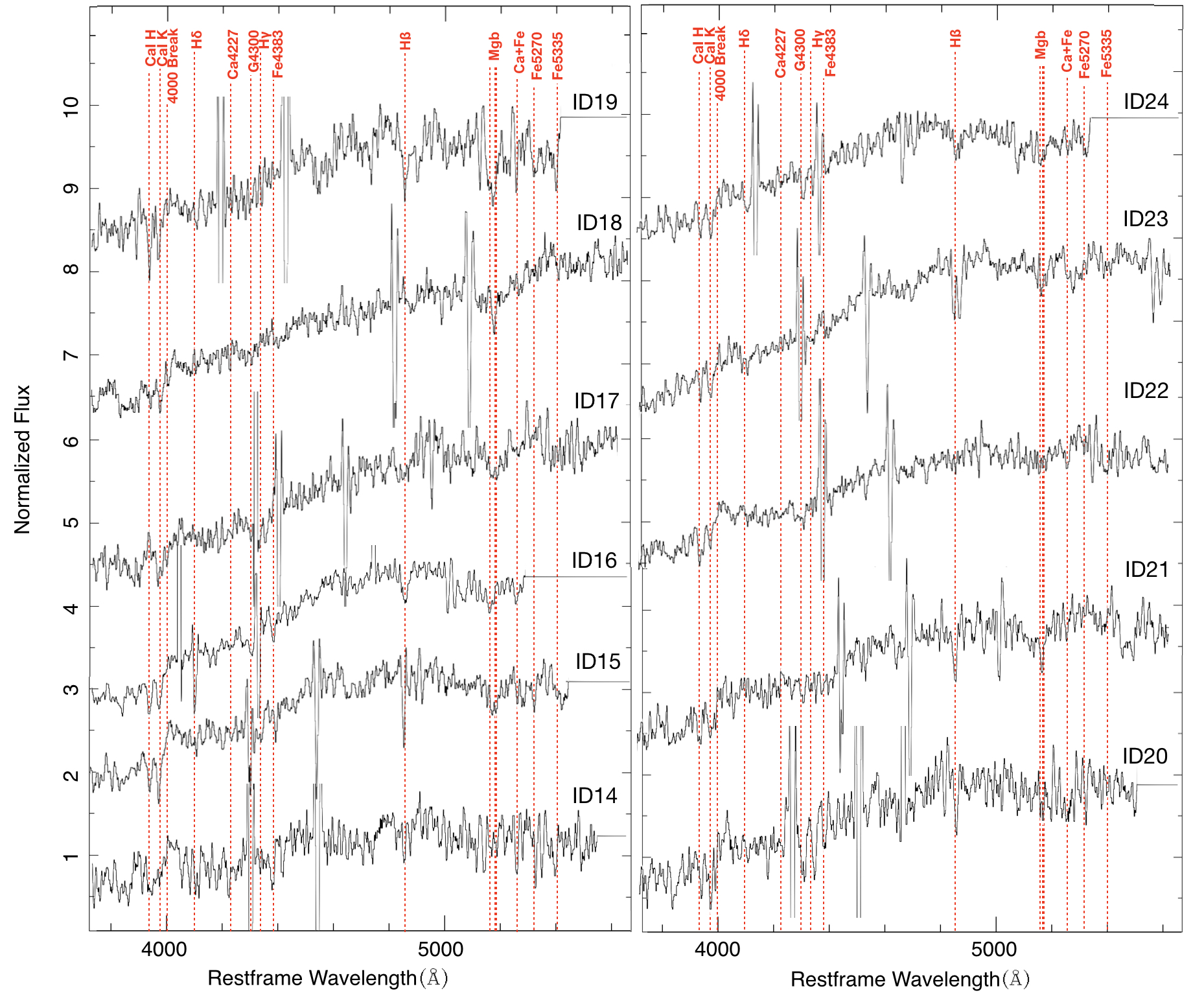}
\caption{Same as Figure \ref{fig:INT}, but for the 11 candidates observed in our spectroscopic campaign with TNG ($\textsc{ucmg\_tng\_2017}$), for which we obtain a spectroscopic redshift estimation. }\label{fig:TNG_A34_spectra}
\end{figure*}

\begin{figure}
\centering
\includegraphics[width=0.46\textwidth]{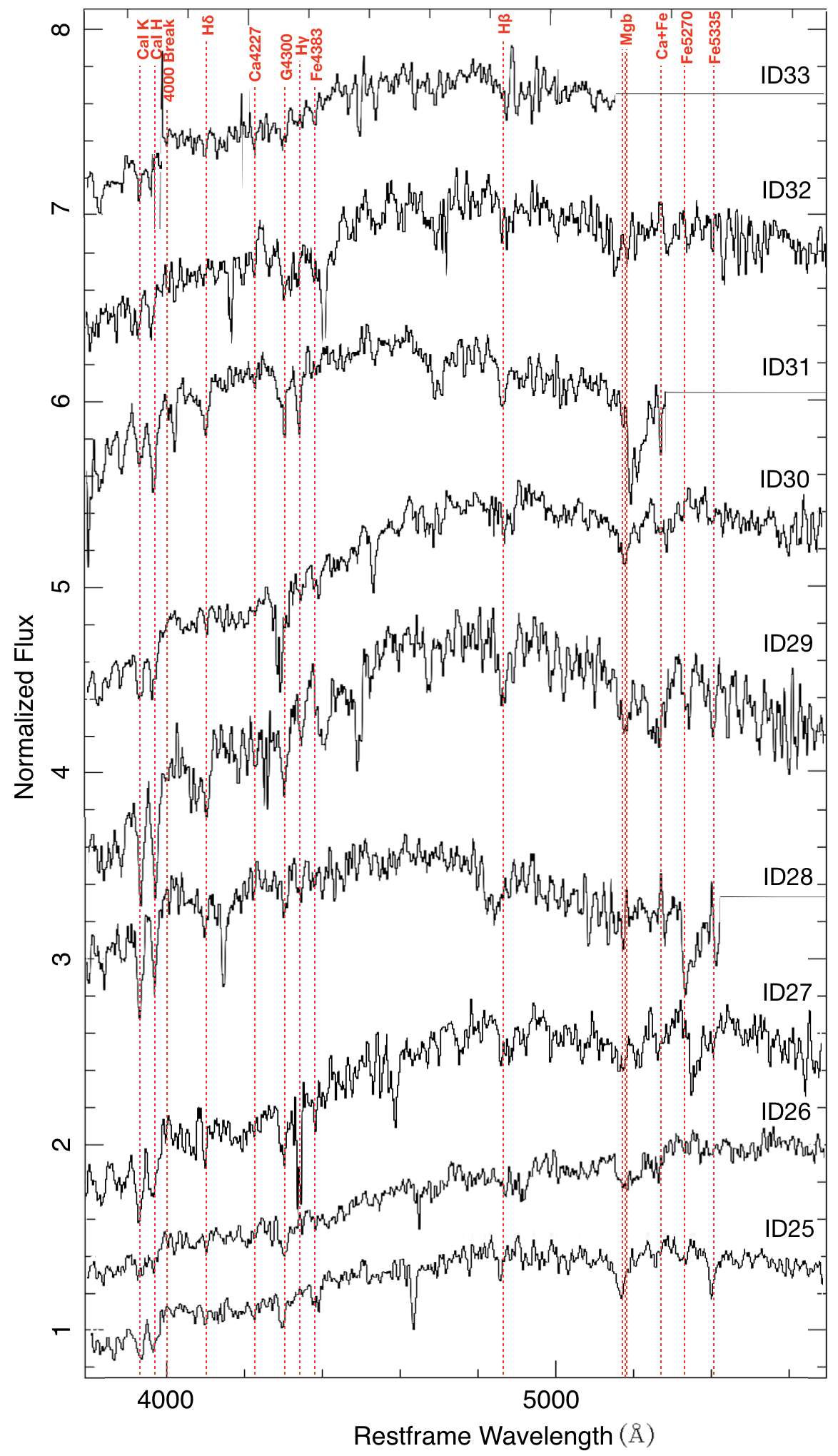}
\caption{Same as Figure \ref{fig:INT}, but for the 9 candidates observed in our spectroscopic campaign with TNG ($\textsc{ucmg\_tng\_2018}$), for which we obtain a spectroscopic redshift estimation.}
\label{fig:A36}
\end{figure}

\subsection{INT spectroscopy}
\label{INT_spec} 
Data on 13 luminous \UCMG\ candidates belonging to the $\textsc{ucmg\_int\_2017}$ sample have been obtained with the IDS spectrograph during 6 nights at the INT telescope, in visitor mode (PI: C. Tortora, ID: 17AN005).
The observations have been carried on with the RED+2 detector and the low resolution grating R400V, covering the wavelength range from 4000 to 8000 \AA. 
The spectra have been acquired with long-slits of $1.6\arcsec$ or $2\arcsec$ width, providing a spectral resolution of $\Delta \lambda /\lambda = 560$, a dispersion of 1.55 \AA/pixel, and a pixel scale of $0.33$ arcsec/pixel. The average seeing during the observing run was FWHM $\sim1.5\arcsec$, the single exposure time ranged between 600 and 1200 seconds and from 1 up to 5 single exposures have been obtained per target, depending on their magnitudes.

Data reduction has been performed using \textsc{iraf}\footnote{\textsc{iraf} is distributed by the National Optical Astronomy Observatories, which is operated by the Associated Universities for Research in Astronomy, Inc. under cooperative agreement with the National Science Foundation.} image processing packages. 
The main data reduction steps include dark subtraction, flat-fielding correction and  sky subtraction. The wavelength calibration has been performed by means of comparison spectra of CuAr$+$CuNe lamps acquired for each observing night using the \textsc{identify} task. 
A sky spectrum has been extracted from the outer edges of the slit, and subtracted from each row of the two dimensional spectra using the \textsc{iraf} task \textsc{background} in the \textsc{twodspec.longslit} package. 
The sky-subtracted frames have been co-added to averaged 2D spectra and then the 1D spectra, which have been used to derive the spectroscopic redshifts, have been obtained extracting and summing up the lines with higher \SN\ using the task \textsc{scopy}.

The 1D reduced spectra are showed in \Fig\ref{fig:INT}. They are plotted in rest-framed wavelength from $\sim$ 3600 to $\sim$ 5600 \AA\ and units of normalized flux (each spectrum has been divided by its median). The spectra are vertically shifted for better visualization.  Vertical red dotted lines show absorption spectral features typical of an old stellar population. 


\subsection{TNG spectroscopy}\label{TNG_spec} 

The 20 spectra of \UCMG\ candidates in the $\textsc{ucmg\_tng\_}$ $\textsc{2017}$ and $\textsc{ucmg\_tng\_2018}$ samples have been collected using the Device Optimized for the Low RESolution (DOLORES) spectrograph mounted on the 3.5m TNG, during 6 nights in 2017 and 2018 (PI: N.R. Napolitano,  ID: A34TAC\_22 and A36TAC\_20).
The instrument has a 2k $\times$ 2k CCD detector with a pixel scale of 0.252 arcsec/pixel. The observations for both subsamples have been carried out with the LR-B grism with dispersion of 2.52 \AA/pixel and resolution of 585 (calculated for a slit width of $1\arcsec$), covering the wavelength range from 4000 to 8000 \AA.
As in the previous case, we have obtained from 1 to 5 single exposures per target, each with exposure time ranging between 600 and 1200 seconds. 
Following T18, the DOLORES 2D spectra have been flat-fielded, sky-subtracted and wavelength calibrated using the HgNe arc lamps. Then, the 1D spectra have been extracted by integrating over the source spatial profile. All these procedures have been performed using the same standard \textsc{iraf} tasks, as explained in \Sec\ref{INT_spec}. 
The TNG spectra are showed in \Figs\ref{fig:TNG_A34_spectra} and \ref{fig:A36}, using the same units and scale of \Fig\ref{fig:INT}. Similarly to the previous case, the main stellar absorption features are highlighted with vertical red dotted lines. 

\subsection{Spectroscopic signal-to-noise ratio determination}
\label{subsec:SN}
To calculate the signal-to-noise ratio \textbf{($S/N_{spec}$)} of the integrated spectra we use the IDL code \texttt{}{DER\_SNR}\footnote{The code is written by Felix Stoehr and published on the ST-ECF Newsletter, Issue num. 42. The software is available here: \url{www.stecf.org/software/ASTROsoft/DER_SNR/}; the Newsletter can be found here: \url{www.spacetelescope.org/about/further_information/stecfnewsletters/hst_stecf_0042/}.}. 
The code estimates the derived \SN\ from the flux under the assumptions that the noise is uncorrelated in wavelength bins spaced two pixels apart and that it is approximately Gaussian distributed.
The biggest advantages of using this code are that it is very simple and robust and, above all, it computes the \SN\ from the data alone.  
In fact, the noise is calculated directly from the flux using the following equation:
\begin{equation}
N  =   \dfrac{1.482602}{\sqrt6 \times \langle |2S(i) - S(i-2) - S(i+2)| \rangle}  
\end{equation}
where $S$ is the signal (taken to be the flux of the continuum level), the index $i$ runs over the pixels, and the ``$\langle \rangle$" symbol indicates a median calculation done over all the non-zero pixels in the restframe wavelength range 3600 $-$ 4600 \AA, which is the common wavelength range for all the spectra, including the T18 ones, for which we determine, in the next section, also the velocity dispersion. 
We note that these signal-to-noise ratio estimates have to be interpreted as lower limit for the whole spectrum, since they are calculated over a rather blue wavelength range, whereas the light of early-type galaxies is expected to be strong in redder regions. 
This arises clearly from the comparison of these \textbf{$S/N_{spec}$} with the ones we will describe in the next section, which are computed, for each galaxy, over the region used for the kinematic fit and are systematically larger. Both of them will be used in Section \ref{FJr} as one of the proxy of the reliability of the velocity dispersion ($\sigma$) measurements.

\subsection{Redshift and velocity dispersion measurements}
\label{redshift and velocity}
Redshift and velocity dispersion values have been measured with the Optimised Modelling of Early-type Galaxy Aperture Kinematics pipeline (OMEGA-K, \citealt{dago+18proc}), a Python wrapper based on the Penalized Pixel-Fitting code (\textsc{pPXF}, \citealt{Cappellari17}). 

OMEGA-K comprises a graphical user interface (PPGUI, written by G. D'Ago and to be distributed soon) that allows the user to visualize and inspect the observed spectrum in order to easily set the \textsc{pPXF} fitting parameters (i.e., template libraries, noise level, polynomials, fit wavelength range and custom pixel masks). We use PPGUI to restframe the spectra and obtain a first guess of the redshift, initially based on the \zp. 

The aim of OMEGA-K is to automatically retrieve an optimal pixel mask and noise level (1$\sigma$ noise spectrum) for the observed spectrum, and to find a robust estimate of the galaxy kinematics together with its uncertainties by randomizing the initial condition for \textsc{pPXF} and running it hundreds of times on the same observed spectrum, to which a Gaussian noise is randomly added. 

As templates for the fitting we use a selection of 156 MILES simple single stellar population (SSP) models from \citet{Vazdekis10}, covering a wide range of metallicities ($0.02 \leq Z/\Zsun \leq 1.58$) and ages (between 3 Gyr and 13 Gyr). We also perform the fitting using single stars (268 empirical stars from MILES library, uniformly sampling effective temperature, metallicity and surface gravity of the full catalogue of templates) and also including templates with ages $ < 3$ Gyr.

The results do not change and are always consistent within the errors, demonstrating that the choice of the templates does not influence the fitting results.\footnote{We note that the stellar templates are used only to infer the kinematics, i.e. to measure the shift and the broadening of the stellar absorption lines. Given the low S/N of our spectra, we do not perform any spectroscopic stellar population analysis.} 
Finally, an additive polynomial is also applied in order to take into account possible template shape and continuum mismatches and correct for imperfect sky subtraction or scattered light. 

For a general description of the OMEGA-K pipeline, we refer the reader to above mentioned reference (see also \citealt{dago+18proc}) and the paper in preparation (D'Ago et al. in prep.). Here, we list the main steps of the OMEGA-K run specifically adopted for this work on a single observed spectrum:

\begin{enumerate}
\item the observed spectrum and the template libraries are ingested; 
\item the optimal 1$\sigma$ noise spectrum and pixel mask are automatically tuned;
\item 256 Monte Carlo re-samplings of the observed spectrum using a random Gaussian noise from the 1$\sigma$ noise spectrum are produced;
\item 256 sets of initial guesses (for the redshift and the velocity dispersion) and of fitting parameters (additive polynomial degree, number of momenta of the line-of-sight velocity distribution to be fitted, random shift of the fitting wavelength range) are produced in order to allow for a complete bootstrap approach within the parameter space, and to avoid internal biases in the pipeline;
\item 256 \textsc{pPXF} runs are performed in parallel and the results from each run are stored (outliers and too noisy reproductions of the observed spectra are automatically discarded);
\item the final redshift and velocity dispersion for each observed spectrum, together with their error are defined as the mean and the standard deviation of the result distribution from the accepted fits.
\end{enumerate}

Among the 257 fits performed on each spectrum (256 from the OMEGA-K bootstrap stage, plus the fit on the original observed spectrum), we discard the ones for which the best fit fails to converge or the measured kinematics is unrealistically low or unrealistically high. As a lower and upper limit on the velocity, we choose thresholds of 110 and 500 \kms, respectively. The low limit is slightly smaller than the typical velocity scale of the instrument, which we measure to be $\sim$120 \kms. On the other hand, we used 500 \kms as a high upper limit in order to incorporate any possible source of uncertainty related to the pipeline, without artificially reduce the errors on our estimates.

We define the success rate (SR) as the ratio between the number of accepted  fits over the total 257 attempts. 

Finally, OMEGA-K derives a mean spectrum of the accepted  fits and performs a measurement of the S/N on its residuals ($(S/N)_{\rm O-K}$). \citet{dago+18proc} showed, using mock data, a large sample of SDSS spectra and the entire GAMA DR3 spectroscopic database, that kinematics values with SR $>65\%$ and $(S/N)_{\rm O-K}>5/$px can be considered totally reliable. This S/N ratio is also consistent with what found in \citet[][and reference therein]{Hopkins+13}. 

The uncertainties on our measures are unfortunately very large. To assess the effect of such large errors on our findings, we separate the \UCMGs\ in two groups: those with ``high-quality" (HQ) velocity dispersion measurements and those with ``low-quality" (LQ) ones. For this purpose, we use a combination of three ``quality criteria": the aforementioned SR, the spectral S/N calculated on a common wavelength range covered by all the spectra (see Section \ref{subsec:SN}) and the $(S/N)_{\rm O-K}$ from the OMEGA-K pipeline (calculated over different wavelength ranges for different spectra). 
We visually inspect the spectra and their fit one by one in order to set reliable thresholds for these criteria. We set-up the following quality low limits: SR $=0.3$, $S/N_{spec}=3.5$ and $(S/N)_{\rm O-K}=6.5$/px. 
We then classify as HQ objects, the ones above these limits. 

\begin{figure}
\includegraphics[width=0.47\textwidth]{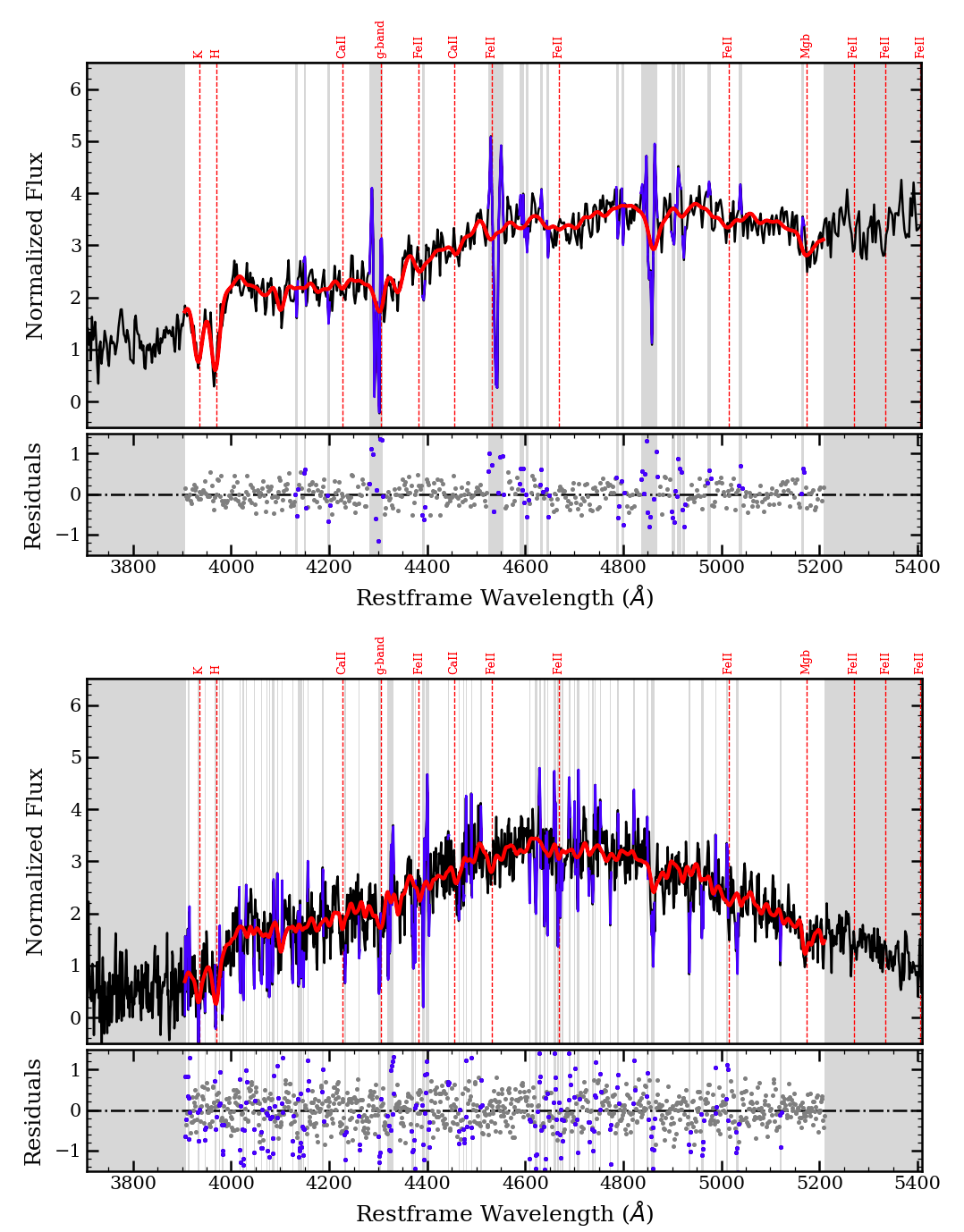}
\caption{Two examples of ppxf fits obtained with OMEGA-K on the spectra of two different \UCMGs, one of the best HQ system and one of the worse LQ system, thus representative of the whole sample, observed with two different telescopes. For each panel we plot the galaxy spectrum in black, the best template fit in red and the regions excluded from the fit as blue lines. We note that the fit is performed only outside the gray shaded regions. Finally, we highlight stellar absorption lines in red and show the residuals of the plot below each panel.}
\label{fig:ppxf}
\end{figure}

In Figure \ref{fig:ppxf} we show two examples of the ppxf fit obtained with OMEGA-K on the spectra of two different objects from the sample of the 33 \UCMG\ candidates for which we obtain new spectroscopy in this paper. These two spectra are representative of the full sample since they have been observed with two different instruments and one is classified as high-quality while the other as low-quality. 
The upper panel shows the galaxy KIDS J090412.45-001819.75 (ID = $15$),  from the  $\textsc{ucmg\_tng\_2017}$ sample, which is classified as HQ and has a large velocity dispersion  ($\sigma=412 \pm 81$ \kms). Instead, the lower panel shows the spectrum of the galaxy KIDS J085700.29--010844.55 (ID = $1$), which belongs to $\textsc{ucmg\_int\_2017}$. This object, classified as LQ, has a relatively lower velocity dispersion ($\sigma = 187\pm 85$ \kms) and is one of the worse cases with very low spectral S/N.

In addition to the 33 new \UCMG\ candidates presented in this paper, we also apply the same kinematics procedure to the 28 \UCMG\ candidates from T18, 6 observed with TNG and 22 with NTT, to which we refer to as $\textsc{ucmg\_tng\_t18}$ and $\textsc{ucmg\_ntt\_t18}$ sample, respectively. 

In general, the velocity dispersion values from OMEGA-K are derived from 1D spectra using various slit widths and extracted using different number of pixels along the slit length. This means that the velocity dispersion values are computed integrating light in apertures with different sizes. The ranges of aperture and slit widths for the new 33 objects presented here and the 28 \UCMG\ candidates from T18 are $1.8\arcsec-3.2\arcsec$ and $1.2\arcsec-2\arcsec$, respectively. This is not an ideal situation if we want to compare velocity dispersion values among different systems and use these measurements to derive scaling relations. We will come back to this specific topic in \Sec \ref{FJr}.
Briefly, in order to uniform the estimates, and correct the velocity dispersion values for the different apertures, 
we first convert the rectangular aperture adopted to extract the \UCMG\ 1D spectra to an equivalent circular aperture of radius $R = 1.025 \sqrt{(\delta x \delta y /\pi)}$, where $\delta x$ and $\delta y$ are the width and length used to extract the spectrum\footnote{The same formula was adopted in \cite{Tortora+14_DMevol}, but reported with a typo in the printed copy of the paper.}. Then, we use the average velocity dispersion profile in \cite{Cappellari+06}, to extrapolate this equivalent velocity dispersion to the effective radius. 

\Tab\ref{tab:redshift_sigma1} and \Tab\ref{tab:redshift_sigma2}  list the results of the fitting procedure for our sample and that of T18. We report the measured spectroscopic redshifts and the velocity dispersion values, each with associate error, the velocity dispersion values corrected to the effective radii ($\sigma_{\rm e}$) and, the equivalent circular apertures for the whole sample of 61 \UCMGs.
We also present the photometric redshifts to provide a direct comparison with the spectroscopic. Finally, the four following columns indicate the three parameters we use to split the sample in high- and low-quality and the resulting classification for each object.  

In addition, we correct the value of the spectroscopic redshift for the object with ID number 46 (corresponding to ID 13 in T18) respect to the wrong one reported in T18. Although this changes the value of \Re\, the result of the spectroscopic validation remains unchanged and the galaxy is still a confirmed \UCMG. The 28 galaxies from T18 are reported in the same order as the previous paper but continuing the numeration (in terms of ID) of this paper.


\begin{table*}
\centering \caption{Results of the fitting procedure on the spectra belonging to the three observational runs presented here: $\textsc{ucmg\_int\_2017}$, $\textsc{ucmg\_tng\_2017}$, $\textsc{ucmg\_tng\_2018}$. The columns from left to right list: the galaxy ID, the photometric redshift, the measured spectroscopic redshift with its error, the measured velocity dispersion in \kms\ with its error, the corrected velocity dispersion to the effective radius, the equivalent circular aperture in arcsec. Finally, in the final four columns, we also report the success rate, the signal-to-noise ratio per pixel calculated in the range 3600 $-$ 4600 \AA, the  signal-to-noise ratio per pixel calculated over the region used for the fit by OMEGA-K, and the quality level of the velocity dispersion estimates, based on these three quality parameters.}

\label{tab:redshift_sigma1}

{
\begin{tabular}{lccccccccc}
\hline

ID & $\zp$ & $\zs \pm \Delta z_{\rm spec} $ & $\sigma \pm \Delta \sigma$ & $\sigma_{\rm e}$ & Aperture  & SR &  $(S/N)_{\rm spec}$ & $(S/N)_{\rm O-K}$ & Quality level  \\

\hline
\hline
$	1	$	&	$	0.28	$	&	$	0.2696	\pm	0.0002	$	&	$	197	\pm	85	$	&	$	211	$	&	$	0.97	$	&	$	0.62	$	&	$	1.99	$	&	$	6.13	$	&	$	LQ	$	\\
$	2	$	&	$	0.26	$	&	$	0.3158	\pm	0.0002	$	&	$	195	\pm	52	$	&	$	210	$	&	$	0.97	$	&	$	0.77	$	&	$	3.21	$	&	$	5.69	$	&	$	LQ	$	\\
$	3	$	&	$	0.26	$	&	$	0.2995	\pm	0.0003	$	&	$	268	\pm	76	$	&	$	291	$	&	$	1.21	$	&	$	0.79	$	&	$	2.50	$	&	$	6.19	$	&	$	LQ	$	\\
$	4	$	&	$	0.45	$	&	$	0.3084	\pm	0.0005	$	&	$	234	\pm	86	$	&	$	281	$	&	$	0.97	$	&	$	0.30	$	&	$	2.18	$	&	$	4.23	$	&	$	LQ	$	\\
$	5	$	&	$	0.37	$	&	$	0.4401	\pm	0.0003	$	&	$	142	\pm	33	$	&	$	161	$	&	$	0.97	$	&	$	0.07	$	&	$	4.00	$	&	$	6.87	$	&	$	LQ	$	\\
$	6	$	&	$	0.22	$	&	$	0.2988	\pm	0.0002	$	&	$	202	\pm	48	$	&	$	217	$	&	$	1.21	$	&	$	0.75	$	&	$	2.42	$	&	$	7.27	$	&	$	LQ	$	\\
$	7	$	&	$	0.23	$	&	$	0.3221	\pm	0.0002	$	&	$	208	\pm	84	$	&	$	224	$	&	$	0.97	$	&	$	0.15	$	&	$	2.96	$	&	$	6.71	$	&	$	LQ	$	\\
$	8	$	&	$	0.24	$	&	$	0.2976	\pm	0.0002	$	&	$	241	\pm	100	$	&	$	257	$	&	$	0.97	$	&	$	0.59	$	&	$	3.06	$	&	$	6.31	$	&	$	LQ	$	\\
$	9	$	&	$	0.34	$	&	$	0.2915	\pm	0.0001	$	&	$	227	\pm	84	$	&	$	251	$	&	$	0.97	$	&	$	0.21	$	&	$	4.07	$	&	$	6.04	$	&	$	LQ	$	\\
$	10	$	&	$	0.32	$	&	$	0.3590	\pm	0.0004	$	&	$	265	\pm	100	$	&	$	293	$	&	$	0.97	$	&	$	0.12	$	&	$	2.00	$	&	$	2.05	$	&	$	LQ	$	\\
$	11	$	&	$	0.24	$	&	$	0.2797	\pm	0.0003	$	&	$	260	\pm	94	$	&	$	286	$	&	$	0.97	$	&	$	0.85	$	&	$	1.40	$	&	$	4.58	$	&	$	LQ	$	\\
$	12	$	&	$	0.28	$	&	$	0.3312	\pm	0.0002	$	&	$	202	\pm	59	$	&	$	218	$	&	$	0.97	$	&	$	0.73	$	&	$	2.70	$	&	$	6.76	$	&	$	LQ	$	\\
$	13	$	&	$	0.23	$	&	$	0.2668	\pm	0.0007	$	&	$	259	\pm	113	$	&	$	274	$	&	$	0.97	$	&	$	0.23	$	&	$	1.77	$	&	$	2.89	$	&	$	LQ	$	\\
\hline
$	14	$	&	$	0.35	$	&	$	0.2946	\pm	0.0003	$	&	$	340	\pm	99	$	&	$	369	$	&	$	0.94	$	&	$	0.66	$	&	$	2.01	$	&	$	3.97	$	&	$	LQ	$	\\
$	15	$	&	$	0.27	$	&	$	0.2974	\pm	0.0002	$	&	$	412	\pm	81	$	&	$	451	$	&	$	1.07	$	&	$	0.69	$	&	$	6.90	$	&	$	13.25	$	&	$	HQ	$	\\
$	16	$	&	$	0.33	$	&	$	0.3594	\pm	0.0001	$	&	$	268	\pm	84	$	&	$	292	$	&	$	1.01	$	&	$	0.84	$	&	$	6.87	$	&	$	14.32	$	&	$	HQ	$	\\
$	17	$	&	$	0.33	$	&	$	0.2656	\pm	0.0006	$	&	$	321	\pm	93	$	&	$	347	$	&	$	1.01	$	&	$	0.43	$	&	$	1.95	$	&	$	8.20	$	&	$	LQ	$	\\
$	18	$	&	$	0.32	$	&	$	0.1586	\pm	0.0002	$	&	$	253	\pm	92	$	&	$	276	$	&	$	1.01	$	&	$	0.70	$	&	$	2.93	$	&	$	12.76	$	&	$	LQ	$	\\
$	19	$	&	$	0.30	$	&	$	0.3281	\pm	0.0002	$	&	$	230	\pm	91	$	&	$	251	$	&	$	1.18	$	&	$	0.30	$	&	$	2.97	$	&	$	6.27	$	&	$	LQ	$	\\
$	20	$	&	$	0.30	$	&	$	0.2728	\pm	0.0003	$	&	$	331	\pm	92	$	&	$	361	$	&	$	1.12	$	&	$	0.21	$	&	$	2.85	$	&	$	5.58	$	&	$	LQ	$	\\
$	21	$	&	$	0.33	$	&	$	0.2523	\pm	0.0003	$	&	$	323	\pm	95	$	&	$	366	$	&	$	1.12	$	&	$	0.85	$	&	$	2.62	$	&	$	9.93	$	&	$	LQ	$	\\
$	22	$	&	$	0.28	$	&	$	0.2719	\pm	0.0002	$	&	$	355	\pm	99	$	&	$	413	$	&	$	1.18	$	&	$	0.66	$	&	$	5.91	$	&	$	12.72	$	&	$	HQ	$	\\
$	23	$	&	$	0.25	$	&	$	0.2971	\pm	0.0002	$	&	$	407	\pm	56	$	&	$	443	$	&	$	1.12	$	&	$	0.79	$	&	$	6.18	$	&	$	17.38	$	&	$	HQ	$	\\
$	24	$	&	$	0.33	$	&	$	0.3491	\pm	0.0002	$	&	$	194	\pm	64	$	&	$	215	$	&	$	1.07	$	&	$	0.23	$	&	$	5.79	$	&	$	11.15	$	&	$	LQ	$	\\
\hline
$	25	$	&	$	0.28	$	&	$	0.2703	\pm	0.0002	$	&	$	274	\pm	57	$	&	$	298	$	&	$	1.12	$	&	$	0.91	$	&	$	6.80	$	&	$	18.11	$	&	$	HQ	$	\\
$	26	$	&	$	0.32	$	&	$	0.1984	\pm	0.0002	$	&	$	287	\pm	57	$	&	$	316	$	&	$	1.18	$	&	$	0.89	$	&	$	3.96	$	&	$	17.92	$	&	$	HQ	$	\\
$	27	$	&	$	0.33	$	&	$	0.2843	\pm	0.0002	$	&	$	241	\pm	53	$	&	$	267	$	&	$	1.23	$	&	$	0.91	$	&	$	5.08	$	&	$	15.85	$	&	$	HQ	$	\\
$	28	$	&	$	0.33	$	&	$	0.4203	\pm	0.0002	$	&	$	172	\pm	63	$	&	$	191	$	&	$	1.18	$	&	$	0.02	$	&	$	6.59	$	&	$	11.69	$	&	$	LQ	$	\\
$	29	$	&	$	0.29	$	&	$	0.3116	\pm	0.0002	$	&	$	164	\pm	39	$	&	$	177	$	&	$	1.01	$	&	$	0.52	$	&	$	7.74	$	&	$	15.65	$	&	$	HQ	$	\\
$	30	$	&	$	0.39	$	&	$	0.2994	\pm	0.0002	$	&	$	289	\pm	52	$	&	$	319	$	&	$	1.12	$	&	$	1.00	$	&	$	8.53	$	&	$	24.59	$	&	$	HQ	$	\\
$	31	$	&	$	0.41	$	&	$	0.4655	\pm	0.0001	$	&	$	253	\pm	57	$	&	$	280	$	&	$	1.18	$	&	$	0.98	$	&	$	9.18	$	&	$	18.13	$	&	$	HQ	$	\\
$	32	$	&	$	0.29	$	&	$	0.3382	\pm	0.0003	$	&	$	277	\pm	85	$	&	$	301	$	&	$	1.18	$	&	$	0.88	$	&	$	3.51	$	&	$	9.73	$	&	$	HQ	$	\\
$	33	$	&	$	0.43	$	&	$	0.4028	\pm	0.0003	$	&	$	299	\pm	91	$	&	$	335	$	&	$	1.28	$	&	$	0.84	$	&	$	4.96	$	&	$	9.16	$	&	$	HQ	$	\\

\hline
\end{tabular}
}
\end{table*}

\begin{table*}

\centering \caption{Same as \Tab\ref{tab:redshift_sigma1}, but for the samples $\textsc{ucmg\_tng\_t18}$ and $\textsc{ucmg\_ntt\_t18}$.}

\label{tab:redshift_sigma2}
{\begin{tabular}{lccccccccc}
\hline

ID & $z_{\rm phot}$ & $z_{\rm spec} \pm \Delta z_{\rm spec} $ & $\sigma \pm \Delta \sigma$ & $\sigma_{\rm e}$ & Aperture  & SR &  $(S/N)_{\rm spec}$ & $(S/N)_{\rm O-K}$ & Quality level  \\

\hline
\hline
$	34	$	&	$	0.29	$	&	$	0.3705	\pm	0.0001	$	&	$	361	\pm	63	$	&	$	392	$	&	$	1.12	$	&	$	0.98	$	&	$	15.05	$	&	$	22.41	$	&	$	HQ	$	\\
$	35	$	&	$	0.22	$	&	$	0.2175	\pm	0.0004	$	&	$	404	\pm	101	$	&	$	446	$	&	$	1.59	$	&	$	0.31	$	&	$	7.68	$	&	$	14.62	$	&	$	HQ	$	\\
$	36	$	&	$	0.35	$	&	$	0.4078	\pm	0.0002	$	&	$	366	\pm	79	$	&	$	412	$	&	$	1.33	$	&	$	0.93	$	&	$	6.70	$	&	$	14.33	$	&	$	HQ	$	\\
$	37	$	&	$	0.31	$	&	$	0.3341	\pm	0.0002	$	&	$	218	\pm	54	$	&	$	242	$	&	$	1.12	$	&	$	0.92	$	&	$	7.84	$	&	$	17.82	$	&	$	HQ	$	\\
$	38	$	&	$	0.42	$	&	$	0.3988	\pm	0.0003	$	&	$	390	\pm	71	$	&	$	448	$	&	$	1.01	$	&	$	0.75	$	&	$	5.33	$	&	$	12.67	$	&	$	HQ	$	\\
$	39	$	&	$	0.36	$	&	$	0.3190	\pm	0.0004	$	&	$	226	\pm	65	$	&	$	245	$	&	$	1.01	$	&	$	0.82	$	&	$	4.14	$	&	$	10.20	$	&	$	HQ	$	\\
\hline
$	40	$	&	$	0.20	$	&	$	0.3019	\pm	0.0002	$	&	$	432	\pm	41	$	&	$	464	$	&	$	0.69	$	&	$	0.73	$	&	$	2.09	$	&	$	6.75	$	&	$	LQ	$	\\
$	41	$	&	$	0.35	$	&	$	0.3853	\pm	0.0001	$	&	$	211	\pm	40	$	&	$	223	$	&	$	0.69	$	&	$	0.98	$	&	$	3.69	$	&	$	10.92	$	&	$	HQ	$	\\
$	42	$	&	$	0.28	$	&	$	0.2367	\pm	0.0003	$	&	$	225	\pm	34	$	&	$	235	$	&	$	0.69	$	&	$	1.00	$	&	$	2.38	$	&	$	9.30	$	&	$	LQ	$	\\
$	43	$	&	$	0.29	$	&	$	0.2801	\pm	0.0001	$	&	$	196	\pm	39	$	&	$	214	$	&	$	0.69	$	&	$	0.94	$	&	$	2.77	$	&	$	9.55	$	&	$	LQ	$	\\
$	44	$	&	$	0.31	$	&	$	0.2789	\pm	0.0001	$	&	$	218	\pm	34	$	&	$	235	$	&	$	0.69	$	&	$	1.00	$	&	$	3.67	$	&	$	12.46	$	&	$	HQ	$	\\
$	45	$	&	$	0.27	$	&	$	0.2888	\pm	0.0001	$	&	$	195	\pm	46	$	&	$	216	$	&	$	0.69	$	&	$	0.94	$	&	$	3.09	$	&	$	9.30	$	&	$	LQ	$	\\
$	46	$	&	$	0.31	$	&	$	0.3618	\pm	0.0053	$	&	$	181	\pm	68	$	&	$	196	$	&	$	0.69	$	&	$	0.09	$	&	$	1.39	$	&	$	4.08	$	&	$	LQ	$	\\
$	47	$	&	$	0.25	$	&	$	0.2622	\pm	0.0003	$	&	$	340	\pm	53	$	&	$	363	$	&	$	0.69	$	&	$	0.99	$	&	$	2.31	$	&	$	7.65	$	&	$	LQ	$	\\
$	48	$	&	$	0.27	$	&	$	0.2949	\pm	0.0003	$	&	$	280	\pm	50	$	&	$	295	$	&	$	0.69	$	&	$	1.00	$	&	$	3.79	$	&	$	10.53	$	&	$	HQ	$	\\
$	49	$	&	$	0.28	$	&	$	0.2974	\pm	0.0001	$	&	$	142	\pm	22	$	&	$	149	$	&	$	0.69	$	&	$	0.58	$	&	$	3.54	$	&	$	10.01	$	&	$	HQ	$	\\
$	50	$	&	$	0.29	$	&	$	0.3188	\pm	0.0001	$	&	$	387	\pm	63	$	&	$	408	$	&	$	0.69	$	&	$	0.96	$	&	$	3.88	$	&	$	11.85	$	&	$	HQ	$	\\
$	51	$	&	$	0.34	$	&	$	0.3151	\pm	0.0001	$	&	$	154	\pm	29	$	&	$	166	$	&	$	0.69	$	&	$	0.66	$	&	$	3.82	$	&	$	11.69	$	&	$	HQ	$	\\
$	52	$	&	$	0.22	$	&	$	0.2124	\pm	0.0001	$	&	$	252	\pm	43	$	&	$	265	$	&	$	0.69	$	&	$	1.00	$	&	$	1.64	$	&	$	9.19	$	&	$	LQ	$	\\
$	53	$	&	$	0.25	$	&	$	0.2578	\pm	0.0002	$	&	$	183	\pm	48	$	&	$	194	$	&	$	0.69	$	&	$	0.68	$	&	$	2.37	$	&	$	9.73	$	&	$	LQ	$	\\
$	54	$	&	$	0.34	$	&	$	0.3024	\pm	0.0009	$	&	$	214	\pm	66	$	&	$	226	$	&	$	0.69	$	&	$	0.70	$	&	$	1.97	$	&	$	4.14	$	&	$	LQ	$	\\
$	55	$	&	$	0.31	$	&	$	0.3667	\pm	0.0001	$	&	$	244	\pm	30	$	&	$	262	$	&	$	0.69	$	&	$	1.00	$	&	$	4.99	$	&	$	13.10	$	&	$	HQ	$	\\
$	56	$	&	$	0.32	$	&	$	0.4070	\pm	0.0001	$	&	$	322	\pm	54	$	&	$	342	$	&	$	0.69	$	&	$	1.00	$	&	$	4.82	$	&	$	10.60	$	&	$	HQ	$	\\
$	57	$	&	$	0.33	$	&	$	0.2612	\pm	0.0001	$	&	$	219	\pm	44	$	&	$	233	$	&	$	0.69	$	&	$	0.99	$	&	$	3.00	$	&	$	10.88	$	&	$	LQ	$	\\
$	58	$	&	$	0.27	$	&	$	0.2818	\pm	0.0002	$	&	$	218	\pm	64	$	&	$	227	$	&	$	0.69	$	&	$	0.92	$	&	$	2.41	$	&	$	7.38	$	&	$	LQ	$	\\
$	59	$	&	$	0.23	$	&	$	0.2889	\pm	0.0002	$	&	$	209	\pm	52	$	&	$	221	$	&	$	0.69	$	&	$	0.95	$	&	$	2.80	$	&	$	9.99	$	&	$	LQ	$	\\
$	60	$	&	$	0.34	$	&	$	0.3393	\pm	0.0001	$	&	$	155	\pm	30	$	&	$	167	$	&	$	0.69	$	&	$	0.73	$	&	$	4.59	$	&	$	10.78	$	&	$	HQ	$	\\
$	61	$	&	$	0.31	$	&	$	0.2889	\pm	0.0001	$	&	$	220	\pm	33	$	&	$	236	$	&	$	0.69	$	&	$	1.00	$	&	$	2.47	$	&	$	8.67	$	&	$	LQ	$	\\
\hline
\end{tabular}}
\end{table*}

\section{Results}
\label{sec:results}
\begin{figure}
\includegraphics[width=0.45\textwidth]{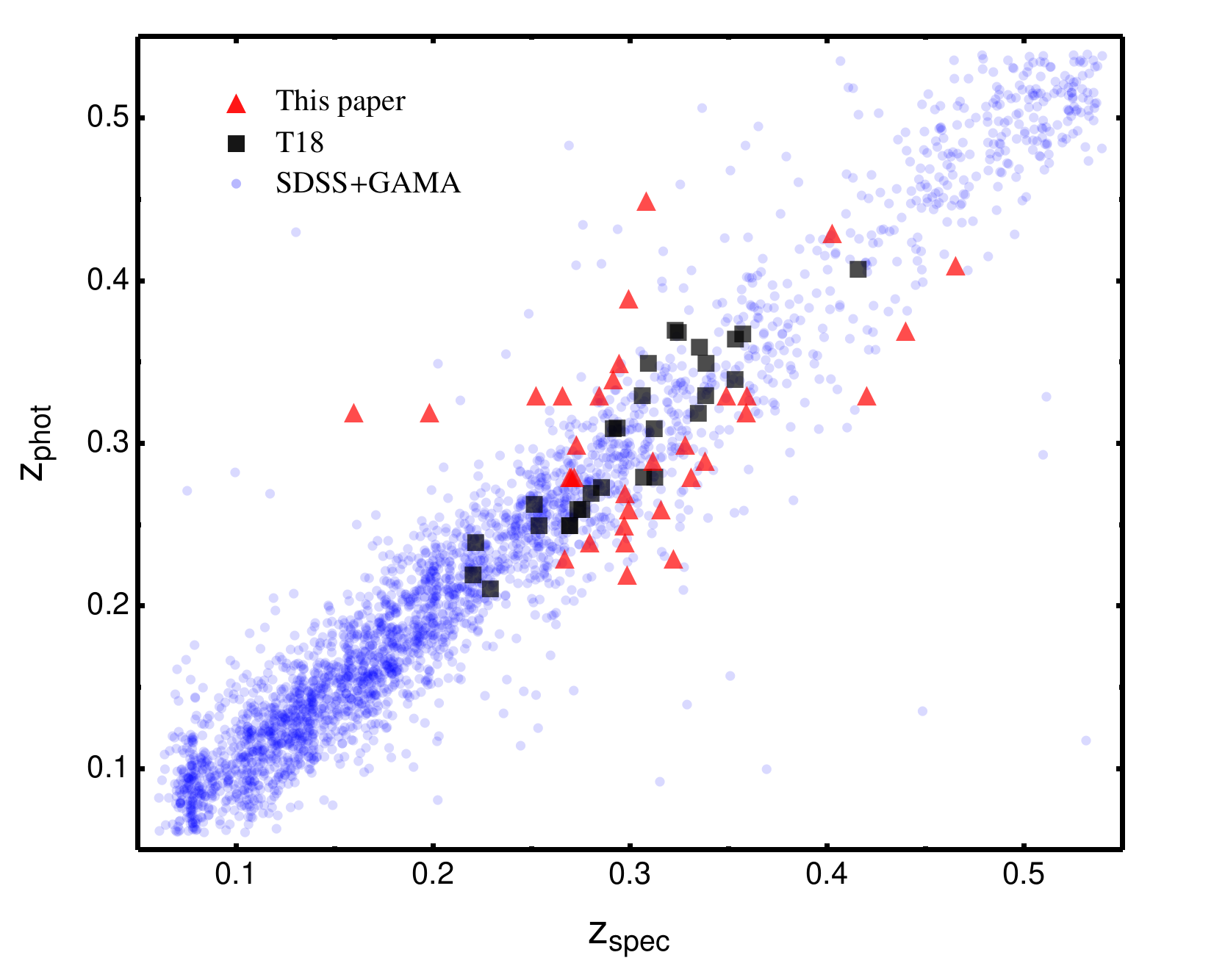}
\caption{Spectroscopic vs photometric redshifts. Red triangles are for the new sample of 33 \UCMG\ candidates analyzed in this paper with redshifts
measured from observations at INT and TNG. Black squares
are relative to the set of 28 \UCMG\ KiDS candidates with redshifts
measured from observations at TNG and NTT presented in T18. Blue points are for a parent sample of galaxies with SDSS and GAMA spectroscopy (extracted from \KiDSspec), used by \citet{Cavuoti+15_KIDS_I} as a test set for the validation of the photometric redshift determination. 
We find a good agreement with the 1-to-1 relation for most of the objects in For all the all the datasets.}\label{fig: Z_SPEC_ZPHOT}
\end{figure}

Although the photometric redshifts generally reproduce quite well the spectroscopic ones (Figure \ref{fig: Z_SPEC_ZPHOT}), small variations
in $z_{\rm phot}$ can induce variations in \Re\ and \mst{} large enough to bring them outside the limits for our definition of \UCMG\ (i.e., it might happens that $\Re > 1.5 \, \rm kpc$ and/or $\mst < 8 \times 10^{10} \, \rm \Msun$). Thus, having obtained the spectroscopic redshifts, we are now able to re-calculate both \Re\ and \mst, and find how many candidates are still ultra-compact and massive according to our definition. 

Following the analysis of T18, in the next subsections we study the success rate of our selection and systematics in \UCMG\ abundances. We then quantify the \UCMG\ number counts, comparing our new results with the ones in the literature. We finally show where the final sample of spectroscopically confirmed objects (i.e., the ones presented in T18 plus the ones presented here) locates on the $M_* - \sigma$ plane, to establish some basis for future analysis of the scaling relation.

\subsection{\UCMGs\ validation}\label{subsec:confirm}
In \Fig\ref{fig: Z_SPEC_ZPHOT} we compare the spectroscopic redshifts measured for the candidates of this paper with the photometric redshift values (red triangles). The results are also compared with the 28 \UCMG\ from T18 (black squares) and with a sample of galaxies with SDSS and GAMA spectroscopy (blue points) from KiDS--DR2 (\citealt{Cavuoti+15_KIDS_I}). 
As one can clearly see from the figure, the distribution of the new redshifts is generally consistent with what found using the full sample of galaxies included in KiDS--DR3, on average reproducing well the spectroscopic redshifts. 

The agreement on the redshifts can be better quantified by using statistical indicators (\citealt{Cavuoti+15_KIDS_I}; T18).
Following the analysis of T18, we define this quantity:
\begin{equation}
\Delta z \equiv \dfrac{\zs -\zp}{1+\zs},
\end{equation}
then we interpret the scatter as the standard deviation of $\Delta z$, and bias as the absolute value of the mean of $\Delta z$. We find a bias of 0.0008 and a scatter of 0.0516 for our 33 systems.
These estimates show a larger scatter of the new sample with respect to the sample of galaxies in T18, for which we found a bias of 0.0045 and a standard deviation of 0.028. 

Since we use a new stellar mass calculation set-up with respect to the one in T18, we recalculate sizes and masses, with both $z_{\rm phot}$ and $z_{\rm spec}$ for the final, total, spectroscopic sample of $61$ systems. The results are provided in Table \ref{tab:validation} and Table \ref{tab:validation2}, where we also report, in the last column, the \UCMGs\ spectral validation. 

Using the face values for masses and sizes inferred from the spectroscopic redshifts, we confirm as \UCMGs\ $19$ out of $33$ new \UCMG\ candidates. This corresponds to a success rate of $58\%$, a number which is fully consistent with the $50-60\%$ estimate 
found in T18. Moreover, using the new mass set-up, $27$ out the $28$ objects of T18 are still \UCMG\ candidates according to the mass selection using the photometric redshift values, and $18$ are spectroscopically confirmed \UCMGs. This corresponds to a success rate of  $67\%$. In total, we confirmed $37$ out of $61$ \UCMGs, with a success rate of  $60\%$.
Considering only the new $19/33$ confirmed \UCMGs, we find a bias of $0.016$ and a scatter of $0.037$ in the $z_{\rm phot}$ -- $z_{\rm spec}$ plot. This reflects the expectation that the objects with a larger scatter after the validation do not result compact and massive anymore according to our formal definition.

A very important point to stress here is that in the validation process we do not propagate the error on the photometric and spectroscopic redshifts into masses and sizes errors. We simply use the face values and include/exclude galaxies on the basis of the resulting nominal size and mass values. This might lead us to loose some galaxies at the edges, but it simplifies the analysis of the systematics, necessary to correct the number density (see Section \ref{subsec:number_counts}). 
If we take into account the average statistical $1\sigma$-level uncertainties for the measured effective radii and the stellar masses calculated in T18 (see the paper), i.e. $\delta R_{\rm e} \sim 20\% $ and $\delta \log_{10}( \mst/M_\odot) \sim 0.15$, we confirm as \UCMGs\ $57$ out of $61$ \UCMG\ candidates ($\sim93\%$). If we consider, instead, the $3\sigma$-level uncertainties, all the candidates are statistically consistent with the \UCMG\ definition. In the following, we analyze the systematics considering the face values for \Re\ and \mst\ in the selection.

\begin{table}
\centering
\caption{Photometric and spectroscopic parameters (redshifts, median effective radii in kpc and stellar masses) for the validation of the new samples:  $\textsc{ucmg\_int\_2017}$, $\textsc{ucmg\_tng\_2017}$, and $\textsc{ucmg\_tng\_2018}$. The last column indicates the spectral validation response: ``Y'' if the candidate is a confirmed \UCMG, (i.e. $\log_{10}( \mst/M_\odot)>10.9$ and  $R_{\rm e}<1.5$ kpc), ``N'' if it is not.}
\label{tab:validation}
\begin{tabular}{cccccccc} 
\hline

  ID  & \multicolumn{2}{c}{$z$} & \multicolumn{2}{c}{$R_{\rm e}$} & \multicolumn{2}{c}{$\log_{10}( \mst/M_\odot$)} & Spec. \\
 \cmidrule(lr){2-3} \cmidrule(lr){4-5} \cmidrule(lr){6-7}

 & $\rm phot$ & $\rm spec$ & $\rm phot$ & $\rm spec$ & $\rm phot$ & $\rm spec$  &   Valid. \\
\hline
\hline
$	1	$	&	$	0.28	$	&	$	0.27	$	&	$	1.43	$	&	$	1.39	$	&	$	11.03	$	&	$	11.00	$	&		$	Y	$	\\
$	2	$	&	$	0.26	$	&	$	0.32	$	&	$	1.23	$	&	$	1.43	$	&	$	10.94	$	&	$	11.07	$	&		$	Y	$	\\
$	3	$	&	$	0.26	$	&	$	0.30	$	&	$	1.36	$	&	$	1.51	$	&	$	10.92	$	&	$	11.21	$	&		$	N	$	\\
$	4	$	&	$	0.45	$	&	$	0.31	$	&	$	0.35	$	&	$	0.28	$	&	$	11.29	$	&	$	10.83	$	&		$	N	$	\\
$	5	$	&	$	0.37	$	&	$	0.44	$	&	$	0.71	$	&	$	0.79	$	&	$	11.32	$	&	$	11.24	$	&		$	Y	$	\\
$	6	$	&	$	0.22	$	&	$	0.30	$	&	$	1.44	$	&	$	1.81	$	&	$	10.93	$	&	$	11.20	$	&		$	N	$	\\
$	7	$	&	$	0.23	$	&	$	0.32	$	&	$	1.12	$	&	$	1.42	$	&	$	10.92	$	&	$	11.27	$	&		$	Y	$	\\
$	8	$	&	$	0.24	$	&	$	0.30	$	&	$	1.36	$	&	$	1.60	$	&	$	10.93	$	&	$	11.06	$	&		$	N	$	\\
$	9	$	&	$	0.34	$	&	$	0.29	$	&	$	1.04	$	&	$	0.94	$	&	$	10.92	$	&	$	10.73	$	&		$	N	$	\\
$	10	$	&	$	0.32	$	&	$	0.36	$	&	$	0.98	$	&	$	1.06	$	&	$	11.21	$	&	$	11.19	$	&		$	Y	$	\\
$	11	$	&	$	0.24	$	&	$	0.28	$	&	$	1.31	$	&	$	0.96	$	&	$	10.98	$	&	$	10.99	$	&		$	Y	$	\\
$	12	$	&	$	0.28	$	&	$	0.33	$	&	$	1.30	$	&	$	1.45	$	&	$	10.95	$	&	$	11.07	$	&		$	Y	$	\\
$	13	$	&	$	0.23	$	&	$	0.27	$	&	$	1.50	$	&	$	1.69	$	&	$	11.03	$	&	$	11.03	$	&		$	N	$	\\
\hline
$	14	$	&	$	0.35	$	&	$	0.29	$	&	$	1.37	$	&	$	1.20	$	&	$	11.08	$	&	$	10.96	$	&		$	Y	$	\\
$	15	$	&	$	0.27	$	&	$	0.30	$	&	$	1.13	$	&	$	1.22	$	&	$	11.08	$	&	$	11.10	$	&		$	Y	$	\\
$	16	$	&	$	0.33	$	&	$	0.36	$	&	$	1.28	$	&	$	1.36	$	&	$	11.25	$	&	$	11.34	$	&		$	Y	$	\\
$	17	$	&	$	0.33	$	&	$	0.27	$	&	$	1.47	$	&	$	1.28	$	&	$	11.16	$	&	$	10.97	$	&		$	Y	$	\\
$	18	$	&	$	0.32	$	&	$	0.16	$	&	$	1.25	$	&	$	0.74	$	&	$	10.98	$	&	$	10.61	$	&		$	N	$	\\
$	19	$	&	$	0.30	$	&	$	0.33	$	&	$	1.38	$	&	$	1.47	$	&	$	11.01	$	&	$	10.83	$	&		$	N	$	\\
$	20	$	&	$	0.30	$	&	$	0.27	$	&	$	1.37	$	&	$	1.27	$	&	$	10.95	$	&	$	10.97	$	&		$	Y	$	\\
$	21	$	&	$	0.33	$	&	$	0.25	$	&	$	0.81	$	&	$	0.67	$	&	$	10.99	$	&	$	10.82	$	&		$	N	$	\\
$	22	$	&	$	0.28	$	&	$	0.27	$	&	$	0.50	$	&	$	0.49	$	&	$	11.01	$	&	$	10.85	$	&		$	N	$	\\
$	23	$	&	$	0.25	$	&	$	0.30	$	&	$	1.22	$	&	$	1.39	$	&	$	11.12	$	&	$	11.26	$	&		$	Y	$	\\
$	24	$	&	$	0.33	$	&	$	0.35	$	&	$	1.07	$	&	$	1.11	$	&	$	11.01	$	&	$	11.06	$	&		$	Y	$	\\
\hline

$	25	$	&	$	0.28	$	&	$	0.27	$	&	$	1.30	$	&	$	1.27	$	&	$	10.97	$	&	$	10.94	$	&		$	Y	$	\\
$	26	$	&	$	0.32	$	&	$	0.20	$	&	$	1.28	$	&	$	0.90	$	&	$	10.92	$	&	$	10.46	$	&		$	N	$	\\
$	27	$	&	$	0.33	$	&	$	0.28	$	&	$	1.26	$	&	$	1.12	$	&	$	10.97	$	&	$	10.85	$	&		$	N	$	\\
$	28	$	&	$	0.33	$	&	$	0.42	$	&	$	1.14	$	&	$	1.32	$	&	$	11.00	$	&	$	11.25	$	&		$	Y	$	\\
$	29	$	&	$	0.29	$	&	$	0.31	$	&	$	1.42	$	&	$	1.49	$	&	$	10.99	$	&	$	10.99	$	&		$	Y	$	\\
$	30	$	&	$	0.39	$	&	$	0.30	$	&	$	1.35	$	&	$	1.14	$	&	$	11.02	$	&	$	10.78	$	&		$	N	$	\\
$	31	$	&	$	0.41	$	&	$	0.47	$	&	$	1.37	$	&	$	1.49	$	&	$	10.93	$	&	$	11.03	$	&		$	Y	$	\\
$	32	$	&	$	0.29	$	&	$	0.34	$	&	$	1.48	$	&	$	1.65	$	&	$	11.06	$	&	$	11.18	$	&		$	N	$	\\
$	33	$	&	$	0.43	$	&	$	0.40	$	&	$	1.30	$	&	$	1.24	$	&	$	11.31	$	&	$	11.24	$	&		$	Y	$	\\
\hline
\end{tabular}
\end{table} 

\begin{table}
\centering \caption{Same as \Tab\ref{tab:validation}, but for the $\textsc{ucmg\_tng\_t18}$ and $\textsc{ucmg\_ntt\_T18}$ samples.}
\label{tab:validation2}
\begin{tabular}{cccccccc} 
\hline

  ID  & \multicolumn{2}{c}{$z$} & \multicolumn{2}{c}{$R_{\rm e}$} & \multicolumn{2}{c}{$\log_{10}( \mst/M_\odot$)} & Spec. \\

 \cmidrule(lr){2-3} \cmidrule(lr){4-5} \cmidrule(lr){6-7}

 & $\rm phot$ & $\rm spec$ & $\rm phot$ & $\rm spec$ & $\rm phot$ & $\rm spec$  &   Valid. \\

\hline
\hline

$	34	$	&	$	0.29	$	&	$	0.37	$	&	$	1.43	$	&	$	1.68	$	&	$	10.97	$	&	$	11.35	$	&	$	N	$	\\
$	35	$	&	$	0.22	$	&	$	0.22	$	&	$	1.28	$	&	$	1.27	$	&	$	11.12	$	&	$	11.11	$	&	$	Y	$	\\
$	36	$	&	$	0.35	$	&	$	0.41	$	&	$	1.09	$	&	$	1.19	$	&	$	10.92	$	&	$	10.97	$	&	$	Y	$	\\
$	37	$	&	$	0.31	$	&	$	0.33	$	&	$	1.06	$	&	$	1.10	$	&	$	10.73	$	&	$	10.80	$	&	$	N	$	\\
$	38	$	&	$	0.42	$	&	$	0.40	$	&	$	0.67	$	&	$	0.66	$	&	$	10.98	$	&	$	10.94	$	&	$	Y	$	\\
$	39	$	&	$	0.36	$	&	$	0.32	$	&	$	1.46	$	&	$	1.36	$	&	$	10.99	$	&	$	10.87	$	&	$	N	$	\\
\hline

$	40	$	&	$	0.2	$	&	$	0.30	$	&	$	1.11	$	&	$	1.06	$	&	$	10.94	$	&	$	10.94	$	&	$	Y	$	\\
$	41	$	&	$	0.35	$	&	$	0.39	$	&	$	1.45	$	&	$	1.54	$	&	$	11.37	$	&	$	11.43	$	&	$	N	$	\\
$	42	$	&	$	0.28	$	&	$	0.24	$	&	$	1.47	$	&	$	1.32	$	&	$	10.91	$	&	$	10.84	$	&	$	N	$	\\
$	43	$	&	$	0.29	$	&	$	0.28	$	&	$	0.81	$	&	$	0.80	$	&	$	11.01	$	&	$	10.99	$	&	$	Y	$	\\
$	44	$	&	$	0.31	$	&	$	0.28	$	&	$	1.01	$	&	$	0.95	$	&	$	11.01	$	&	$	10.77	$	&	$	N	$	\\
$	45	$	&	$	0.27	$	&	$	0.29	$	&	$	0.62	$	&	$	0.65	$	&	$	10.99	$	&	$	11.00	$	&	$	Y	$	\\
$	46	$	&	$	0.31	$	&	$	0.36	$	&	$	0.92	$	&	$	1.01	$	&	$	10.95	$	&	$	10.94	$	&	$	Y	$	\\
$	47	$	&	$	0.25	$	&	$	0.26	$	&	$	1.02	$	&	$	1.04	$	&	$	10.97	$	&	$	10.94	$	&	$	Y	$	\\
$	48	$	&	$	0.27	$	&	$	0.29	$	&	$	1.29	$	&	$	1.36	$	&	$	11.04	$	&	$	11.09	$	&	$	Y	$	\\
$	49	$	&	$	0.28	$	&	$	0.30	$	&	$	1.36	$	&	$	1.42	$	&	$	10.91	$	&	$	10.97	$	&	$	Y	$	\\
$	50	$	&	$	0.29	$	&	$	0.32	$	&	$	1.36	$	&	$	1.43	$	&	$	11.02	$	&	$	11.04	$	&	$	Y	$	\\
$	51	$	&	$	0.34	$	&	$	0.32	$	&	$	1.04	$	&	$	0.99	$	&	$	10.98	$	&	$	10.89	$	&	$	N	$	\\
$	52	$	&	$	0.22	$	&	$	0.21	$	&	$	1.11	$	&	$	1.08	$	&	$	10.96	$	&	$	10.70	$	&	$	N	$	\\
$	53	$	&	$	0.25	$	&	$	0.26	$	&	$	1.15	$	&	$	1.16	$	&	$	10.95	$	&	$	10.97	$	&	$	Y	$	\\
$	54	$	&	$	0.34	$	&	$	0.30	$	&	$	1.47	$	&	$	1.37	$	&	$	11.03	$	&	$	10.93	$	&	$	Y	$	\\
$	55	$	&	$	0.31	$	&	$	0.37	$	&	$	1.10	$	&	$	1.24	$	&	$	10.96	$	&	$	11.13	$	&	$	Y	$	\\
$	56	$	&	$	0.32	$	&	$	0.41	$	&	$	1.29	$	&	$	1.50	$	&	$	11.22	$	&	$	11.20	$	&	$	Y	$	\\
$	57	$	&	$	0.33	$	&	$	0.26	$	&	$	1.27	$	&	$	1.07	$	&	$	10.96	$	&	$	10.81	$	&	$	N	$	\\
$	58	$	&	$	0.27	$	&	$	0.28	$	&	$	1.49	$	&	$	1.54	$	&	$	11.00	$	&	$	11.04	$	&	$	N	$	\\
$	59	$	&	$	0.23	$	&	$	0.29	$	&	$	1.10	$	&	$	1.30	$	&	$	10.94	$	&	$	11.12	$	&	$	Y	$	\\
$	60	$	&	$	0.34	$	&	$	0.34	$	&	$	1.05	$	&	$	1.05	$	&	$	10.99	$	&	$	10.99	$	&	$	Y	$	\\
$	61	$	&	$	0.31	$	&	$	0.29	$	&	$	1.08	$	&	$	1.03	$	&	$	11.09	$	&	$	11.03	$	&	$	Y	$	\\
\hline
\end{tabular}
\end{table} 

\subsection{Contamination and incompleteness}
\label{subsec:systematic}
One of the main aims of our spectroscopic campaigns is to quantify the impact of systematics on the \UCMG\ photometric selection. 
Because of the uncertain photometric redshifts, the candidate selection: 1) includes ``contaminants" (or false-positives), i.e., galaxies which are selected as \UCMGs\ according to their photometric redshifts, but would not result ultra-compact and massive when recalculating the masses on the basis of the more accurate spectroscopic redshift values (see T16 and T18) and 2) ``missed" systems (or false-negatives), i.e., those galaxies which are not selected as \UCMGs\ according to their photometric redshifts, but would be selected using the spectroscopic values instead (i.e., they are real \UCMGs\ that our selection excluded). Thus, following T18,  we define the {\it contamination factor}, \CF\ the inverse of the success rate discussed in the previous subsection, to account for the number of ``contaminants" and the {\it incompleteness factor}, \IF\ the difference between the number of \UCMG\ candidates using \zs{} and \zp{}, to estimate the incompleteness of the sample, i.e. quantifying the number of ``missing" objects. 

In this section we only report the average values for these factors across the full redshift range. We use instead different values calculated in different redshift bins to correct the abundances presented in \Sec{\ref{subsec:number_counts}}.
\smallskip 
To estimate the fraction of contaminants, we need \UCMG\ samples selected using the photometric redshifts, but for which we have also spectroscopic redshifts available. Thus, we evaluate \CF\ using three different photometrically selected samples with $\zp < 0.5$:
\begin{itemize}
\item [a)] the new sample of $33$ \UCMG\ candidates presented in this paper and discussed in \Sec\ref{sec:observations_and_analysis}, 
\item [b)] $27$ (out of $28$) \UCMG\ candidates from T18, that satisfy the new mass and size selection based on $z_{\rm phot}$ using the new set-up for stellar masses adopted here, and 
\item [c)] the sample of $50$ photometrically selected galaxies introduced in \Sec\ref{subsec:UCMG_criteria}, \UCMGPhotSpec\ with measured spectroscopic redshifts from SDSS, GAMA and 2dFLenS, similar to the one presented in T18 but selected with the new mass set-up.
\end{itemize}

For a), the new sample of \UCMGs\ presented in this paper, we obtain a $\CF = 1.72$ (corresponding to a success rate of $58\%$, see Section \ref{subsec:confirm}). Considering the samples in b) and c), we find $\CF = 1.50$ and $1.72$, respectively. Joining these three samples, we collect a sample of $110$ \UCMG\ candidates, of which $68$ have been validated after spectroscopy, implying a cumulative success rate of $62\%$ or $\CF = 1.62$.

\medskip 
To quantify how many real \UCMGs\ are missing from the photometric selection (incompleteness), we need to use objects with spectroscopic redshifts available from the literature. 
Thus, to determine \IF, we use \UCMGSpec: the sample of spectroscopically validated \UCMGs\ with spectroscopic redshifts from SDSS, GAMA and 2dFLenS. 
This sample updates and complements the one already presented in T18 (Tables C1 and C2) and consists of $54$ galaxies between $0.15 < z < 0.5$. The basic photometric and structural parameters for these \UCMGs\ in the spectroscopically selected sample is given in the Appendix. Only $29$ out of $54$ galaxies, i.e., $54\%$, would have been selected as candidates using \zp\ instead of \zs, which  corresponds to $\IF = 1.86$.

Having estimated contaminants and incompleteness, we can now obtain the correction factor for the number counts, as $\IF / \CF$. 
In conclusion, we find that the true number counts for \UCMGs\ at $z<0.5$ would be $\sim 15\%$ higher than the values one would find in a photometrically selected sample, on average. This is valid for the whole redshift range we consider here. 
In the next section, instead, we calculate a correction in each single redshift bin to minimize the errors on number counts.

\subsection{\UCMG\ number counts}
\label{subsec:number_counts}

\UCMG\ number counts are calculated following the procedure outlined in T18. For completeness, we report here some details. 

\begin{figure}
\centering
\includegraphics[width=0.49\textwidth]{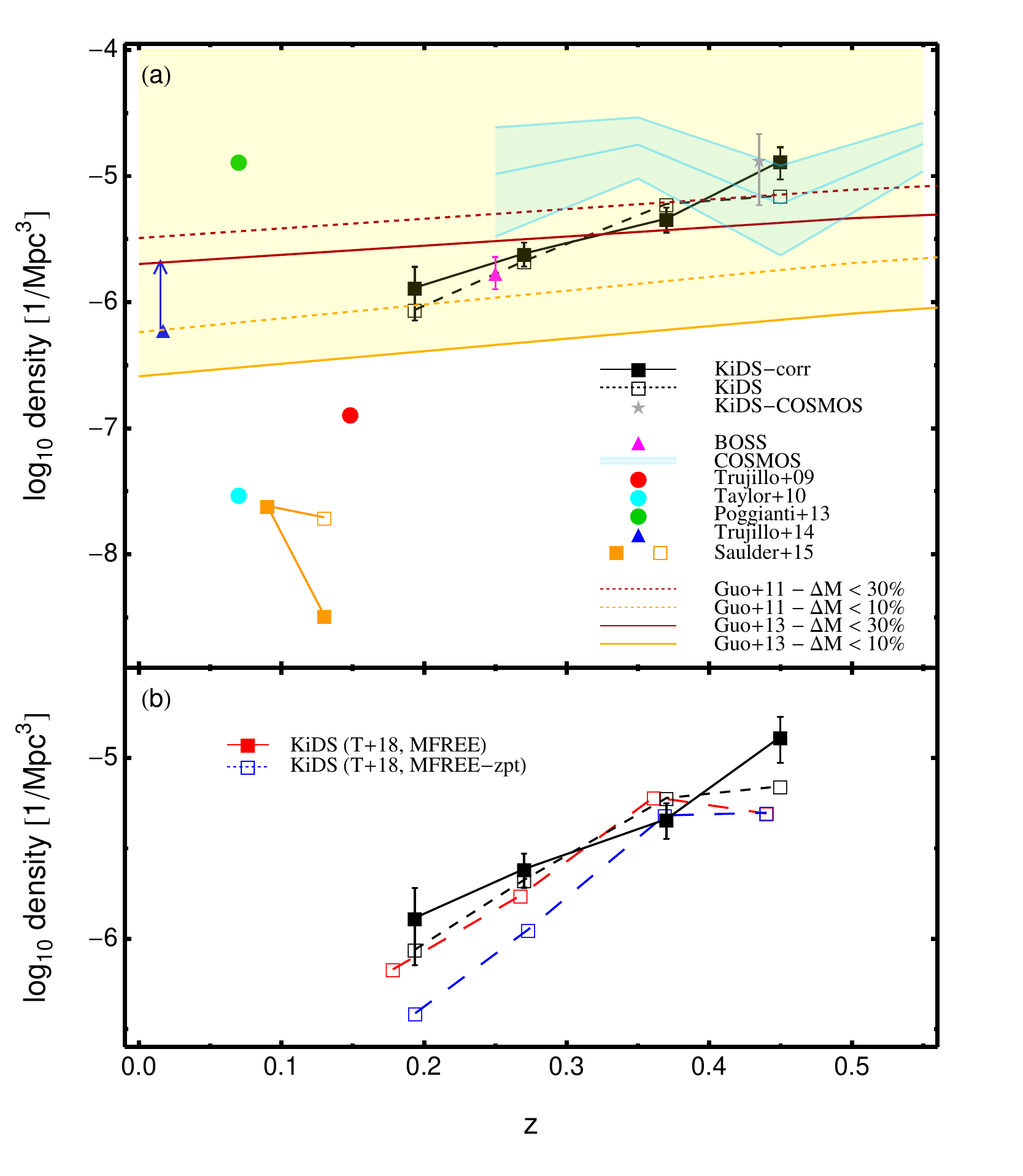}
\caption{{\it Panel a.} Filled (open) black squares, with solid (dashed) line, quoted as KiDS-corr(KiDS) in the legend, plot the number density after (before) correction for systematics, for the selected sample assuming reference masses. Error bars denote $1\sigma$ uncertainties, taking into account Poisson noise, cosmic variance and errors on \mst\ and \Re\ (see the text for more details). The gray star is for the \UCMG\ candidates at $z<0.5$ found in the tile KIDS\_150.1\_2.2 centered on COSMOS field. Other colored symbols are number density obtained from other papers, as described in the caption.
{\it Panel b.} Number counts obtained here are compared with those  presented in T18, named MFREE and MFREE--zpt, 
see the text for more details.}
\label{fig:number_densities}
\end{figure}

Taking into account the two systematic effects discussed in \Sec\ref{subsec:systematic}, we correct the number counts of the $1221$ candidates in \UCMGfull. In \Fig\ref{fig:number_densities} we plot the uncorrected and corrected counts as open squares/dashed line and filled squares/solid line respectively. We bin galaxies in four redshift bins ($z \in (0.15,0.2)$, $(0.2,0.3)$, $(0.3,0.4)$, $(0.4,0.5)$) and normalize to the comoving volume corresponding to the observed KiDS effective sky area of $333$ deg$^2$ (see T18 for further details). 
The errors on number counts take into account fluctuations due to Poisson noise, as well as those due to large-scale structure, i.e. the cosmic variance\footnote{These sources of errors are calculated with the public cosmic variance calculator available at \url{ http://casa.colorado.edu/$\sim$trenti/CosmicVariance.html} (\citealt{Trenti_Stiavelli08})}. For this calculation, we use the number of spectroscopically validated \UCMGs\ in each redshift bin. The uncertainties in stellar mass and effective radius measurements are also included in the error budget (as discussed in T18). The number density expectation for the KiDS tile centered on the COSMOS field is also plotted as a gray star.
Increasing the number of confirmed objects, thanks to the validation presented in this paper, we are able to reduce the error budget from cosmic variance and Poisson noise of $5-25\%$, in the four redshift bins.

The final result is fully consistent with the one found in T18 and shows a decrease of number counts with cosmic time, from $\sim 9\times 10^{-6} \, \rm Mpc^{-3}$ at $z\sim 0.5$, to $\sim 10^{-6}\, \rm Mpc^{-3}$ at $z \sim 0.15$. The number of \UCMGs\ decreases by a factor of $\sim 9$ in about $3$ Gyr. 

Following T18, we also compare our findings to lower redshift analyses  (\citealt{Trujillo+09_superdense}; \citealt{Poggianti+13_low_z}; \citealt{Taylor+10_compacts}; \citealt{Trujillo+14}; \citealt{Saulder+15_compacts}), and to other intermediate redshifts studies (\citealt[BOSS]{Damjanov+14_compacts}; \citealt[COSMOS]{Damjanov+15_compacts}). 
The reader is referred to T18 for a more detailed comparison between the different literature results and a detailed discussion on the impact of the different thresholds and selection criteria that different publications have used. 
In particular, we do not plot here the results obtained in \cite{Charbonnier+17_compact_galaxies} and \cite{Buitrago+18_compacts}, since these authors use less restrictive size criterion ($\Re  < 2 \, \rm kpc$).  However, including these results, we would have a perfect agreement with number densities reported in \cite{Charbonnier+17_compact_galaxies} in terms of normalization and evolution with redshift.

Finally, we also make a comparison with the results presented in \cite{Quilis_Trujillo13}, who have determined the evolution of the number counts of compact galaxies from Semi-analytical models, based on the Millennium N-body simulations by \cite{Guo+11_sims, Guo+13_sims}. They define ``relic compacts'' those galaxies with mass changing less than $10$ and $30$ per cent, from $z \sim 2$. The redshift evolution predicted by these simulations is milder than that obtained with our data, which are in agreement with COSMOS selection at $z \sim 0.5$ instead (\citealt{Damjanov+15_compacts}), and with the most recent number density determination in local environment made by \cite{Trujillo+14}.

In the bottom panel of \Fig\ref{fig:number_densities} we directly compare our uncorrected and corrected (for systematics) counts with those found in T18, where we used two different set-ups for the stellar mass derivation,  both of them without any constraints on ages and metallicity (which we instead set here in this paper, as described in Section \ref{subsec:selectionprocedure}). 
In particular, the MFREE masses (red lines and points in the plot) do not include zero-point calibration errors, while MFREE--zpt ones (blue points) include such contribution. 
Our results are in a good agreement with the reference T18 results assuming MFREE, and consistent within $2\sigma$ with the T18 results assuming MFREE--zpt.

It is important to remark that in Figure \ref{fig:number_densities} we obtain number counts for all the \UCMGs, without any distinction between relics (old stellar population) and non relics (young stellar population). Unfortunately the spectra obtained here and in our previous runs (T18) do not reach a signal-to-noise high enough to allow us to perform an in-depth stellar population analysis. This is however a \textit{conditio-sine qua non} to isolate these compact and massive galaxies whose stellar population is as old as the Universe and has been formed ``in situ" during the first phase of the two-phase formation scenario (\citealt{Oser 2010}).  We will thus postpone this more detailed analysis and the redefinition of the obtained number densities to a future publication where we will remove the non-relic contaminants thanks to spectroscopic stellar population modeling. 

\subsection{Relationship between stellar mass and velocity dispersion}
\label{FJr}
The correlation between luminosity (or stellar mass) and velocity dispersion in elliptical galaxies is a well established scaling relation (\citealt{FJ76}; \citealt{HB09_curv}). 
The location of \UCMGs\ in a mass--velocity dispersion diagram ($ M_*-\sigma$) can give remarkable insights on their intrinsic properties (\citealt{Saulder+15_compacts}). Indeed, given the compact sizes of \UCMGs, the virial theorem predicts larger velocity dispersions with respect to normal-sized galaxies of similar mass. This has also been directly confirmed with deep spectroscopy of a handful of these objects at high redshift  \citep{vanDokkum+9, Toft+12} and of the three local relics \citep{Ferre-Mateu+17}. 
Therefore, \UCMGs\ should segregate in this parameter space, having a mass--velocity dispersion correlation that deviates from the one of normal-size galaxies. 
Also, being this $M_*- \sigma$ relation intimately connected to the assembly of baryons and dark matter, it can also provide important constraints on our understanding of the formation and evolution of these systems. This might be particularly important in the specific case of relics.

In this section we present a preliminary result on the $ M_*- \sigma $ relation, based on the velocity dispersion measurements presented in Section \ref{redshift and velocity}. 

\begin{figure}
\centering
\includegraphics[width=0.48\textwidth]{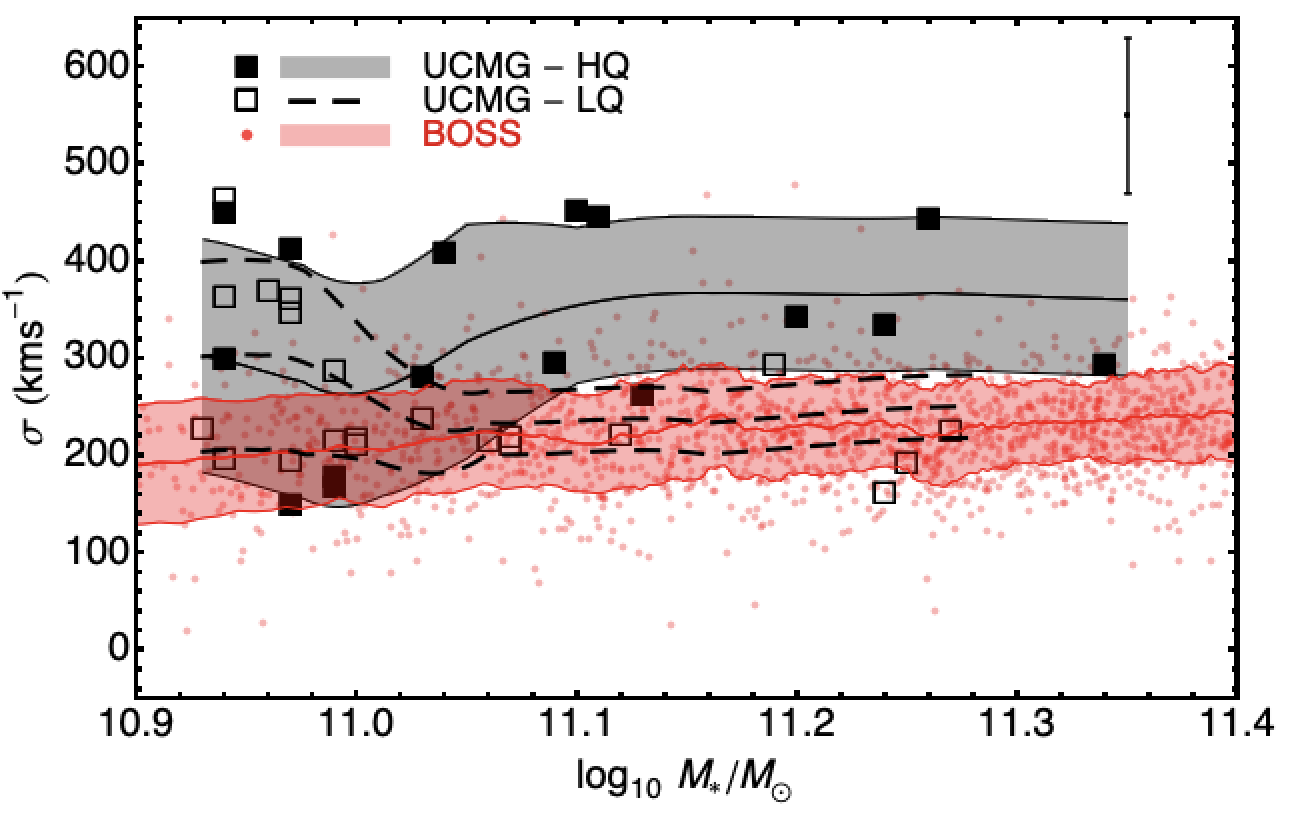}
\caption{{Distribution on the $M_{*} - \sigma$ plane for the 37 confirmed \UCMGs\ compared with a sample of elliptical galaxies (red symbols) from the BOSS survey. Filled squared symbols are \UCMGs\ classified as high quality (HQ), with spectra that satisfy at the same time the three conditions SR $\ge 0.3$, $S/N_{spec} \ge 3.5$ and $(S/N)_{\rm O-K} \ge 6.5$. Empty squares are instead classified as low-quality (LQ) since their spectra do not satisfy one or more of the aforementioned criteria. For each sample, running means and $1\sigma$ scatter are overplotted. In the top-right corner we show the mean error bar for the \UCMG\ velocity dispersions. For bot} \UCMGs\ and the sample of ellipticals, velocity dispersions are calculated within 1 effective radius, as explained in the text.}
\label{fig: fj}
\end{figure}

In \Fig\ref{fig: fj} we plot the $M_*- \sigma$ distribution of the $37$ \footnote{We have in total $19$ confirmed systems from the three new spectroscopic runs and $18$ from the runs presented in T18 and confirmed on the basis of the new mass-calculation set-up.} confirmed \UCMGs\ (squared symbols).  

For comparison, we overplot a sample of normal-sized ETGs (red small dots) analyzed in \cite{Tortora+18_KiDS_DMevol} and derived from SDSS-III/BOSS (Baryon 
Oscillation Spectroscopic Survey) Data Release 10\footnote{The data catalogs are
available from \url{http://www.sdss3.org/dr10/spectro/galaxy\_portsmouth.php}.} (DR10, \citealt{Ahn+14_SDSS_DR10}). We restrict the BOSS sample to the redshift 
range $0.15 \lesssim z \lesssim 0.5$, to provide a direct comparison with the sample of \UCMGs. 
For these systems, in \cite{Tortora+18_KiDS_DMevol} we have derived stellar masses using the same set-up adopted in this paper, while the velocity dispersion values are originally measured in a circular aperture of radius 1 arcsec.

The distribution of all the confirmed \UCMGs\ presents a large scatter, 
which is mainly the consequence of the large errors on the velocity dispersion values (see typical error bars on top right corner of the figure). We plot with full squares \UCMGs\ classified in the high-quality group and open squares the ones belonging to the low quality group, according to the definition given in Section \ref{redshift and velocity}. 

Finally, in order to highlight significant patterns in this figure,  we also plot the running mean and $1\sigma$ scatter for the \UCMGs\ and BOSS galaxies. The running means obtained from the \UCMGs\ in the HQ subsample (i.e. the grey shaded region in the figure) and that obtained for all the normal-sized BOSS galaxies (i.e. red region) differ significantly. 
The \UCMGs\ have systematically larger velocity dispersions at any fixed mass, especially above $\log M_{*}/M_{\odot} = 11.05$, and this result is consistent with other studies of high--z systems \citep{vanDokkum+9, Toft+12} and local massive relics \citep{Ferre-Mateu+17}.
The offset almost disappears when including the LQ \UCMGs, which, at least for larger masses, are scattered toward lower $\sigma$ and are consistent with the normal ETG distribution within the (large) errors.

We consider the offset between BOSS and HQ \UCMGs\ robust and statistically significant, although we anticipate that with better data we will be able to improve the measurement errors and also increase the size of the sample. 
Nevertheless, taking these finding at face value, one can speculate about possible explanations for this offset. The first possibility is that more compact massive galaxies host a more massive black hole (e.g. \citealt{vdb_2012, Bosch_2015, Ferre-Mateu+17}) which might influence the kinematics in the innermost region. Another possibility is that the IMF in very massive galaxies can be different from an universal Milky Way-like IMF. However, whereas for larger galaxies the bottom-heavy IMF is restricted only in the very central region ($\sim 0.2$ -- $0.3 \Re$), for relics the IMF is heavier than Salpeter everywhere up to few effective radii. One physical scenario able to explain this difference would be that only the ``in situ'' stars formed during the first phase of the assembly of massive ETGs form with a dwarf-rich IMF, while accreted stars (only present in normal-sized ETGs) form with a standard IMF \citep{Chabrier_2014}. 

We will investigate these possibilities in a dedicated paper, already in preparation. There, we will compare these (and new) measurements with theoretically motivated predictions, including more than one galaxy formation recipe. We will check whether the $M_*-$  $\sigma$ relation preserves the footprints of the stellar and dark assembly of these systems, trying to quantify the dynamical contribution of a central supermassive black hole and a bottom-heavy IMF.

In conclusion, given the large uncertainties on the velocity dispersion measurements and the fact that we cannot yet distinguish between relics and non-relics, we provide here only some preliminary speculative explanations. In the future, we aim at consolidating this result with a larger number of systems, to increase the statistics, and using spectroscopic data of better quality, in order to have more robust velocity dispersion estimates.
With new, better spectroscopic data we will also be able to constrain the age of the systems, which is the crucial ingredient to identify relics among the confirmed \UCMGs.


\section{Conclusions}\label{sec:conclusions}
The existence of ultra-compact massive galaxies (\UCMGs) at $z<1$ and their evolution up to the local Universe challenges the currently accepted galaxy formation models. 
In the effort of ``bridging the gap'' between the high redshift red nuggets and the local relics, we have started a census of \UCMGs\ at intermediate redshifts. 
In particular, in the first paper of this series \citep{Tortora+16_compacts_KiDS}, we have demonstrated that the high image quality, the large area covered, the excellent spatial resolution and the exquisite seeing of the Kilo Degree Survey (KiDS) make this survey perfect to find \UCMGs\ candidates. In the second paper \citep{Tortora+18_UCMGs} we have started a multi-site and multi-telescope spectroscopic observational campaign to confirm as many candidates as possible with the final goal of building the largest spectroscopically confirmed sample of \UCMGs\ in the redshift range $0.15 \lesssim z \lesssim 0.5$.

In this third paper of the series, we have continued in this direction and we have:
\begin{itemize}
\item spectroscopically followed up a sample of $33$ \UCMG\ candidates at redshifts  $0.15 \lesssim z \lesssim 0.5$, found in $333$ deg$^2$ of KiDS. 
We have provided details on how the galaxies have been photometrically selected and discussed the spectroscopic campaign on  the INT and TNG telescopes, including also the main data reduction steps for each instrument;

\item obtained the spectroscopic redshift and velocity dispersion values for these objects and for the $28$ objects already presented in T18. To this purpose, we have used the Optimised Modelling of Early-type Galaxy Aperture Kinematics pipeline  (OMEGA-K, D'Ago et al., in prep.);

\item confirmed $19$ out of $33$ as \UCMGs, with the newly spectroscopically based masses and effective radii.  This translates into a success rate of $58\%$, in good agreement with the one reported in T18. In addition, using the new mass set-up we have confirmed $18$ out of $27$ \UCMGs\ from T18, corresponding to a success rate of $67\%$. One galaxy from T18 did not qualify as \UCMG\ candidate when re-computing its mass with the newly defined set-up. 
Thus, in total, we confirm as \UCMGs\ $37$ out of $61$ candidates, implying a success rate of  $60\%$. Allowing a tolerance at the $1\sigma$-level ($3\sigma$-level) on the effective radii and stellar masses inferred from the spectroscopic redshifts, we confirm as \UCMGs\ 57 (61) out of $61$ \UCMG\ {candidates with a success rate of $\sim93\%$ ($100\%$); }

\item quantified the effect of contamination and incompleteness due to the difference in redshift between the photometric and spectroscopic values. We have found that the true number counts for \UCMGs\ at $z<0.5$ is $\sim 15$ per cent higher than the values found in a photometrically selected sample;

\item obtained the \UCMG\ number counts, after correcting them with the incompleteness and the contamination factors, and their evolution with redshift in the range $0.15<z<0.5$. We have also compared our new results with these obtained in T18, using a different set-up for the mass inference and in the literature. We have confirmed the clear decrease of the number counts with the cosmic time already found in T18, from $\sim 9\times 10^{-6} \, \rm Mpc^{-3}$ at $z\sim 0.5$, to $\sim 10^{-6}\, \rm Mpc^{-3}$ at $z \sim 0.15$, $\sim 9$ times less in about 3 Gyr;

\item shown the distribution of the $37$ confirmed \UCMGs\ in the $M_* - \sigma$ plane. We have corrected the sigma values to a common aperture of one effective radius in order to compare the \UCMGs\ distribution with that of a sample of normal-sized  ellipticals from the BOSS Survey. 
Despite the large uncertainties on the velocity dispersion measurements, due to the low signal-to-noise of the spectra, we  found tentative evidence suggesting that the \UCMGs\ have larger values compared to regular ETGs of same mass. This seems to be statistically significant at least for the high-quality sample and large masses. This preliminary result, in agreement with what expected from the evolution of massive and compact galaxies, will be checked again once new, higher resolution spectroscopy (already awarded) will be obtained.
\end{itemize}

After KiDS will be completed, we expect to at least double the number of confirmed \UCMGs, and thus reduce by a factor $\sim40\%$ the uncertainties on the number counts, while keeping the systematics under full control. 

In the future, we also plan to continue to enlarge the 
sample of spectroscopically confirmed \UCMGs\ at low and intermediate redshifts, based on photometric candidates from the KiDS survey. 
Moreover, thanks to already awarded spectroscopic data with much higher S/N, which will allow us to perform a detailed stellar population analysis, we will separate relics from younger \UCMGs. 
With the higher signal-to-noise spectra that we will soon have at our disposal, we aim at unambiguously demonstrate that the majority the objects in our sample are indeed red and dead, as already indicated by their photometric colors and that they have formed their baryonic matter early on in cosmic time, with a fast and ``bursty" star formation episode. In this way, we will be able to unambiguously confirm the two-phase formation scenario proposed for the mass assembly of massive/giant ETGs \citep{Oser 2010}. 

Relics, \UCMGs\ as old as the Universe, are the only systems that, with current observing facilities, allow us to study the physical processes that shaped the mass assembly of massive galaxies in the high-z universe with the amount of details currently reachable only for the near-by Universe.

\section*{Acknowledgements}
DS is a member of the International Max Planck Research School (IMPRS) for Astronomy and Astrophysics at the Universities of Bonn and Cologne. CT acknowledges funding from the INAF PRIN-SKA 2017 program 1.05.01.88.04.
MS acknowledges financial support from the VST project (PI: P. Schipani). CS has received funding from the European Union's Horizon 2020 research and innovation programme under the Marie Sk\l odowska-Curie actions grant agreement No 664931. GD acknowledges support from CONICYT project Basal AFB-170002. MB acknowledges the \textit{INAF PRIN-SKA 2017 program 1.05.01.88.04}, the funding from \textit{MIUR Premiale 2016: MITIC}
and the financial contribution from the agreement ASI/INAF nr. 2018-23-HH.0 \textit{Euclid mission scientific activities - Phase D}. The INT is operated on the island of La Palma by the Isaac Newton Group of Telescopes in the Spanish Observatorio del Roque de los Muchachos of the Instituto de Astrof\'isica de Canarias. Based on observations made with the Italian Telescopio Nazionale Galileo (TNG) and Isaac Netwton (INT) telescopes operated by the Fundaci\'on Galileo Galilei of the INAF (Istituto Nazionale di Astrofisica) and the Isaac Newton Group of Telescopes, both of them installed in the Spanish Observatorio del Roque de los Muchachos of the Instituto de Astrof\'isica de Canarias. Based on data products from observations made with ESO Telescopes at the La Silla Paranal Observatory under programme IDs 177.A-3016, 177.A-3017 and 177.A-3018, andon data products produced by Target/OmegaCEN, INAF-OACN, INAF-OAPD and the KiDS production team, on behalf of the KiDS consortium. OmegaCEN and the KiDS production team acknowledge support by NOVA and NWO-M grants. Members of INAF-OAPD and INAF-OACN also acknowledge the support from the Department of Physics \& Astronomy of the University of Padova, and of the Department of Physics of Univ. Federico II (Naples).



\section*{Appendix}
In order to quantify the impact of the systematics on the \UCMG\ selection, we have created, in Section 4.2 the \UCMGSpec\ sample: a sample of $55$ \UCMGs\ with spectroscopic redshifts from the literature, similar to the sample used in T18 but selected with the new mass set-up. 
We have gathered these spectroscopic redshifts from SDSS \citep{Ahn+12_SDSS_DR9, Ahn+14_SDSS_DR10}, GAMA \citep{Driver+11_GAMA}, which overlap the KiDS fields in the Northern cap, and 2dFLenS \citep{Blake+16_2dflens}, observed in the Southern hemisphere. 
Here in this Appendix we provide the basic photometric and structural parameters for such $55$ \UCMGs\ in the spectroscopically selected sample. In particular, in Table \ref{tab:ucmg_liter} we show $r$-band Kron magnitude, aperture magnitudes used in the SED fitting, spectroscopic redshifts and stellar masses (in decimal logarithm). S\'ersic structural parameters from the \twodphot\ fit of $g$-, $r$-, and $i$-band KiDS surface photometry, as such as $\chi^{2}$s and \SN\ values, are instead presented in Table \ref{tab:struc_parameters_ucmg_liter}.

\begin{table*}[h]
\centering 
\caption{Integrated photometry for the 55 systems in the \UCMGSpec\ sample. From left we show: a) progressive ID number; b) KIDS identification name; c) $r$-band KiDS $\texttt{MAG\_AUTO}$; d-g) $u$-, $g$-, $r$- and $i$-band KiDS magnitudes measured in
an aperture of 6 arcsec of diameter with 1$\sigma$ errors; h) spectroscopic redshift. All the magnitudes have been corrected for Galactic extinction using \citet{Schlafly_Finkbeiner11} maps.} 
\label{tab:ucmg_liter}
\begin{tabular}{ccccccccc} 
\hline
$\rm ID$ & $\rm name$ &  $\texttt{MAG\_AUTO\_r}$ & $u_{6''}$ & $ g_{6''}$ & $ r_{6''}$ & $i_{6''}$  & \zp & $\log_{10}\mst/M_\odot$ \\ 
\hline
\hline

$\rm{L}1	$	&	\rm	KIDS J025942.84--315933.74	&	$	18.96	$	&	$	22.61	\pm	0.13	$	&	$	20.39	\pm	0.008	$	&	$	18.97	\pm	0.003	$	&	$	18.52	\pm	0.007	$	&	$	0.29	$	&	$	11.00	$	\\
$\rm{L}2	$	&	\rm	KIDS J032700.87--300112.34	&	$	20.37	$	&	$	23.02	\pm	0.21	$	&	$	21.93	\pm	0.04	$	&	$	20.34	\pm	0.009	$	&	$	19.43	\pm	0.01	$	&	$	0.33	$	&	$	11.00	$	\\
$\rm{L}	3	$	&	\rm	KIDS J084320.59--000543.77	&	$	18.52	$	&	$	21.55	\pm	0.06	$	&	$	19.71	\pm	0.005	$	&	$	18.53	\pm	0.002	$	&	$	18.12	\pm	0.005	$	&	$	0.24	$	&	$	10.93	$	\\
$\rm{L}	4	$	&	\rm	KIDS J084738.70+011220.57	&	$	18.41	$	&	$	21.78	\pm	0.12	$	&	$	19.70	\pm	0.006	$	&	$	18.44	\pm	0.002	$	&	$	18.02	\pm	0.005	$	&	$	0.18	$	&	$	11.00	$	\\
$\rm{L}	5	$	&	\rm	KIDS J085335.58+001805.97	&	$	18.84	$	&	$	21.67	\pm	0.09	$	&	$	20.13	\pm	0.009	$	&	$	18.95	\pm	0.003	$	&	$	18.63	\pm	0.008	$	&	$	0.33	$	&	$	10.94	$	\\
$\rm{L}	6	$	&	\rm	KIDS J085344.88+024948.47	&	$	18.49	$	&	$	21.63	\pm	0.07	$	&	$	19.70	\pm	0.005	$	&	$	18.50	\pm	0.002	$	&	$	18.08	\pm	0.005	$	&	$	0.23	$	&	$	10.93	$	\\
$\rm{L}	7	$	&	\rm	KIDS J090324.20+022645.50	&	$	17.25	$	&	$	20.24	\pm	0.02	$	&	$	18.34	\pm	0.002	$	&	$	17.34	\pm	0.001	$	&	$	16.98	\pm	0.001	$	&	$	0.19	$	&	$	11.21	$	\\
$\rm{L}	8	$	&	\rm	KIDS J090935.74+014716.81	&	$	18.68	$	&	$	22.52	\pm	0.17	$	&	$	20.15	\pm	0.008	$	&	$	18.75	\pm	0.002	$	&	$	18.23	\pm	0.006	$	&	$	0.22	$	&	$	11.02	$	\\
$\rm{L}	9	$	&	\rm	KIDS J092055.70+021245.66	&	$	18.87	$	&	$	22.80	\pm	0.21	$	&	$	20.37	\pm	0.01	$	&	$	18.89	\pm	0.003	$	&	$	18.46	\pm	0.005	$	&	$	0.28	$	&	$	11.01	$	\\
$\rm{L}	10	$	&	\rm	KIDS J102653.56+003329.15	&	$	17.39	$	&	$	20.49	\pm	0.02	$	&	$	18.52	\pm	0.002	$	&	$	17.45	\pm	0.001	$	&	$	17.04	\pm	0.002	$	&	$	0.17	$	&	$	11.17	$	\\
$\rm{L}	11	$	&	\rm	KIDS J112825.16--015303.29	&	$	20.94	$	&	$	23.90	\pm	0.57	$	&	$	22.56	\pm	0.06	$	&	$	20.91	\pm	0.01	$	&	$	20.19	\pm	0.04	$	&	$	0.46	$	&	$	10.94	$	\\
$\rm{L}	12	$	&	\rm	KIDS J113612.68+010316.86	&	$	19.01	$	&	$	22.07	\pm	0.08	$	&	$	20.26	\pm	0.007	$	&	$	19.02	\pm	0.003	$	&	$	18.59	\pm	0.005	$	&	$	0.22	$	&	$	10.97	$	\\
$\rm{L}	13	$	&	\rm	KIDS J114248.56+001215.63	&	$	17.02	$	&	$	19.72	\pm	0.01	$	&	$	17.95	\pm	0.002	$	&	$	17.14	\pm	0.0008	$	&	$	16.71	\pm	0.001	$	&	$	0.11	$	&	$	10.58	$	\\
$\rm{L}14	$	&	\rm	KIDS J115652.47--002340.77	&	$	18.83	$	&	$	21.98	\pm	0.09	$	&	$	20.06	\pm	0.007	$	&	$	18.83	\pm	0.003	$	&	$	18.08	\pm	0.006	$	&	$	0.26	$	&	$	11.14	$	\\
$\rm{L}	15	$	&	\rm	KIDS J120251.61+013825.15	&	$	17.89	$	&	$	20.69	\pm	0.03	$	&	$	19.39	\pm	0.003	$	&	$	18.04	\pm	0.001	$	&	$	17.75	\pm	0.003	$	&	$	0.20	$	&	$	11.04	$	\\
$\rm{L}	16	$	&	\rm	KIDS J120818.93+004600.16	&	$	17.74	$	&	$	20.65	\pm	0.03	$	&	$	18.88	\pm	0.004	$	&	$	17.93	\pm	0.001	$	&	$	17.56	\pm	0.002	$	&	$	0.18	$	&	$	10.92	$	\\
$\rm{L}	17	$	&	\rm	KIDS J120902.53--010503.08	&	$	18.83	$	&	$	22.68	\pm	0.21	$	&	$	20.16	\pm	0.008	$	&	$	18.82	\pm	0.003	$	&	$	18.36	\pm	0.008	$	&	$	0.27	$	&	$	11.04	$	\\
$\rm{L}	18	$	&	\rm	KIDS J121152.97--014439.23	&	$	18.60	$	&	$	21.64	\pm	0.08	$	&	$	19.79	\pm	0.006	$	&	$	18.65	\pm	0.003	$	&	$	18.23	\pm	0.005	$	&	$	0.23	$	&	$	10.96	$	\\
$\rm{L}	19	$	&	\rm	KIDS J121555.27+022828.13	&	$	20.56	$	&	$	23.36	\pm	0.32	$	&	$	22.21	\pm	0.04	$	&	$	20.53	\pm	0.01	$	&	$	19.81	\pm	0.02	$	&	$	0.47	$	&	$	10.97	$	\\
$\rm{L}	20	$	&	\rm	KIDS J140620.09+010643.00	&	$	19.16	$	&	$	22.55	\pm	0.13	$	&	$	20.68	\pm	0.01	$	&	$	19.19	\pm	0.004	$	&	$	18.70	\pm	0.009	$	&	$	0.37	$	&	$	11.28	$	\\
$\rm{L}	21	$	&	\rm	KIDS J141108.94--003647.51	&	$	19.22	$	&	$	22.27	\pm	0.14	$	&	$	20.57	\pm	0.01	$	&	$	19.20	\pm	0.004	$	&	$	18.74	\pm	0.01	$	&	$	0.29	$	&	$	10.98	$	\\
$\rm{L}	22	$	&	\rm	KIDS J141200.92--002038.65	&	$	19.19	$	&	$	22.94	\pm	0.27	$	&	$	20.76	\pm	0.02	$	&	$	19.21	\pm	0.005	$	&	$	18.69	\pm	0.02	$	&	$	0.28	$	&	$	11.08	$	\\
$\rm{L}	23	$	&	\rm	KIDS J141213.62+021202.06	&	$	18.37	$	&	$	19.30	\pm	0.01	$	&	$	19.14	\pm	0.004	$	&	$	18.38	\pm	0.002	$	&	$	18.16	\pm	0.005	$	&	$	0.30	$	&	$	11.06	$	\\
$\rm{L}	24	$	&	\rm	KIDS J141415.53+000451.51	&	$	18.99	$	&	$	22.86	\pm	0.17	$	&	$	20.41	\pm	0.009	$	&	$	19.00	\pm	0.003	$	&	$	18.50	\pm	0.006	$	&	$	0.30	$	&	$	11.07	$	\\
$\rm{L}	25	$	&	\rm	KIDS J141417.33+002910.20	&	$	18.77	$	&	$	21.73	\pm	0.07	$	&	$	20.04	\pm	0.007	$	&	$	18.77	\pm	0.003	$	&	$	18.34	\pm	0.006	$	&	$	0.30	$	&	$	11.03	$	\\
$\rm{L}	26	$	&	\rm	KIDS J141728.44+010626.61	&	$	17.90	$	&	$	20.94	\pm	0.04	$	&	$	19.06	\pm	0.004	$	&	$	17.98	\pm	0.002	$	&	$	17.59	\pm	0.003	$	&	$	0.18	$	&	$	10.96	$	\\
$\rm{L}	27	$	&	\rm	KIDS J141828.24--013436.27	&	$	18.82	$	&	$	21.13	\pm	0.07	$	&	$	19.90	\pm	0.006	$	&	$	18.80	\pm	0.003	$	&	$	18.39	\pm	0.005	$	&	$	0.43	$	&	$	11.26	$	\\
$\rm{L}	28	$	&	\rm	KIDS J142033.15+012650.38	&	$	19.38	$	&	$	23.58	\pm	0.38	$	&	$	20.79	\pm	0.02	$	&	$	19.37	\pm	0.005	$	&	$	18.89	\pm	0.01	$	&	$	0.32	$	&	$	10.92	$	\\
$\rm{L}	29	$	&	\rm	KIDS J142041.17--003511.27	&	$	18.95	$	&	$	22.40	\pm	0.14	$	&	$	20.37	\pm	0.009	$	&	$	19.01	\pm	0.003	$	&	$	18.51	\pm	0.005	$	&	$	0.25	$	&	$	11.00	$	\\
$\rm{L}	30	$	&	\rm	KIDS J142235.50--014207.95	&	$	19.24	$	&	$	23.10	\pm	0.27	$	&	$	20.65	\pm	0.01	$	&	$	19.27	\pm	0.004	$	&	$	18.82	\pm	0.009	$	&	$	0.28	$	&	$	10.92	$	\\

\hline
\end{tabular}
\end{table*} 

\newpage

\begin{table*}

\addtocounter{table}{-1}
\caption{-- continued from previous page} 

\label{tab:ucmg_liter}
\begin{tabular}{ccccccccc} 
\hline
$\rm ID$ & $\rm name$ &  $\texttt{MAG\_AUTO\_r}$ & $u_{6''}$ & $ g_{6''}$ & $ r_{6''}$ & $i_{6''}$  & \zp & $\log_{10}\mst/M_\odot$ \\ 
\hline
$\rm{L}	31	$	&	\rm	KIDS J142606.67+015719.28	&	$	19.33	$	&	$	22.97	\pm	0.22	$	&	$	20.69	\pm	0.01	$	&	$	19.30	\pm	0.005	$	&	$	18.86	\pm	0.01	$	&	$	0.35	$	&	$	11.14	$	\\
$\rm{L}	32	$	&	\rm	KIDS J142800.20--001026.87	&	$	18.75	$	&	$	19.42	\pm	0.01	$	&	$	19.33	\pm	0.004	$	&	$	18.83	\pm	0.003	$	&	$	18.56	\pm	0.009	$	&	$	0.33	$	&	$	11.05	$	\\
$\rm{L}	33	$	&	\rm	KIDS J142922.11+011450.00	&	$	18.69	$	&	$	21.95	\pm	0.12	$	&	$	20.09	\pm	0.008	$	&	$	18.69	\pm	0.003	$	&	$	18.35	\pm	0.007	$	&	$	0.37	$	&	$	11.10	$	\\
$\rm{L}	34	$	&	\rm	KIDS J143025.44--023311.23	&	$	18.80	$	&	$	19.25	\pm	0.01	$	&	$	19.13	\pm	0.005	$	&	$	18.79	\pm	0.003	$	&	$	18.49	\pm	0.007	$	&	$	0.40	$	&	$	11.15	$	\\
$\rm{L}	35	$	&	\rm	KIDS J143155.56--000358.65	&	$	19.34	$	&	$	22.74	\pm	0.18	$	&	$	20.73	\pm	0.02	$	&	$	19.32	\pm	0.004	$	&	$	18.82	\pm	0.007	$	&	$	0.34	$	&	$	11.04	$	\\
$\rm{L}	36	$	&	\rm	KIDS J143419.53--005231.62	&	$	19.14	$	&	$	22.64	\pm	0.17	$	&	$	20.79	\pm	0.01	$	&	$	19.13	\pm	0.004	$	&	$	18.57	\pm	0.005	$	&	$	0.46	$	&	$	11.20	$	\\
$\rm{L}	37	$	&	\rm	KIDS J143459.11--010154.63	&	$	19.37	$	&	$	22.95	\pm	0.25	$	&	$	20.70	\pm	0.01	$	&	$	19.36	\pm	0.004	$	&	$	18.88	\pm	0.01	$	&	$	0.28	$	&	$	10.96	$	\\
$\rm{L}	38	$	&	\rm	KIDS J143528.88+013055.39	&	$	19.31	$	&	$	22.82	\pm	0.33	$	&	$	20.65	\pm	0.02	$	&	$	19.31	\pm	0.004	$	&	$	18.81	\pm	0.01	$	&	$	0.28	$	&	$	10.91	$	\\
$\rm{L}	39	$	&	\rm	KIDS J143607.24+003902.15	&	$	19.18	$	&	$	22.87	\pm	0.23	$	&	$	20.64	\pm	0.01	$	&	$	19.17	\pm	0.004	$	&	$	18.72	\pm	0.008	$	&	$	0.30	$	&	$	10.92	$	\\
$\rm{L}	40	$	&	\rm	KIDS J143611.55+000718.29	&	$	18.27	$	&	$	21.53	\pm	0.06	$	&	$	19.57	\pm	0.004	$	&	$	18.29	\pm	0.002	$	&	$	17.87	\pm	0.004	$	&	$	0.22	$	&	$	11.06	$	\\
$\rm{L}	41	$	&	\rm	KIDS J143616.24+004801.40	&	$	19.24	$	&	$	22.78	\pm	0.25	$	&	$	20.62	\pm	0.01	$	&	$	19.24	\pm	0.004	$	&	$	18.76	\pm	0.009	$	&	$	0.29	$	&	$	11.08	$	\\
$\rm{L}	42	$	&	\rm	KIDS J143805.25--012729.78	&	$	19.29	$	&	$	22.74	\pm	0.19	$	&	$	20.64	\pm	0.01	$	&	$	19.29	\pm	0.004	$	&	$	18.73	\pm	0.007	$	&	$	0.29	$	&	$	10.94	$	\\
$\rm{L}	43	$	&	\rm	KIDS J144138.27--011840.93	&	$	19.35	$	&	$	23.62	\pm	0.48	$	&	$	20.78	\pm	0.01	$	&	$	19.35	\pm	0.004	$	&	$	18.83	\pm	0.008	$	&	$	0.29	$	&	$	11.00	$	\\
$\rm{L}	44	$	&	\rm	KIDS J144557.12--013510.24	&	$	19.16	$	&	$	22.12	\pm	0.13	$	&	$	20.45	\pm	0.009	$	&	$	19.15	\pm	0.004	$	&	$	18.73	\pm	0.009	$	&	$	0.29	$	&	$	10.92	$	\\
$\rm{L}	45	$	&	\rm	KIDS J144751.78--014927.41	&	$	18.61	$	&	$	21.88	\pm	0.11	$	&	$	19.87	\pm	0.007	$	&	$	18.63	\pm	0.003	$	&	$	18.17	\pm	0.005	$	&	$	0.21	$	&	$	10.93	$	\\
$\rm{L}	46	$	&	\rm	KIDS J144924.11--013845.59	&	$	19.40	$	&	$	22.79	\pm	0.24	$	&	$	20.82	\pm	0.02	$	&	$	19.39	\pm	0.005	$	&	$	18.89	\pm	0.009	$	&	$	0.27	$	&	$	11.01	$	\\
$\rm{L}	47	$	&	\rm	KIDS J145245.48+025321.32	&	$	17.69	$	&	$	20.60	\pm	0.03	$	&	$	18.74	\pm	0.002	$	&	$	17.77	\pm	0.001	$	&	$	17.50	\pm	0.003	$	&	$	0.26	$	&	$	11.18	$	\\
$\rm{L}	48	$	&	\rm	KIDS J145356.13+001849.32	&	$	20.32	$	&	$	23.24	\pm	0.30	$	&	$	22.06	\pm	0.04	$	&	$	20.32	\pm	0.009	$	&	$	19.68	\pm	0.03	$	&	$	0.42	$	&	$	11.16	$	\\
$\rm{L}	49	$	&	\rm	KIDS J145638.63+010933.24	&	$	19.66	$	&	$	23.21	\pm	0.26	$	&	$	21.31	\pm	0.02	$	&	$	19.63	\pm	0.006	$	&	$	19.09	\pm	0.01	$	&	$	0.42	$	&	$	11.18	$	\\
$\rm{L}	50	$	&	\rm	KIDS J153936.50--003904.58	&	$	20.15	$	&	$	21.46	\pm	0.09	$	&	$	20.76	\pm	0.02	$	&	$	20.11	\pm	0.01	$	&	$	19.70	\pm	0.02	$	&	$	0.47	$	&	$	10.99	$	\\
$\rm{L}	51	$	&	\rm	KIDS J154949.48--003655.52	&	$	19.02	$	&	$	19.38	\pm	0.01	$	&	$	19.19	\pm	0.004	$	&	$	19.02	\pm	0.004	$	&	$	18.86	\pm	0.01	$	&	$	0.47	$	&	$	11.30	$	\\
$\rm{L}	52	$	&	\rm	KIDS J155133.16+005709.77	&	$	19.37	$	&	$	24.82	\pm	1.76	$	&	$	20.95	\pm	0.02	$	&	$	19.34	\pm	0.005	$	&	$	18.86	\pm	0.01	$	&	$	0.42	$	&	$	11.29	$	\\
$\rm{L}	53	$	&	\rm	KIDS J220453.48--311200.94	&	$	19.32	$	&	$	22.90	\pm	0.23	$	&	$	20.84	\pm	0.01	$	&	$	19.34	\pm	0.004	$	&	$	18.87	\pm	0.005	$	&	$	0.26	$	&	$	10.96	$	\\
$\rm{L}	54	$	&	\rm	KIDS J231410.93--324101.31	&	$	19.26	$	&	$	22.59	\pm	0.16	$	&	$	20.56	\pm	0.009	$	&	$	19.26	\pm	0.004	$	&	$	18.75	\pm	0.006	$	&	$	0.29	$	&	$	11.01	$	\\
$\rm{L}	55	$	&	\rm	KIDS J233148.39--333402.05	&	$	20.46	$	&	$	24.47	\pm	0.74	$	&	$	22.12	\pm	0.04	$	&	$	20.44	\pm	0.009	$	&	$	19.78	\pm	0.02	$	&	$	0.48	$	&	$	11.09	$	\\
\hline
\end{tabular}
\end{table*}


\newpage
\begin{table*}
\centering \caption{Structural parameters derived running \twodphot\ on $g$-, $r$- and $i$-bands for the 55 in the \UCMGSpec\ sample. For each band we show: a)
circularized effective radius $\Theta_{\rm e}$, measured in arcsec, b)
circularized effective radius $R_{\rm e}$, measured in kpc (calculated
using \zp{} values listed in Table \ref{tab:photometry}), c)
S\'ersic index $n$, d) axis ratio q, e) $\chi^{2}$ of the surface
photometry fit, f) $\chi^{\prime 2}$ of the surface photometry fit
including only central pixels and g) the  signal-to-noise ratio
\SN\ of the photometric images, defined as the inverse of the error on $\texttt{MAG\_AUTO}$.}

\label{tab:struc_parameters_ucmg_liter} 

\hspace{-4.8cm}
\resizebox{1.25\textwidth}{!}{
\begin{tabular}{cccccccccccccccccccccc}
\hline

$ $ & \multicolumn{7}{c}{$g$-band} & \multicolumn{7}{c}{$r$-band} & \multicolumn{7}{c}{$i$-band} \\
\cmidrule(rl){2-8} \cmidrule(rl){9-15} \cmidrule(rl){16-22}

$ID$ & $\Theta_{e}$ & $R_{e}$ & n & q & $\chi^{2}$ & $\chi^{\prime 2}$ & $\SN$ & $\Theta_{e}$ & $R_{e}$ & n & q & $\chi^{2}$ & $\chi^{\prime 2}$ & $\SN$ & $\Theta_{e}$ & $R_{e}$ & n & q & $\chi^{2}$ & $\chi^{\prime 2}$ & $\SN$ \\
\hline
\hline
$\rm{L}	1	$	&	$	0.25	$	&	$	1.07	$	&	$	4.11	$	&	$	0.34	$	&	$	1.02	$	&	$	1.05	$	&	$	123	$	&	$	0.27	$	&	$	1.19	$	&	$	4.54	$	&	$	0.37	$	&	$	1.07	$	&	$	1.35	$	&	$	367	$	&	$	0.29	$	&	$	1.27	$	&	$	5.62	$	&	$	0.39	$	&	$	1.00	$	&	$	0.93	$	&	$	147	$	\\
$\rm{L}	2	$	&	$	0.05	$	&	$	0.24	$	&	$	4.66	$	&	$	0.12	$	&	$	1.01	$	&	$	0.98	$	&	$	32	$	&	$	0.16	$	&	$	0.74	$	&	$	4.03	$	&	$	0.33	$	&	$	1.02	$	&	$	1.01	$	&	$	127	$	&	$	0.24	$	&	$	1.12	$	&	$	2.96	$	&	$	0.42	$	&	$	0.98	$	&	$	0.87	$	&	$	96	$	\\
$\rm{L}	3	$	&	$	0.29	$	&	$	1.12	$	&	$	4.40	$	&	$	0.58	$	&	$	1.03	$	&	$	1.06	$	&	$	190	$	&	$	0.26	$	&	$	1.01	$	&	$	5.59	$	&	$	0.61	$	&	$	1.20	$	&	$	1.72	$	&	$	506	$	&	$	0.33	$	&	$	1.25	$	&	$	8.48	$	&	$	0.68	$	&	$	1.01	$	&	$	0.95	$	&	$	203	$	\\
$\rm{L}	4	$	&	$	0.46	$	&	$	1.36	$	&	$	3.06	$	&	$	0.27	$	&	$	1.01	$	&	$	1.04	$	&	$	165	$	&	$	0.46	$	&	$	1.39	$	&	$	4.38	$	&	$	0.25	$	&	$	1.07	$	&	$	1.42	$	&	$	462	$	&	$	0.45	$	&	$	1.35	$	&	$	3.33	$	&	$	0.27	$	&	$	1.01	$	&	$	0.96	$	&	$	177	$	\\
$\rm{L}	5	$	&	$	0.56	$	&	$	2.65	$	&	$	10.56	$	&	$	0.75	$	&	$	1.02	$	&	$	0.99	$	&	$	94	$	&	$	0.23	$	&	$	1.09	$	&	$	9.84	$	&	$	0.80	$	&	$	1.12	$	&	$	1.87	$	&	$	279	$	&	$	0.22	$	&	$	1.03	$	&	$	9.27	$	&	$	0.73	$	&	$	1.02	$	&	$	1.04	$	&	$	99	$	\\
$\rm{L}	6	$	&	$	0.39	$	&	$	1.44	$	&	$	3.83	$	&	$	0.46	$	&	$	1.02	$	&	$	1.03	$	&	$	185	$	&	$	0.34	$	&	$	1.25	$	&	$	4.13	$	&	$	0.44	$	&	$	1.08	$	&	$	1.47	$	&	$	443	$	&	$	0.34	$	&	$	1.26	$	&	$	4.00	$	&	$	0.42	$	&	$	1.05	$	&	$	1.10	$	&	$	190	$	\\
$\rm{L}	7	$	&	$	0.46	$	&	$	1.45	$	&	$	4.34	$	&	$	0.24	$	&	$	1.05	$	&	$	1.40	$	&	$	492	$	&	$	0.23	$	&	$	0.73	$	&	$	7.04	$	&	$	0.29	$	&	$	1.34	$	&	$	2.89	$	&	$	1003	$	&	$	0.54	$	&	$	1.70	$	&	$	4.82	$	&	$	0.26	$	&	$	1.06	$	&	$	1.32	$	&	$	641	$	\\
$\rm{L}	8	$	&	$	0.56	$	&	$	1.96	$	&	$	9.95	$	&	$	0.81	$	&	$	0.83	$	&	$	0.86	$	&	$	110	$	&	$	0.14	$	&	$	0.48	$	&	$	10.07	$	&	$	0.76	$	&	$	1.13	$	&	$	1.84	$	&	$	357	$	&	$	0.30	$	&	$	1.05	$	&	$	9.97	$	&	$	0.77	$	&	$	1.01	$	&	$	1.00	$	&	$	152	$	\\
$\rm{L}	9	$	&	$	0.41	$	&	$	1.76	$	&	$	1.97	$	&	$	0.34	$	&	$	1.03	$	&	$	1.13	$	&	$	95	$	&	$	0.34	$	&	$	1.46	$	&	$	1.99	$	&	$	0.32	$	&	$	1.04	$	&	$	1.34	$	&	$	351	$	&	$	0.28	$	&	$	1.20	$	&	$	3.02	$	&	$	0.30	$	&	$	1.00	$	&	$	1.04	$	&	$	206	$	\\
$\rm{L}	10	$	&	$	0.43	$	&	$	1.26	$	&	$	2.70	$	&	$	0.29	$	&	$	1.07	$	&	$	11.51	$	&	$	360	$	&	$	0.32	$	&	$	0.95	$	&	$	3.64	$	&	$	0.29	$	&	$	1.12	$	&	$	25.78	$	&	$	1092	$	&	$	0.34	$	&	$	1.01	$	&	$	3.18	$	&	$	0.29	$	&	$	1.03	$	&	$	9.58	$	&	$	464	$	\\
$\rm{L}	11	$	&	$	0.31	$	&	$	1.78	$	&	$	8.80	$	&	$	0.21	$	&	$	0.99	$	&	$	1.10	$	&	$	16	$	&	$	0.25	$	&	$	1.46	$	&	$	8.54	$	&	$	0.44	$	&	$	1.03	$	&	$	0.99	$	&	$	74	$	&	$	0.21	$	&	$	1.22	$	&	$	3.66	$	&	$	0.59	$	&	$	1.01	$	&	$	1.28	$	&	$	32	$	\\
$\rm{L}	12	$	&	$	0.29	$	&	$	1.02	$	&	$	4.03	$	&	$	0.26	$	&	$	1.07	$	&	$	1.03	$	&	$	130	$	&	$	0.14	$	&	$	0.48	$	&	$	7.96	$	&	$	0.27	$	&	$	1.05	$	&	$	1.20	$	&	$	327	$	&	$	0.11	$	&	$	0.40	$	&	$	8.07	$	&	$	0.25	$	&	$	1.02	$	&	$	0.96	$	&	$	188	$	\\
$\rm{L}	13	$	&	$	0.37	$	&	$	1.39	$	&	$	4.79	$	&	$	0.38	$	&	$	1.03	$	&	$	0.99	$	&	$	602	$	&	$	0.20	$	&	$	1.26	$	&	$	6.53	$	&	$	0.40	$	&	$	1.03	$	&	$	1.18	$	&	$	1109	$	&	$	0.26	$	&	$	1.39	$	&	$	8.63	$	&	$	0.38	$	&	$	1.01	$	&	$	0.94	$	&	$	618	$	\\
$\rm{L}	14	$	&	$	0.71	$	&	$	1.47	$	&	$	3.60	$	&	$	0.22	$	&	$	1.12	$	&	$	1.46	$	&	$	140	$	&	$	0.64	$	&	$	0.79	$	&	$	5.26	$	&	$	0.23	$	&	$	1.40	$	&	$	2.32	$	&	$	381	$	&	$	0.70	$	&	$	1.03	$	&	$	3.48	$	&	$	0.24	$	&	$	1.04	$	&	$	1.19	$	&	$	163	$	\\
$\rm{L}	15	$	&	$	0.28	$	&	$	0.93	$	&	$	9.55	$	&	$	0.72	$	&	$	1.08	$	&	$	1.54	$	&	$	275	$	&	$	0.35	$	&	$	1.15	$	&	$	7.85	$	&	$	0.64	$	&	$	1.31	$	&	$	3.19	$	&	$	621	$	&	$	0.45$	&	$	1.50	$	&	$	10.06	$	&	$	0.73	$	&	$	1.04	$	&	$	1.13	$	&	$	239	$	\\
$\rm{L}	16	$	&	$	0.50	$	&	$	1.49	$	&	$	7.65	$	&	$	0.38	$	&	$	1.02	$	&	$	7.99	$	&	$	210	$	&	$	0.45	$	&	$	1.34	$	&	$	7.52	$	&	$	0.41	$	&	$	1.10	$	&	$	23.21	$	&	$	673	$	&	$	0.72	$	&	$	2.14	$	&	$	7.51	$	&	$	0.45	$	&	$	1.04	$	&	$	11.05	$	&	$	357	$	\\
$\rm{L}	17	$	&	$	0.36	$	&	$	1.49	$	&	$	2.64	$	&	$	0.30	$	&	$	1.01	$	&	$	0.91	$	&	$	127	$	&	$	0.35	$	&	$	1.47	$	&	$	2.88	$	&	$	0.28	$	&	$	1.12	$	&	$	1.51	$	&	$	410	$	&	$	0.35	$	&	$	1.46	$	&	$	2.42	$	&	$	0.27	$	&	$	1.01	$	&	$	0.94	$	&	$	128	$	\\
$\rm{L}	18	$	&	$	0.52	$	&	$	1.94	$	&	$	8.65	$	&	$	0.52	$	&	$	1.04	$	&	$	1.14	$	&	$	154	$	&	$	0.38	$	&	$	1.42	$	&	$	7.59	$	&	$	0.61	$	&	$	1.03	$	&	$	1.35	$	&	$	363	$	&	$	0.25	$	&	$	0.93	$	&	$	8.95	$	&	$	0.59	$	&	$	1.04	$	&	$	1.04	$	&	$	193	$	\\
$\rm{L}	19	$	&	$	0.17	$	&	$	1.01	$	&	$	0.69	$	&	$	0.14	$	&	$	0.98	$	&	$	0.95	$	&	$	29	$	&	$	0.20	$	&	$	1.19	$	&	$	3.60	$	&	$	0.51	$	&	$	0.98	$	&	$	0.98	$	&	$	97	$	&	$	0.17	$	&	$	1.04	$	&	$	4.96	$	&	$	0.49	$	&	$	1.00	$	&	$	1.01	$	&	$	69	$	\\
$\rm{L}	20	$	&	$	0.32	$	&	$	1.64	$	&	$	6.76	$	&	$	0.29	$	&	$	1.02	$	&	$	1.21	$	&	$	85	$	&	$	0.26	$	&	$	1.36	$	&	$	7.52	$	&	$	0.33	$	&	$	1.07	$	&	$	1.56	$	&	$	276	$	&	$	0.25	$	&	$	1.27	$	&	$	9.23	$	&	$	0.35	$	&	$	1.02	$	&	$	1.25	$	&	$	115	$	\\
$\rm{L}	21	$	&	$	0.40	$	&	$	1.76	$	&	$	2.80	$	&	$	0.56	$	&	$	1.04	$	&	$	1.07	$	&	$	76	$	&	$	0.30	$	&	$	1.32	$	&	$	3.13	$	&	$	0.45	$	&	$	1.01	$	&	$	1.09	$	&	$	261	$	&	$	0.25	$	&	$	1.10	$	&	$	4.71	$	&	$	0.40	$	&	$	0.99	$	&	$	0.86	$	&	$	75	$	\\
$\rm{L}	22	$	&	$	0.34	$	&	$	1.44	$	&	$	5.00	$	&	$	0.33	$	&	$	0.99	$	&	$	0.93	$	&	$	52	$	&	$	0.32	$	&	$	1.35	$	&	$	6.30	$	&	$	0.39	$	&	$	1.01	$	&	$	1.02	$	&	$	217	$	&	$	0.33	$	&	$	1.41	$	&	$	6.13	$	&	$	0.42	$	&	$	1.02	$	&	$	0.99	$	&	$	66	$	\\
$\rm{L}	23	$	&	$	0.18	$	&	$	0.81	$	&	$	7.66	$	&	$	0.61	$	&	$	1.06	$	&	$	1.24	$	&	$	265	$	&	$	0.28$	&	$	1.21	$	&	$	7.51	$	&	$	0.58	$	&	$	1.28	$	&	$	2.27	$	&	$	507	$	&	$	0.76	$	&	$	3.33	$	&	$	3.62	$	&	$	0.60	$	&	$	1.05	$	&	$	1.17	$	&	$	175	$	\\
$\rm{L}	24	$	&	$	0.38	$	&	$	1.69	$	&	$	3.99	$	&	$	0.46	$	&	$	1.02	$	&	$	1.01	$	&	$	108	$	&	$	0.31	$	&	$	1.40	$	&	$	4.26	$	&	$	0.42	$	&	$	1.04	$	&	$	1.21	$	&	$	316	$	&	$	0.30	$	&	$	1.33	$	&	$	5.03	$	&	$	0.42	$	&	$	0.99	$	&	$	0.89	$	&	$	169	$	\\
$\rm{L}	25	$	&	$	0.31	$	&	$	1.36	$	&	$	5.12	$	&	$	0.81	$	&	$	1.02	$	&	$	0.97	$	&	$	142	$	&	$	0.32	$	&	$	1.41	$	&	$	4.72	$	&	$	0.85	$	&	$	1.04	$	&	$	1.22	$	&	$	383	$	&	$	0.27	$	&	$	1.18	$	&	$	7.84	$	&	$	0.88	$	&	$	1.02	$	&	$	0.96	$	&	$	173	$	\\
$\rm{L}	26	$	&	$	0.54	$	&	$	1.63	$	&	$	3.35	$	&	$	0.35	$	&	$	1.03	$	&	$	1.08	$	&	$	244	$	&	$	0.48	$	&	$	1.47	$	&	$	3.92	$	&	$	0.31	$	&	$	1.07	$	&	$	1.53	$	&	$	555	$	&	$	0.45	$	&	$	1.36	$	&	$	4.74	$	&	$	0.33	$	&	$	1.03	$	&	$	1.08	$	&	$	294	$	\\
$\rm{L}	27	$	&	$	0.22	$	&	$	1.22	$	&	$	3.66	$	&	$	0.52	$	&	$	1.02	$	&	$	1.83	$	&	$	168	$	&	$	0.23	$	&	$	1.30	$	&	$	3.95	$	&	$	0.58	$	&	$	1.02	$	&	$	6.89	$	&	$	399	$	&	$	0.24	$	&	$	1.36	$	&	$	3.15	$	&	$	0.56	$	&	$	1.05	$	&	$	2.85	$	&	$	232	$	\\
$\rm{L}	28	$	&	$	0.19	$	&	$	0.90	$	&	$	3.87	$	&	$	0.15	$	&	$	1.02	$	&	$	0.89	$	&	$	72	$	&	$	0.22	$	&	$	1.02	$	&	$	4.04	$	&	$	0.17	$	&	$	1.01	$	&	$	1.12	$	&	$	237	$	&	$	0.23	$	&	$	1.07	$	&	$	3.67	$	&	$	0.21	$	&	$	1.02	$	&	$	1.04	$	&	$	100	$	\\
$\rm{L}	29	$	&	$	0.37	$	&	$	1.42	$	&	$	6.64	$	&	$	0.64	$	&	$	1.08	$	&	$	1.04	$	&	$	94	$	&	$	0.31	$	&	$	1.23	$	&	$	4.76	$	&	$	0.62	$	&	$	1.03	$	&	$	1.25	$	&	$	299	$	&	$	0.34	$	&	$	1.34	$	&	$	5.67	$	&	$	0.61	$	&	$	1.01	$	&	$	0.94	$	&	$	156	$	\\
$\rm{L}	30	$	&	$	0.29	$	&	$	1.22	$	&	$	1.67	$	&	$	0.82	$	&	$	1.04	$	&	$	1.54	$	&	$	63	$	&	$	0.35	$	&	$	1.48	$	&	$	5.28	$	&	$	0.46	$	&	$	1.05	$	&	$	1.08	$	&	$	206	$	&	$	0.41	$	&	$	1.74	$	&	$	5.10	$	&	$	0.55	$	&	$	1.01	$	&	$	0.90	$	&	$	106	$	\\
$\rm{L}	31	$	&	$	0.28	$	&	$	1.39	$	&	$	7.43	$	&	$	0.35	$	&	$	1.01	$	&	$	1.02	$	&	$	77	$	&	$	0.18	$	&	$	0.89	$	&	$	8.44	$	&	$	0.30	$	&	$	1.50	$	&	$	1.17	$	&	$	244	$	&	$	0.28	$	&	$	1.37	$	&	$	6.47	$	&	$	0.25	$	&	$	1.00	$	&	$	0.94	$	&	$	115	$	\\
$\rm{L}	32	$	&	$	0.16	$	&	$	0.77	$	&	$	9.50	$	&	$	0.70	$	&	$	1.03	$	&	$	1.11	$	&	$	201	$	&	$	0.24	$	&	$	1.16	$	&	$	10.64	$	&	$	0.70	$	&	$	1.26	$	&	$	2.43	$	&	$	327	$	&	$	1.24	$	&	$	5.95	$	&	$	6.79	$	&	$	0.58	$	&	$	1.01	$	&	$	0.99	$	&	$	95	$	\\
$\rm{L}	33	$	&	$	0.29	$	&	$	1.51	$	&	$	5.90	$	&	$	0.87	$	&	$	1.01	$	&	$	0.99	$	&	$	126	$	&	$	0.26	$	&	$	1.36	$	&	$	3.77	$	&	$	0.88	$	&	$	1.08	$	&	$	1.33	$	&	$	368	$	&	$	0.26	$	&	$	1.32	$	&	$	4.10	$	&	$	0.86	$	&	$	1.00	$	&	$	0.91	$	&	$	139	$	\\
$\rm{L}	34	$	&	$	0.06	$	&	$	0.31	$	&	$	7.63	$	&	$	0.92	$	&	$	1.07	$	&	$	1.04	$	&	$	229	$	&	$	0.21	$	&	$	1.11	$	&	$	6.13	$	&	$	0.87	$	&	$	1.03	$	&	$	1.20	$	&	$	345	$	&	$	0.13	$	&	$	0.71	$	&	$	8.56	$	&	$	0.89	$	&	$	1.01	$	&	$	0.91	$	&	$	151	$	\\
$\rm{L}	35	$	&	$	0.26	$	&	$	1.26	$	&	$	4.24	$	&	$	0.70	$	&	$	0.95	$	&	$	0.87	$	&	$	69	$	&	$	0.28	$	&	$	1.36	$	&	$	3.31	$	&	$	0.78	$	&	$	1.02	$	&	$	1.11	$	&	$	272	$	&	$	0.30	$	&	$	1.47	$	&	$	2.89	$	&	$	0.70	$	&	$	1.00	$	&	$	0.90	$	&	$	174	$	\\
$\rm{L}	36	$	&	$	0.27	$	&	$	1.56	$	&	$	2.84	$	&	$	0.29	$	&	$	1.03	$	&	$	1.01	$	&	$	83	$	&	$	0.23	$	&	$	1.37	$	&	$	3.21	$	&	$	0.26	$	&	$	1.25	$	&	$	1.23	$	&	$	297	$	&	$	0.20	$	&	$	1.20	$	&	$	3.29	$	&	$	0.30	$	&	$	1.03	$	&	$	0.96	$	&	$	199	$	\\
$\rm{L}	37	$	&	$	0.17	$	&	$	0.71	$	&	$	6.34	$	&	$	0.53	$	&	$	1.01	$	&	$	0.98	$	&	$	82	$	&	$	0.19	$	&	$	0.84	$	&	$	5.21	$	&	$	0.50	$	&	$	1.02	$	&	$	1.06	$	&	$	249	$	&	$	0.18	$	&	$	0.80	$	&	$	7.52	$	&	$	0.34	$	&	$	1.01	$	&	$	0.98	$	&	$	72	$	\\
$\rm{L}	38	$	&	$	0.39	$	&	$	1.67	$	&	$	4.09	$	&	$	0.39	$	&	$	1.02	$	&	$	0.96	$	&	$	64	$	&	$	0.35	$	&	$	1.49	$	&	$	4.18	$	&	$	0.38	$	&	$	1.00	$	&	$	1.05	$	&	$	232	$	&	$	0.24	$	&	$	1.03	$	&	$	6.96	$	&	$	0.37	$	&	$	1.01	$	&	$	1.06	$	&	$	79	$	\\
$\rm{L}	39	$	&	$	0.36	$	&	$	1.62	$	&	$	3.16	$	&	$	0.27	$	&	$	1.00	$	&	$	1.01	$	&	$	96	$	&	$	0.33	$	&	$	1.47	$	&	$	3.77	$	&	$	0.32	$	&	$	1.10	$	&	$	1.56	$	&	$	311	$	&	$	0.30	$	&	$	1.32	$	&	$	3.25	$	&	$	0.31	$	&	$	1.01	$	&	$	0.90	$	&	$	132	$	\\
$\rm{L}	40	$	&	$	0.40	$	&	$	1.42	$	&	$	2.55	$	&	$	0.20	$	&	$	1.04	$	&	$	1.11	$	&	$	232	$	&	$	0.39	$	&	$	1.39	$	&	$	2.65	$	&	$	0.19	$	&	$	1.15	$	&	$	1.58	$	&	$	597	$	&	$	0.34	$	&	$	1.22	$	&	$	2.77	$	&	$	0.17	$	&	$	1.02	$	&	$	0.97	$	&	$	260	$	\\
$\rm{L}	41	$	&	$	0.51	$	&	$	2.26	$	&	$	5.63	$	&	$	0.53	$	&	$	0.97	$	&	$	0.95	$	&	$	81	$	&	$	0.33	$	&	$	1.47	$	&	$	7.59	$	&	$	0.56	$	&	$	1.03	$	&	$	1.33	$	&	$	255	$	&	$	0.30	$	&	$	1.33	$	&	$	8.73	$	&	$	0.50	$	&	$	0.99	$	&	$	0.91	$	&	$	108	$	\\
$\rm{L}	42	$	&	$	0.37	$	&	$	1.60	$	&	$	4.80	$	&	$	0.37	$	&	$	0.99	$	&	$	1.08	$	&	$	95	$	&	$	0.28	$	&	$	1.19	$	&	$	4.07	$	&	$	0.38	$	&	$	1.02	$	&	$	1.42	$	&	$	259	$	&	$	0.26	$	&	$	1.11	$	&	$	4.11	$	&	$	0.38	$	&	$	1.03	$	&	$	1.46	$	&	$	149	$	\\
$\rm{L}	43	$	&	$	0.37	$	&	$	1.61	$	&	$	6.28	$	&	$	0.28	$	&	$	1.00	$	&	$	0.92	$	&	$	89	$	&	$	0.32	$	&	$	1.40	$	&	$	4.73	$	&	$	0.29	$	&	$	1.03	$	&	$	1.25	$	&	$	246	$	&	$	0.32	$	&	$	1.42	$	&	$	6.48	$	&	$	0.29	$	&	$	1.03	$	&	$	0.90	$	&	$	137	$	\\
$\rm{L}	44	$	&	$	0.32	$	&	$	1.39	$	&	$	6.67	$	&	$	0.73	$	&	$	1.05	$	&	$	1.00	$	&	$	110	$	&	$	0.18	$	&	$	0.79	$	&	$	7.31	$	&	$	0.84	$	&	$	1.02	$	&	$	1.08	$	&	$	263	$	&	$	0.29	$	&	$	1.25	$	&	$	8.04	$	&	$	0.82	$	&	$	1.02	$	&	$	0.92	$	&	$	121	$	\\
$\rm{L}	45	$	&	$	0.44	$	&	$	1.50	$	&	$	2.93	$	&	$	0.47	$	&	$	1.03	$	&	$	1.04	$	&	$	143	$	&	$	0.39	$	&	$	1.31	$	&	$	3.06	$	&	$	0.40	$	&	$	1.06	$	&	$	1.45	$	&	$	349	$	&	$	0.4	$	&	$	1.53	$	&	$	3.08	$	&	$	0.45	$	&	$	1.00	$	&	$	1.00	$	&	$	195	$	\\
$\rm{L}	46	$	&	$	0.35	$	&	$	1.43	$	&	$	5.48	$	&	$	0.23	$	&	$	1.05	$	&	$	1.08	$	&	$	74	$	&	$	0.27	$	&	$	1.12	$	&	$	6.38	$	&	$	0.39	$	&	$	1.06	$	&	$	1.73	$	&	$	216	$	&	$	0.37	$	&	$	1.51	$	&	$	5.81	$	&	$	0.33	$	&	$	1.04	$	&	$	1.16	$	&	$	128	$	\\
$\rm{L}	47	$	&	$	0.31	$	&	$	1.27	$	&	$	8.90	$	&	$	0.67	$	&	$	1.13	$	&	$	1.46	$	&	$	357	$	&	$	0.33	$	&	$	1.33	$	&	$	9.29	$	&	$	0.63	$	&	$	1.19	$	&	$	1.85	$	&	$	694	$	&	$	0.29	$	&	$	1.19	$	&	$	9.37	$	&	$	0.72	$	&	$	1.04	$	&	$	1.07	$	&	$	282	$	\\
$\rm{L}	48	$	&	$	0.22	$	&	$	1.20	$	&	$	6.55	$	&	$	0.33	$	&	$	1.00	$	&	$	0.93	$	&	$	23	$	&	$	0.36	$	&	$	1.99	$	&	$	7.11	$	&	$	0.47	$	&	$	1.01	$	&	$	1.03	$	&	$	109	$	&	$	0.23	$	&	$	1.30	$	&	$	6.66	$	&	$	0.44	$	&	$	1.01	$	&	$	1.00	$	&	$	39	$	\\
$\rm{L}	49	$	&	$	0.29	$	&	$	1.60	$	&	$	5.37	$	&	$	0.54	$	&	$	0.99	$	&	$	1.00	$	&	$	56	$	&	$	0.14	$	&	$	0.78	$	&	$	6.90	$	&	$	0.41	$	&	$	1.04	$	&	$	1.29	$	&	$	198	$	&	$	0.22	$	&	$	1.23	$	&	$	3.24	$	&	$	0.51	$	&	$	1.03	$	&	$	0.93	$	&	$	107	$	\\
$\rm{L}	50	$	&	$	0.08	$	&	$	0.47	$	&	$	6.49	$	&	$	0.52	$	&	$	0.99	$	&	$	0.91	$	&	$	65	$	&	$	0.19	$	&	$	1.11	$	&	$	9.04	$	&	$	0.48	$	&	$	1.04	$	&	$	1.11	$	&	$	114	$	&	$	0.61	$	&	$	3.63	$	&	$	1.10	$	&	$	0.65	$	&	$	1.06	$	&	$	1.20	$	&	$	72	$	\\
$\rm{L}	51	$	&	$	0.06	$	&	$	0.35	$	&	$	5.39	$	&	$	0.64	$	&	$	1.09	$	&	$	1.17	$	&	$	265	$	&	$	0.13	$	&	$	0.74	$	&	$	9.21	$	&	$	0.89	$	&	$	1.04	$	&	$	1.09	$	&	$	272	$	&	$	0.26	$	&	$	1.51	$	&	$	7.55	$	&	$	0.86	$	&	$	1.02	$	&	$	1.00	$	&	$	89	$	\\
$\rm{L}	52	$	&	$	0.14	$	&	$	0.76	$	&	$	6.14	$	&	$	0.28	$	&	$	1.05	$	&	$	1.02	$	&	$	54	$	&	$	0.09	$	&	$	0.51	$	&	$	4.83	$	&	$	0.32	$	&	$	1.04	$	&	$	1.25	$	&	$	239	$	&	$	0.13	$	&	$	0.74	$	&	$	4.45	$	&	$	0.28	$	&	$	1.02	$	&	$	0.96	$	&	$	105	$	\\
$\rm{L}	53	$	&	$	0.34	$	&	$	1.35	$	&	$	6.48	$	&	$	0.34	$	&	$	1.00	$	&	$	0.99	$	&	$	74	$	&	$	0.34	$	&	$	1.38	$	&	$	6.36	$	&	$	0.31	$	&	$	1.05	$	&	$	1.34	$	&	$	282	$	&	$	0.44	$	&	$	1.76	$	&	$	3.91	$	&	$	0.29	$	&	$	1.00	$	&	$	0.98	$	&	$	207	$	\\
$\rm{L}	54	$	&	$	0.36	$	&	$	1.59	$	&	$	4.71	$	&	$	0.46	$	&	$	1.02	$	&	$	0.94	$	&	$	106	$	&	$	0.29	$	&	$	1.29	$	&	$	5.14	$	&	$	0.43	$	&	$	1.04	$	&	$	1.18	$	&	$	286	$	&	$	0.30	$	&	$	1.34	$	&	$	3.52	$	&	$	0.43	$	&	$	1.03	$	&	$	0.97	$	&	$	159	$	\\

$\rm{L}	55	$	&	$	0.81	$	&	$	4.84	$	&	$	9.20	$	&	$	0.76	$	&	$	0.99	$	&	$	1.01	$	&	$	24	$	&	$	0.18	$	&	$	1.06	$	&	$	9.19	$	&	$	0.61	$	&	$	1.01	$	&	$	1.17	$	&	$	114	$	&	$	0.11	$	&	$	0.69	$	&	$	8.62	$	&	$	0.59	$	&	$	1.00	$	&	$	1.01	$	&	$	50	$	\\

\hline
\end{tabular}
}
\end{table*}


\newpage
\newpage
\begin{thebibliography}{99}



\bibitem[{{Ahn} {et~al}\mbox{.}(2012){Ahn}, {Alexandroff}, {Allende Prieto},
  {Anderson}, {Anderton}, {Andrews}, {Aubourg}, {Bailey}, {Balbinot}, {Barnes},
  \& et~al.}]{Ahn+12_SDSS_DR9}
{Ahn} C.~P.,{Alexandroff}, R., {Allende Prieto}, C., {et~al.}, 2012, \apjs, 203, 21


\bibitem[{{Ahn} {et~al}\mbox{.}(2014){Ahn}, {Alexandroff}, {Allende Prieto},
  {Anders}, {Anderson}, {Anderton}, {Andrews}, {Aubourg}, {Bailey}, {Bastien},
  \& et~al.}]{Ahn+14_SDSS_DR10}
{Ahn} C.~P.,{Alexandroff}, R., {Allende Prieto}, C., {et~al.}, 2014, \apjs, 211, 17


\bibitem[{{Arnouts} {et~al}\mbox{.}(1999){Arnouts}, {Cristiani}, {Moscardini},
  {Matarrese}, {Lucchin}, {Fontana}, \& {Giallongo}}]{Arnouts+99}
{Arnouts} S., {Cristiani} S., {Moscardini} L., {et~al.}, 1999, \mnras, 310, 540

\bibitem[{{Beasley} {et~al}\mbox{.}(2018){Beasley}, {Trujillo}, {Leaman}, \&
  {Montes}}]{Beasley+18}
{Beasley} M.~A., {Trujillo} I., {Leaman} R., {Montes} M., 2018,
\nat, 555, 483


\bibitem[{{Bertin} \& {Arnouts}(1996)}]{Bertin_Arnouts96_SEx}
{Bertin} E. \& {Arnouts} S., 1996, \aaps, 117, 393

\bibitem[{{Blake} {et~al}\mbox{.}(2016){Blake}, {Amon}, {Childress}, {Erben},
  {Glazebrook}, {Harnois-Deraps}, {Heymans}, {Hildebrandt}, {Hinton},
  {Janssens}, {Johnson}, {Joudaki}, {Klaes}, {Kuijken}, {Lidman}, {Marin},
  {Parkinson}, {Poole}, \& {Wolf}}]{Blake+16_2dflens}
{Blake}, C., {Amon}, A., {Childress}, M., {et~al.}, 2016, \mnras, 462, 4240

\bibitem[{{Brescia} {et~al}\mbox{.}(2013){Brescia}, {Cavuoti}, {D'Abrusco},
  {Longo}, \& {Mercurio}}]{Brescia+13}
{Brescia} M., {Cavuoti} S., {D'Abrusco} R., {Longo} G., \& {Mercurio}
A., 2013,
  \apj, 772, 140

\bibitem[{{Brescia} {et~al}\mbox{.}(2014){Brescia}, {Cavuoti}, {Longo}, \& {De
  Stefano}}]{Brescia+14}
{Brescia} M., {Cavuoti} S., {Longo} G., {De Stefano} V., 2014,
\aap, 568, A126

\bibitem[{{Bruzual} \& {Charlot}(2003)}]{BC03}
{Bruzual} G. \& {Charlot} S., 2003, \mnras, 344, 1000

\bibitem[{{Buitrago} {et~al}\mbox{.}(2018){Buitrago}, {Ferreras}, {Kelvin},
  {Baldry}, {Davies}, {Angthopo}, {Khochfar}, {Hopkins}, {Driver}, {Brough},
  {Sabater}, {Conselice}, {Liske}, {Holwerda}, {Bremer}, {Phillipps},
  {Lopez-Sanchez}, {Graham}, \& {Norberg}}]{Buitrago+18_compacts}
{Buitrago}, F., {Ferreras}, I., {Kelvin}, L.~S., {et~al.}, 2018, ArXiv e-prints

\bibitem[{{Capaccioli} \& {Schipani}(2011)}]{Capaccioli_Schipani11}
{Capaccioli} M. \& {Schipani} P., 2011, The Messenger, 146, 2

\bibitem[{Cappellari} {et~al}\mbox{.}(2006)]{Cappellari+06}
{Cappellari}, M., {Bacon}, R.,  {Bureau}, M., et al., 2006, \mnras, 366,  1126-1150

\bibitem[{Cappellari} {et~al}\mbox{.}(2012)]{Cappellari+12}
  {Cappellari}, M., {McDermid}, R.~M., {Alatalo}, K., et al., 2012, \nat, 484
485-488


\bibitem[{{Cappellari}(2017)}]{Cappellari17}
{Cappellari} M., 2017, \mnras, 466, 798



\bibitem[{{Cavuoti} {et~al}\mbox{.}(2015{\natexlab{a}}){Cavuoti}, {Brescia},
  {De Stefano}, \& {Longo}}]{Cavuoti+15_PhotoRApToR}
{Cavuoti} S., {Brescia} M., {De Stefano} V., {Longo} G.,
2015{\natexlab{a}},
  Experimental Astronomy, 39, 45

\bibitem[{{Cavuoti} {et~al}\mbox{.}(2015{\natexlab{b}}){Cavuoti}, {Brescia},
  {Tortora}, {Longo}, {Napolitano}, {Radovich}, {Barbera}, {Capaccioli}, {de
  Jong}, {Getman}, {Grado}, \& {Paolillo}}]{Cavuoti+15_KIDS_I}
{Cavuoti}, S., {Brescia}, M., 
  {Tortora}, C., {et~al.}, 2015{\natexlab{b}}, \mnras, 452, 3100

\bibitem[{{Cavuoti} {et~al}\mbox{.}(2017){Cavuoti}, {Tortora}, {Brescia},
  {Longo}, {Radovich}, {Napolitano}, {Amaro}, {Vellucci}, {La Barbera},
  {Getman}, \& {Grado}}]{Cavuoti+17_KiDS}
{Cavuoti}, S., {Tortora}, C., {Brescia}, M., {et~al.}, 2017, \mnras, 466, 2039



\bibitem[{{Chabrier}(2001)}]{Chabrier01}
{Chabrier} G., 2001, \apj, 554, 1274

\bibitem[{Chabrier} {et~al}\mbox{.}(2014)]{Chabrier_2014}
{Chabrier}, G., {Hennebelle}, P. \& {Charlot}, S., 2014, \apj, 2, 75 


\bibitem[{{Charbonnier} {et~al}\mbox{.}(2017){Charbonnier}, {Huertas-Company},
  {Gon{\c c}alves}, {Men{\'e}ndez-Delmestre}, {Bundy}, {Galliano}, {Moraes},
  {Makler}, {Pereira}, {Erben}, {Hildebrandt}, {Shan}, {Caminha}, {Grossi}, \&
  {Riguccini}}]{Charbonnier+17_compact_galaxies}
{Charbonnier}, A., {Huertas-Company}, M.,
  {Gon{\c c}alves}, T.~S., {et~al.}, 2017, \mnras, 469, 4523


\bibitem[{Conroy} {et~al}\mbox{.}(2012)]{Conroy_vanDokkum12a}
   {Conroy}, C. \& {van Dokkum}, P., 2012, \apj, 747, 69


\bibitem[{D'Ago} {et~al}\mbox{.}(2018)]{dago+18proc}
   {D'Ago}, Giuseppe, {Napolitano}, R.~N., {Tortora}, C.,{Spiniello}, Chiara \& {La Barbera}, Francesco, 2018, VST in the Era of the Large Sky Surveys, 15


\bibitem[{{Daddi} {et~al}\mbox{.}(2005){Daddi}, {Renzini}, {Pirzkal},
  {Cimatti}, {Malhotra}, {Stiavelli}, {Xu}, {Pasquali}, {Rhoads}, {Brusa}, {di
  Serego Alighieri}, {Ferguson}, {Koekemoer}, {Moustakas}, {Panagia}, \&
  {Windhorst}}]{Daddi+05}
{Daddi}, E., {Renzini}, A., {Pirzkal}, N., {et~al.}, 2005, \apj, 626, 680



\bibitem[Damjanov {et~al}\mbox{.}(2009)]{Damjanov+09}
Damjanov, I., Abraham, R.~G., McCarthy P.~J., Glazebrook, K., 2009, Bulletin of the American Astronomical Society, 41, 512



\bibitem[Damjanov {et~al}\mbox{.}(2011)]{Damjanov+11}
Damjanov I., Abraham R., Glazebrook K. et al., 2011, ApJL, 739, L44



\bibitem[{{Damjanov} {et~al}\mbox{.}(2015{\natexlab{a}}){Damjanov}, {Geller},
  {Zahid}, \& {Hwang}}]{Damjanov+15_compacts}
{Damjanov} I., {Geller} M.~J., {Zahid} H.~J., {Hwang} H.~S.,
  2015{\natexlab{a}}, \apj, 806, 158

\bibitem[{{Damjanov} {et~al}\mbox{.}(2014){Damjanov}, {Hwang}, {Geller}, \&
  {Chilingarian}}]{Damjanov+14_compacts}
{Damjanov} I., {Hwang} H.~S., {Geller} M.~J., {Chilingarian} I.,
2014, \apj,
  793, 39

\bibitem[{{Damjanov} {et~al}\mbox{.}(2015{\natexlab{b}}){Damjanov}, {Zahid},
  {Geller}, \& {Hwang}}]{Damjanov+15_env_compacts}
{Damjanov} I., {Zahid} H.~J., {Geller} M.~J., {Hwang} H.~S.,
  2015{\natexlab{b}}, ArXiv e-prints

\bibitem[{{Dawson} {et~al}\mbox{.}(2013){Dawson}, {Schlegel}, {Ahn},
  {Anderson}, {Aubourg}, {Bailey}, {Barkhouser}, {Bautista}, {Beifiori},
  {Berlind}, {Bhardwaj}, {Bizyaev}, {Blake}, {Blanton}, {Blomqvist}, {Bolton},
  {Borde}, {Bovy}, {Brandt}, {Brewington}, {Brinkmann}, {Brown}, {Brownstein},
  {Bundy}, {Busca}, {Carithers}, {Carnero}, {Carr}, {Chen}, {Comparat},
  {Connolly}, {Cope}, {Croft}, {Cuesta}, {da Costa}, {Davenport}, {Delubac},
  {de Putter}, {Dhital}, {Ealet}, {Ebelke}, {Eisenstein}, {Escoffier}, {Fan},
  {Filiz Ak}, {Finley}, {Font-Ribera}, {G{\'e}nova-Santos}, {Gunn}, {Guo},
  {Haggard}, {Hall}, {Hamilton}, {Harris}, {Harris}, {Ho}, {Hogg}, {Holder},
  {Honscheid}, {Huehnerhoff}, {Jordan}, {Jordan}, {Kauffmann}, {Kazin},
  {Kirkby}, {Klaene}, {Kneib}, {Le Goff}, {Lee}, {Long}, {Loomis}, {Lundgren},
  {Lupton}, {Maia}, {Makler}, {Malanushenko}, {Malanushenko}, {Mandelbaum},
  {Manera}, {Maraston}, {Margala}, {Masters}, {McBride}, {McDonald}, {McGreer},
  {McMahon}, {Mena}, {Miralda-Escud{\'e}}, {Montero-Dorta}, {Montesano},
  {Muna}, {Myers}, {Naugle}, {Nichol}, {Noterdaeme}, {Nuza}, {Olmstead},
  {Oravetz}, {Oravetz}, {Owen}, {Padmanabhan}, {Palanque-Delabrouille}, {Pan},
  {Parejko}, {P{\^a}ris}, {Percival}, {P{\'e}rez-Fournon},
  {P{\'e}rez-R{\`a}fols}, {Petitjean}, {Pfaffenberger}, {Pforr}, {Pieri},
  {Prada}, {Price-Whelan}, {Raddick}, {Rebolo}, {Rich}, {Richards}, {Rockosi},
  {Roe}, {Ross}, {Ross}, {Rossi}, {Rubi{\~n}o-Martin}, {Samushia},
  {S{\'a}nchez}, {Sayres}, {Schmidt}, {Schneider}, {Sc{\'o}ccola}, {Seo},
  {Shelden}, {Sheldon}, {Shen}, {Shu}, {Slosar}, {Smee}, {Snedden}, {Stauffer},
  {Steele}, {Strauss}, {Streblyanska}, {Suzuki}, {Swanson}, {Tal}, {Tanaka},
  {Thomas}, {Tinker}, {Tojeiro}, {Tremonti}, {Vargas Maga{\~n}a}, {Verde},
  {Viel}, {Wake}, {Watson}, {Weaver}, {Weinberg}, {Weiner}, {West}, {White},
  {Wood-Vasey}, {Yeche}, {Zehavi}, {Zhao}, \& {Zheng}}]{Dawson+13_GAMA}
{Dawson} K.~S. {et~al.}, 2013, \aj, 145, 10



\bibitem[{{de Jong} {et~al}\mbox{.}(2015){de Jong}, {Verdoes Kleijn},
  {Boxhoorn}, {Buddelmeijer}, {Capaccioli}, {Getman}, {Grado}, {Helmich},
  {Huang}, {Irisarri}, {Kuijken}, {La Barbera}, {McFarland}, {Napolitano},
  {Radovich}, {Sikkema}, {Valentijn}, {Begeman}, {Brescia}, {Cavuoti}, {Choi},
  {Cordes}, {Covone}, {Dall'Ora}, {Hildebrandt}, {Longo}, {Nakajima},
  {Paolillo}, {Puddu}, {Rifatto}, {Tortora}, {van Uitert}, {Buddendiek},
  {Harnois-D{\'e}raps}, {Erben}, {Eriksen}, {Heymans}, {Hoekstra}, {Joachimi},
  {Kitching}, {Klaes}, {Koopmans}, {K{\"o}hlinger}, {Roy}, {Sif{\'o}n},
  {Schneider}, {Sutherland}, {Viola}, \& {Vriend}}]{deJong+15_KiDS_paperI}
{de Jong}, J.~T.~A., {Verdoes Kleijn}, G.~A., {Boxhoorn}, D.~R., {et~al.}, 2015, \aap, 582, A62

\bibitem[{{de Jong} {et~al}\mbox{.}(2017){de Jong}, {Kleijn}, {Erben},
  {Hildebrandt}, {Kuijken}, {Sikkema}, {Brescia}, {Bilicki}, {Napolitano},
  {Amaro}, {Begeman}, {Boxhoorn}, {Buddelmeijer}, {Cavuoti}, {Getman}, {Grado},
  {Helmich}, {Huang}, {Irisarri}, {La Barbera}, {Longo}, {McFarland},
  {Nakajima}, {Paolillo}, {Puddu}, {Radovich}, {Rifatto}, {Tortora},
  {Valentijn}, {Vellucci}, {Vriend}, {Amon}, {Blake}, {Choi}, {Conti}, {Gwyn},
  {Herbonnet}, {Heymans}, {Hoekstra}, {Klaes}, {Merten}, {Miller}, {Schneider},
  \& {Viola}}]{deJong+17_KiDS_DR3}
{de Jong}, J.~T.~A., {Kleijn}, G.~A. {Erben}, T., {et~al.}, 2017, \aap, 604, A134


\bibitem[{{Dekel} \& {Burkert}(2014)}]{Dekel_Burkert14}
{Dekel} A., {Burkert} A., 2014, \mnras, 438, 1870

\bibitem[{{Driver} {et~al}\mbox{.}(2011){Driver}, {Hill}, {Kelvin}, {Robotham},
  {Liske}, {Norberg}, {Baldry}, {Bamford}, {Hopkins}, {Loveday}, {Peacock},
  {Andrae}, {Bland-Hawthorn}, {Brough}, {Brown}, {Cameron}, {Ching}, {Colless},
  {Conselice}, {Croom}, {Cross}, {de Propris}, {Dye}, {Drinkwater}, {Ellis},
  {Graham}, {Grootes}, {Gunawardhana}, {Jones}, {van Kampen}, {Maraston},
  {Nichol}, {Parkinson}, {Phillipps}, {Pimbblet}, {Popescu}, {Prescott},
  {Roseboom}, {Sadler}, {Sansom}, {Sharp}, {Smith}, {Taylor}, {Thomas},
  {Tuffs}, {Wijesinghe}, {Dunne}, {Frenk}, {Jarvis}, {Madore}, {Meyer},
  {Seibert}, {Staveley-Smith}, {Sutherland}, \& {Warren}}]{Driver+11_GAMA}
{Driver}, S.~P., {Hill}, D.~T. {Kelvin}, L.~S., {et~al.}, 2011, \mnras, 413, 971

\bibitem[{{Edge} {et~al}\mbox{.}(2014){Edge}, {Sutherland}, \& {The Viking
  Team}}]{Edge+14_VIKING-DR1}
{Edge} A., {Sutherland} W., {The Viking Team}, 2014, VizieR Online
Data
  Catalog, 2329, 0



\bibitem[{Faber} {et~al}\mbox{.}(1976)]{FJ76}
{Faber}, S.~M. \& {Jackson}, R.~E., 1976, \apj, 204, 668-683


\bibitem[{Fasano} {et~al}\mbox{.}(2006)]{Fasano+2006}  
{Fasano}, G., {Marmo}, C. and {Varela}, J., et al., 2006, \aap, 445, 805-871


\bibitem[{{Ferr{\'e}-Mateu} {et~al}\mbox{.}(2012){Ferr{\'e}-Mateu}, {Vazdekis},
  {Trujillo}, {S{\'a}nchez-Bl{\'a}zquez}, {Ricciardelli}, \& {de la
  Rosa}}]{Ferre-Mateu+12}
{Ferr{\'e}-Mateu} A., {Vazdekis} A., {Trujillo} I.,
{S{\'a}nchez-Bl{\'a}zquez}
  P., {Ricciardelli} E., {de la Rosa} I.~G., 2012, \mnras, 423, 632

\bibitem[{{Ferr{\'e}-Mateu} {et~al}\mbox{.}(2015){Ferr{\'e}-Mateu}, {Mezcua},
  {Trujillo}, {Balcells}, \& {van den Bosch}}]{Ferre-Mateu+15}
{Ferr{\'e}-Mateu} A., {Mezcua} M., {Trujillo} I., {Balcells} M.,
{van den
  Bosch} R.~C.~E., 2015, \apj, 808, 79

\bibitem[{{Ferr{\'e}-Mateu} {et~al}\mbox{.}(2017){Ferr{\'e}-Mateu}, {Trujillo},
  {Mart{\'{\i}}n-Navarro}, {Vazdekis}, {Mezcua}, {Balcells}, \&
  {Dom{\'{\i}}nguez}}]{Ferre-Mateu+17}
{Ferr{\'e}-Mateu} A., {Trujillo} I., {Mart{\'{\i}}n-Navarro} I.,
{Vazdekis} A.,
  {Mezcua} M., {Balcells} M., {Dom{\'{\i}}nguez} L., 2017, \mnras, 467, 1929




\bibitem[{{Gargiulo} {et~al}\mbox{.}(2016{\natexlab{a}}){Gargiulo},
  {Bolzonella}, {Scodeggio}, {Krywult}, {De Lucia}, {Guzzo}, {Garilli},
  {Grannet}, {de la Torre}, {Abbas}, {Adami}, {Arnouts}, {Bottini}, {Cappi},
  {Cucciati}, {Davidzon}, {Franzetti}, {Fritz}, {Haines}, {Hawken}, {Iovino},
  {Le Brun}, {Le F{\`e}vre}, {Maccagni}, {Ma{\l}ek}, {Marulli}, {Moutard},
  {Polletta}, {Pollo}, {Tasca}, {Tojeiro}, {Vergani}, {Zanichelli}, {Zamorani},
  {Bel}, {Branchini}, {Coupon}, {Ilbert}, \& {Moscardini}}]{Gargiulo+17_dense}
{Gargiulo}, A., {Bolzonella}, M., {Scodeggio}, M., {et~al.}, 2016{\natexlab{a}}, ArXiv e-prints





\bibitem[{{Guo} {et~al}\mbox{.}(2011){Guo}, {White}, {Boylan-Kolchin}, {De
  Lucia}, {Kauffmann}, {Lemson}, {Li}, {Springel}, \& {Weinmann}}]{Guo+11_sims}
{Guo}, Q., {White}, S., {Boylan-Kolchin}, M., {et~al.}, 2011, \mnras, 413, 101


\bibitem[{{Guo} {et~al}\mbox{.}(2013){Guo}, {White}, {Angulo}, {Henriques},
  {Lemson}, {Boylan-Kolchin}, {Thomas}, \& {Short}}]{Guo+13_sims}
{Guo} Q., {White} S., {Angulo} R.~E., {Henriques} B., {Lemson} G.,
  {Boylan-Kolchin} M., {Thomas} P., {Short} C., 2013, \mnras, 428, 1351




\bibitem[{Hyde} {et~al}\mbox{.}(2009)]{HB09_curv}
{Hyde}, J.~B. \& {Bernardi}, M., 2009, \mnras, 394,  1978-1990


\bibitem[{{Hopkins} {et~al}\mbox{.}(2009){Hopkins}, {Hernquist}, {Cox},
  {Keres}, \& {Wuyts}}]{Hopkins+09_DELGN_IV}
{Hopkins} P.~F., {Hernquist} L., {Cox} T.~J., {Keres} D., {Wuyts}
S., 2009,
  \apj, 691, 1424
  

%


\bibitem[{Hopkins} {et~al}\mbox{.}(2013)]{Hopkins+13}
{Hopkins}, A.~M., {Driver}, S.~P., {Brough}, S., et al., 2013, \mnras, 430, 2047-2066


\bibitem[{{Hsu} {et~al}\mbox{.}(2014){Hsu}, {Stockton}, \&
  {Shih}}]{Hsu+14_compacts}
{Hsu} L.-Y., {Stockton} A., {Shih} H.-Y., 2014, \apj, 796, 92

\bibitem[{{Ilbert} {et~al}\mbox{.}(2006){Ilbert}, {Arnouts}, {McCracken},
  {Bolzonella}, {Bertin}, {Le F{\`e}vre}, {Mellier}, {Zamorani}, {Pell{\`o}},
  {Iovino}, {Tresse}, {Le Brun}, {Bottini}, {Garilli}, {Maccagni}, {Picat},
  {Scaramella}, {Scodeggio}, {Vettolani}, {Zanichelli}, {Adami}, {Bardelli},
  {Cappi}, {Charlot}, {Ciliegi}, {Contini}, {Cucciati}, {Foucaud}, {Franzetti},
  {Gavignaud}, {Guzzo}, {Marano}, {Marinoni}, {Mazure}, {Meneux}, {Merighi},
  {Paltani}, {Pollo}, {Pozzetti}, {Radovich}, {Zucca}, {Bondi}, {Bongiorno},
  {Busarello}, {de La Torre}, {Gregorini}, {Lamareille}, {Mathez}, {Merluzzi},
  {Ripepi}, {Rizzo}, \& {Vergani}}]{Ilbert+06}
{Ilbert}, O., {Arnouts}, S., {McCracken}, H.~J., {et~al.}, 2006, \aap, 457, 841

%
%


\bibitem[{{Komatsu} {et~al}\mbox{.}(2011){Komatsu}, {Smith}, {Dunkley},
  {Bennett}, {Gold}, {Hinshaw}, {Jarosik}, {Larson}, {Nolta}, {Page},
  {Spergel}, {Halpern}, {Hill}, {Kogut}, {Limon}, {Meyer}, {Odegard}, {Tucker},
  {Weiland}, {Wollack}, \& {Wright}}]{Komatsu+11_WMAP7}
{Komatsu}, E., {Smith}, K.~M., {Dunkley}, J., {et~al.}, 2011, \apjs, 192, 18



\bibitem[{{La Barbera} {et~al}\mbox{.}(2008){La Barbera}, {de Carvalho},
  {Kohl-Moreira}, {Gal}, {Soares-Santos}, {Capaccioli}, {Santos}, \&
  {Sant'anna}}]{LaBarbera_08_2DPHOT}
{La Barbera} F., {de Carvalho} R.~R., {Kohl-Moreira} J.~L., {Gal}
R.~R.,
  {Soares-Santos} M., {Capaccioli} M., {Santos} R., {Sant'anna} N., 2008,
  \pasp, 120, 681
  

\bibitem[{{La Barbera} {et~al}\mbox{.}(2010){La Barbera}, {de Carvalho}, {de La
  Rosa}, {Lopes}, {Kohl-Moreira}, \& {Capelato}}]{SPIDER-I}
{La Barbera} F., {de Carvalho} R.~R., {de La Rosa} I.~G., {Lopes}
P.~A.~A.,
  {Kohl-Moreira} J.~L., {Capelato} H.~V., 2010, \mnras, 408, 1313

\bibitem[{{La Barbera} {et~al}\mbox{.}(2013){La Barbera}, {Ferreras},
  {Vazdekis}, {de la Rosa}, {de Carvalho}, {Trevisan}, {Falc{\'o}n-Barroso}, \&
  {Ricciardelli}}]{LaBarbera+13_SPIDERVIII_IMF}
{La Barbera} F., {Ferreras} I., {Vazdekis} A., {de la Rosa} I.~G.,
{de
  Carvalho} R.~R., {Trevisan} M., {Falc{\'o}n-Barroso} J., {Ricciardelli} E.,
  2013, \mnras, 433, 3017

\bibitem[{{L{\"a}sker} {et~al}\mbox{.}(2013){L{\"a}sker}, {van den Bosch}, {van
  de Ven}, {Ferreras}, {La Barbera}, {Vazdekis}, \&
  {Falc{\'o}n-Barroso}}]{Lasker+13_IMF_compact}
{L{\"a}sker} R., {van den Bosch} R.~C.~E., {van de Ven} G.,
 {et~al.}, 2013, \mnras, 434, L31
%

\bibitem[{{Maraston} {et~al}\mbox{.}(2013){Maraston}, {Pforr}, {Henriques},
  {Thomas}, {Wake}, {Brownstein}, {Capozzi}, {Tinker}, {Bundy}, {Skibba},
  {Beifiori}, {Nichol}, {Edmondson}, {Schneider}, {Chen}, {Masters}, {Steele},
  {Bolton}, {York}, {Weaver}, {Higgs}, {Bizyaev}, {Brewington}, {Malanushenko},
  {Malanushenko}, {Snedden}, {Oravetz}, {Pan}, {Shelden}, \&
  {Simmons}}]{Maraston+13_BOSS}
{Maraston}, C., {Pforr}, J., {Henriques}, B.~M., {et~al.}, 2013, \mnras, 435, 2764

\bibitem[{{Mart\'in-Navarro} {et~al}\mbox{.}(2015){Mart\'in-Navarro}, {La
  Barbera}, {Vazdekis}, {Ferr{\'e}-Mateu}, {Trujillo}, \&
  {Beasley}}]{Martin-Navarro+15_IMF_relic}
{Mart\'in-Navarro} I., {La Barbera} F., {Vazdekis} A., {et~al.}, 2015, \mnras, 451, 1081

%


\bibitem[Oser {et~al}\mbox{.}(2010)]{Oser 2010}
Oser L.,  Ostriker J. P.,  Naab T.,  Johansson, P. H., Burkert, A. , 2010, ApJ, 725, 2312







\bibitem[{Poggianti} {et~al}\mbox{.}(2013a)]{Poggianti+13_evol}
{Poggianti}, B.~M., {Moretti}, A., {Calvi}, R. et al., 2013a, \apj, 77,125


\bibitem[{Poggianti} {et~al}\mbox{.}(2013c)]{Poggianti+13_low_z}
  {Poggianti}, B.~M., {Calvi}, R., {Bindoni}, et al., 2013c, \apj, 762, 77


\bibitem[{{Quilis} \& {Trujillo} (2013)}]{Quilis_Trujillo13}
{Quilis}, V. \& {Trujillo}, I., 2013, \apjl, 773, L8

\bibitem[{{Roy} {et~al}\mbox{.}(2018){Roy}, {Napolitano}, {La Barbera},
  {Tortora}, {Getman}, {Radovich}, {Capaccioli}, {Brescia}, {Cavuoti}, {Longo},
  {Raj}, {Puddu}, {Covone}, {Amaro}, {Vellucci}, {Grado}, {Kuijken}, {Verdoes
  Kleijn}, \& {Valentijn}}]{Roy+18}
{Roy}, N., {Napolitano}, N.~R., {La Barbera}, F., {et~al.}, 2018, \mnras, 80, 1

\bibitem[{{Saulder} {et~al}\mbox{.}(2015){Saulder}, {van den Bosch}, \&
  {Mieske}}]{Saulder+15_compacts}
{Saulder} C., {van den Bosch} R.~C.~E., {Mieske} S., 2015, \aap,
578, A134

\bibitem[{{Schlafly} \& {Finkbeiner}(2011)}]{Schlafly_Finkbeiner11}
{Schlafly} E.~F., {Finkbeiner} D.~P., 2011, \apj, 737, 103

%

\bibitem[{{Shih} \& {Stockton}(2011)}]{Shih_Stockton11}
{Shih} H.-Y., {Stockton} A., 2011, \apj, 733, 45


\bibitem[{Spiniello} {et~al}\mbox{.}(2012)]{Spiniello+12}
{Spiniello}, C., {Trager}, S.~C., {Koopmans}, L.~V.~E. \& {Chen}, Y.~P., 2012, \apjl, 753, L32

\bibitem[{Spiniello} {et~al}\mbox{.}(2014)]{Spiniello+14}
   {Spiniello}, C., {Trager}, S., {Koopmans}, L.~V.~E. \&
    {Conroy}, C., 2014,  \mnras, 438,  1483-1499
 

\bibitem[{Spiniello} {et~al}\mbox{.}(2015)]{Spiniello+15_IMF_vs_density}
{Spiniello}, C., {Barnab{\`e}}, M., {Koopmans}, L.~V.~E. \&
    {Trager}, S.~C., 2015, \mnras, 452,  L21-L25




\bibitem[{{Stockton} {et~al}\mbox{.}(2014){Stockton}, {Shih}, {Larson}, \&
  {Mann}}]{Stockton+14_compacts}
{Stockton} A., {Shih} H.-Y., {Larson} K., {Mann} A.~W., 2014,
\apj, 780, 134

\bibitem[{{Stringer} {et~al}\mbox{.}(2015){Stringer}, {Trujillo}, {Dalla
  Vecchia}, \& {Martinez-Valpuesta}}]{Stringer+15_compacts}
{Stringer} M., {Trujillo} I., {Dalla Vecchia} C.,
{Martinez-Valpuesta} I.,
  2015, \mnras, 449, 2396

\bibitem[{{Taylor} {et~al}\mbox{.}(2010){Taylor}, {Franx}, {Glazebrook},
  {Brinchmann}, {van der Wel}, \& {van Dokkum}}]{Taylor+10_compacts}
{Taylor} E.~N., {Franx} M., {Glazebrook} K., {Brinchmann} J., {van
der Wel} A.,
  {van Dokkum} P.~G., 2010, \apj, 720, 723

\bibitem[{{Thomas} {et~al}\mbox{.}(2005){Thomas}, {Maraston}, {Bender}, \&
  {Mendes de Oliveira}}]{Thomas+05}
{Thomas} D., {Maraston} C., {Bender} R., {Mendes de Oliveira} C.,
2005, \apj,
  621, 673


\bibitem[{Toft} {et~al}\mbox{.}(2012)]{Toft+12}
 {Toft}, S., {Gallazzi}, A., {Zirm}, A. et al., 2012, \apj, 754, 3  

%
  


\bibitem[{{Tortora} {et~al}\mbox{.}(2013){Tortora}, {Romanowsky}, \&
  {Napolitano}}]{TRN13_SPIDER_IMF}
{Tortora} C., {Romanowsky} A.~J., {Napolitano} N.~R., 2013, \apj,
765, 8

  \bibitem[{{Tortora} {et~al}\mbox{.}(2014){Tortora}, {Napolitano}, {Saglia},
  {Romanowsky}, {Covone}, \& {Capaccioli}}]{Tortora+14_DMevol}
{Tortora} C., {Napolitano} N.~R., {Saglia} R.~P., {et~al.}, 2014, \mnras, 445, 162

\bibitem[{{Tortora} {et~al}\mbox{.}(2016){Tortora}, {La Barbera}, {Napolitano},
  {Roy}, {Radovich}, {Cavuoti}, {Brescia}, {Longo}, {Getman}, {Capaccioli},
  {Grado}, {Kuijken}, {de Jong}, {McFarland}, \&
  {Puddu}}]{Tortora+16_compacts_KiDS}
{Tortora}, C., {Napolitano}, N.~R., {La Barbera}, F., {et~al.}, 2016, \mnras, 457, 2845



\bibitem[{{Tortora} {et~al}\mbox{.}(2018){Tortora}, {Napolitano}, {Roy},
  {Radovich}, {Getman}, {Koopmans}, {Verdoes Kleijn}, \&
  {Kuijken}}]{Tortora+18_KiDS_DMevol}
{Tortora}, C., {Napolitano}, N.~R., {Roy}, N. , {et~al.}, 2018a, \mnras,
  473, 969


\bibitem[{Tortora} {et~al}\mbox{.}(2018b)]{Tortora+18_UCMGs}
 {{Tortora}, C., {Napolitano}, N.~R., {Spavone}, M.}, 2018b, \mnras, 481, 4728-4752


\bibitem[{{Trenti} \& {Stiavelli}(2008)}]{Trenti_Stiavelli08}
{Trenti} M. \& {Stiavelli} M., 2008, \apj, 676, 767

\bibitem[{{Trujillo} {et~al}\mbox{.}(2006){Trujillo}, {F{\"o}rster Schreiber},
  {Rudnick}, {Barden}, {Franx}, {Rix}, {Caldwell}, {McIntosh}, {Toft},
  {H{\"a}ussler}, {Zirm}, {van Dokkum}, {Labb{\'e}}, \& {}}]{Trujillo+06}
{Trujillo}, I., {F{\"o}rster Schreiber}, N.~M., {Rudnick}, G., {et~al.}, 2006, \apj, 650, 18




\bibitem[{{Trujillo} {et~al}\mbox{.}(2009){Trujillo}, {Cenarro}, {de
  Lorenzo-C{\'a}ceres}, {Vazdekis}, {de la Rosa}, \&
  {Cava}}]{Trujillo+09_superdense}
{Trujillo} I., {Cenarro} A.~J., {de Lorenzo-C{\'a}ceres} A., {et~al.}, 2009, \apjl, 692, L118


\bibitem[{{Trujillo} {et~al}\mbox{.}(2014){Trujillo}, {Ferr{\'e}-Mateu},
  {Balcells}, {Vazdekis}, \& {S{\'a}nchez-Bl{\'a}zquez}}]{Trujillo+14}
{Trujillo} I., {Ferr{\'e}-Mateu} A., {Balcells} M., {Vazdekis} A.,
  {S{\'a}nchez-Bl{\'a}zquez} P., 2014, \apjl, 780, L20




\bibitem[{{Valentinuzzi} {et~al}\mbox{.}(2010){Valentinuzzi}, {Fritz},
  {Poggianti}, {Cava}, {Bettoni}, {Fasano}, {D'Onofrio}, {Couch}, {Dressler},
  {Moles}, {Moretti}, {Omizzolo}, {Kj{\ae}rgaard}, {Vanzella}, \&
  {Varela}}]{Valentinuzzi+10_WINGS}
{Valentinuzzi}, T., {Fritz}, J.,
  {Poggianti}, B.~M, {et~al.}, 2010, \apj, 712, 226

\bibitem[{van den Bosch} {et~al}\mbox{.}(2012)]{vdb_2012}
{van den Bosch}, R.~C.~E., {Gebhardt}, K., {G\"ultekin}, K. et al., 2012, {Nature}, 491,  729--731

\bibitem[{van den Bosch} {et~al}\mbox{.}(2015)]{Bosch_2015}
{van den Bosch}, R.~C.~E., {Gebhardt}, K., {G\"ultekin}



\bibitem[{{vanDokkum} {et~al}\mbox{.}(2009){van Dokkum}, {Kriek}, \& {Franx}}]{vanDokkum+9}
{van Dokkum}, P.~G., {Kriek}, M. \& {Franx}, M., 2009, \nat, 460, 717



\bibitem[{{van Dokkum} {et~al}\mbox{.}(2010){van Dokkum}, {Whitaker},
  {Brammer}, {Franx}, {Kriek}, {Labb{\'e}}, {Marchesini}, {Quadri}, {Bezanson},
  {Illingworth}, {Muzzin}, {Rudnick}, {Tal}, \& {Wake}}]{vanDokkum+10}
{van Dokkum}, P.~G., {Whitaker}, K.~E.,
  {Brammer}, G., {et~al.}, 2010, \apj, 709, 1018

\bibitem[{{Vazdekis} {et~al}\mbox{.}(2010){Vazdekis},
  {S{\'a}nchez-Bl{\'a}zquez}, {Falc{\'o}n-Barroso}, {Cenarro}, {Beasley},
  {Cardiel}, {Gorgas}, \& {Peletier}}]{Vazdekis10}
{Vazdekis} A., {S{\'a}nchez-Bl{\'a}zquez} P., {Falc{\'o}n-Barroso}
J., {et~al.}, 2010, \mnras, 404, 1639

%


\bibitem[{Wellons} {et~al}\mbox{.}(2015)]{Wellons+15_z2}
 {Wellons}, S., {Torrey}, P., {Ma}, C.~P., et~al., 2015, \mnras, 449, 361-372


\bibitem[{{Wellons} {et~al}\mbox{.}(2016){Wellons}, {Torrey}, {Ma},
  {Rodriguez-Gomez}, {Pillepich}, {Nelson}, {Genel}, {Vogelsberger}, \&
  {Hernquist}}]{Wellons+15_lower_z}
{Wellons}, S., {Torrey}, P., {Ma}, C.~P., et al., 2016, \mnras, 456, 1030

\bibitem[{Werner} {et~al}\mbox{.}(2018)]{Werner+18} 
Werner N., Lakhchaura K., Canning R.~E.~A., Gaspari M., \& Simionescu A., 2018, \mnras, 477, 3886-3891

\bibitem[{Wright} {et~al}\mbox{.}(2018)]{Wright+18}
{Wright}, A.~H., {Hildebrandt}, H., {Kuijken}, K ., {et~al.}, 2018, arXiv e-prints, 1812.06077


\bibitem[{{Y{\i}ld{\i}r{\i}m} {et~al}\mbox{.}(2015){Y{\i}ld{\i}r{\i}m}, {van
  den Bosch}, {van de Ven}, {Husemann}, {Lyubenova}, {Walsh}, {Gebhardt}, \&
  {G{\"u}ltekin}}]{Yildirim+15}
{Y{\i}ld{\i}r{\i}m} A., {van den Bosch} R.~C.~E., {van de Ven} G., {et~al.}, 2015,
  \mnras, 452, 1792


\end{thebibliography}
\end{document}